\documentclass{article}
\usepackage{amsmath}
\usepackage{amssymb}
\usepackage{txfonts}
\usepackage{amsfonts}
\usepackage{latexsym,bm}
\usepackage{overcite}
\usepackage{array}
\usepackage{contour, color, CJK}
\usepackage{multirow}
\usepackage{graphics}
\usepackage{graphicx}
\usepackage{graphicx}
\usepackage[includemp,body={398pt,550pt},footskip=30pt,%
            marginparwidth=60pt,marginparsep=10pt]{geometry}
\usepackage{graphicx}
\usepackage{subfigure}
\usepackage{cite}
\usepackage{bm}
\usepackage{booktabs}
\usepackage{subfigure}

\newtheorem{algorithm}{Algorithm}[section]

\date{  } % ......... Newcommand .................

\newcommand{\no}{\noindent}

\begin{document}
\begin{center}

{\LARGE\bf An explicit multistep method for the Wigner problem} {\vskip 0.5cm}Yunfeng Xiong \footnote{
Email addresses: xiongyf@zju.edu.cn.\\
.} \\
  \small (Department of Mathematics, Zhejiang University, Hangzhou 310027, Zhejiang, P.R.China)\\

 \end{center}{\vskip 0.5cm}

% .......... the text...................

{\no\bf Abstract:}\hspace{0.2cm} An explicit multistep scheme is proposed for solving the initial-value Wigner problem. In this scheme, the integrated form of the Wigner equation is approximated by extrapolation or interpolation polynomials on backwards characteristics, and the pseudo-differential operator is tackled by the spectral collocation method. Since it exploits the exact Lagrangian advection, the time stepping of the multistep scheme is not restricted by the CFL-type condition. It is also demonstrated that the calculations of the Wigner potential can be carried out by two successive FFTs, thereby reducing the computational complexity dramatically. Numerical examples illustrating its accuracy are presented.         

{\no\bf Keywords:}\hspace{0.2cm} Wigner equation; spectral collocation method;  Adams multistep scheme; quantum transport.

%{\no\bf Mathematics Subject Classifications:}\hspace{0.2cm} 65F10,
%65W05

\section{Introduction}
\hspace{0.5cm}The progressive miniaturization of semiconductor devices, and the use of bulk materials other than silicon, necessitate the use of a wide variety of model in semiconductor device simulation\cite{RC}. Among various of quantum mechanical models, the Wigner representation \cite{WE} is a useful tool to describe the quantum transport of charged particles in a solid state medium. Although it is not a real probability function, due to possible negative values, the Wigner function serves the role of a distribution \cite{TV}. Hence it is able to predict macroscopically measurable quantities, such as currents and heat fluxes. Recently, the Wigner function has also been widely applied in non-equilibrium quantum statistical mechanics, optics and the density functional theory \cite{TV,SDW}.

Numerical methods for solving the Wigner problem have been greatly developed in past few decades. The first-order upwind finite difference method (FDM) was first employed by Frensley to simulate the resonant tunneling diode (RTD), with the inflow boundary conditions in open quantum system\cite{FWR}. This was then adapted by Ringhofer, by proposing the spectral collocation method to discretize the pseudo-differential operator\cite{RS}. The operator splitting scheme, first used by plasma physicists to study the Landau damping phenomena of a quantum system\cite{SFB}, was generalized to the Wigner-Poisson system and analyzed thoroughly by Arnold and Ringhofer\cite{ARO}. Several advanced numerical techniques, like adaptive mesh and numerical conservative laws, were also employed in solving the linear collisionless Wigner equation\cite{SLC}. 

However, solving the high dimensional Wigner problem through grid-based methods is still problematic, due to the dramatic growth of sampling points in full phase space\cite{BBA}. It will also cause severe numerical errors when discretizing the hyperbolic operator by finite difference techniques, since the Wigner function oscillates rapidly in phase space\cite{SDS,SDW}. In recent years, a particle-based approach, termed the Particle Monte Carlo (PMC) method, has burgeoned with the developments of the particle affinity and effective quantum potential\cite{SFP}. The PMC method doesn't suffer from the problem connected to the diffusion term. Besides, it allows parallel and distributed implementation, thereby facilitating the device simulations of electron-electron interactions in three dimension\cite{HRA}. On the other hand, the particle-based methods may have some inherent statistical noise, due to the finite number of super-particles. Therefore, it necessitates an efficient grid-based numerical solver, along with a proper treatment of the hyperbolic operator. 

An appropriate formulation of boundary conditions is a major problem in the application of Wigner model. The inflow boundary conditions have been reported to cause spurious numerical reflections of outgoing wave packets\cite{AAN}, which can be resolved by absorbing boundary conditions\cite{AAM}. In particle-based methods, the setting of affinity introduces absorbing boundary conditions in a very easy fashion. Besides, for the coupled Wigner-Poisson problem,  it's not trivial to devise a reasonable boundary condition for the self-consistent electrostatic field, since it should satisfy the requirement that the system asymptotically approaches charge neutrality.  Until recently, there is not a systematic study on how to formulate a reasonable boundary condition for the electrostatic field for grid-based Wigner solvers. In fact, the quantum transport equation is a Cauchy problem, thus one should handle unphysical phenomena carefully when introducing an artificial boundary condition. An ideal grid-based solver should be devised for an initial-value problem and compatible with different types of boundary conditions.

The main purpose of this paper is to derive an explicit multistep scheme for the initial-value Wigner problem, which is an extension of semi-Lagrangian scheme\cite{SRBG}. It exploits the $C_{0}-$semigroup generated by the diffusion term, instead of approximating it by finite difference techniques. The intuition comes from the fact that the Wigner equation can be represented as an abstract ODE, therefore several multistep ODE solvers might be adapted to deal with the quantum transport. The smooth part of the pseudo-differential operator is tackled by the spectral collocation method, while the collision term is approximated by numerical integration techniques. It is demonstrated that the cost of computing the Wigner potential can be reduced dramatically via the fast Fourier transform, thereby facilitating its application in high dimensional case. In addition, an explicit scheme allows parallel and distributed implementation, since all the calculations can be carried out independently. The accuracy of the multistep scheme is demonstrated by simulating the motion of a Gaussian wave packet in several potential barriers, that has been studied in \cite{SLC}. 

The rest of the paper is organized as follows. In Section 2, the Wigner equation and the modeling of quantum transport are briefly reviewed. The explicit multistep scheme for the Wigner problem is demonstrated in Section 3, along with the spectral collocation method. Numerical results are illustrated in Section 4, with a conclusion given in Section 5.   

\section{An introduction to the Wigner equation}
\hspace{0.5cm}
We briefly review the Wigner equation and modeling of quantum plasma. For convenience, we adopt the same notations as in \cite{RC}. Our discussion is independent of the dimension $d\left(d=1,2,3\right)$, as the Wigner equation allows a reduction in dimension.

The Wigner function $w\left(x,k,t\right)$ is defined by the Weyl-Wigner transformation of the density matrix for mixed states \cite{RC,TV},
\begin{equation}
\begin{split}
&\rho\left(r,s,t\right)=\sum_{j}\rho\left(\omega_{j}\right)\psi_{j}\left(r,t\right)^{*}\psi_{j}\left(s,t\right),\\
&w\left(x,k,t\right)=\left(2\pi\right)^{-d}\int_{\mathbb{R}^{d}} d\eta ~~\rho \left(x+\frac{\eta}{2}, x-\frac{\eta}{2},t\right)e^{-i\eta \cdot k},\\
\end{split}
\end{equation}
which satisfies the Fourier transformed quantum Liouville equation, referred to as the (collisionless) Wigner equation
\begin{equation}
\begin{split}
&\partial_{t}w+\frac{\hbar}{m}k \cdot \nabla_{x} w+ \theta\left[V\right] w=0,\\
&\theta\left[V\right]=\delta V\left(x,\frac{1}{2i}\nabla_{k}\right), \quad \delta V\left(x,\frac{\eta}{2}\right)=V\left(x+\frac{\eta}{2}\right)-V\left(x-\frac{\eta}{2}\right),\\
\end{split}
\end{equation}
where $\hbar$ is the reduced Planck constant and $\theta\left[V\right]$ is termed pseudo-differential operator. It is convenient to derive the spectral representation of pseudo-differential operator  through the Stone-Weierstrass theorem.

An equivalent representation of $\theta\left[V\right]$ is given by
\begin{equation}
\theta \left[V\right] w\left(x,k,t\right) =\left(2 \pi\right)^{-d} \int_{\mathbb{R}^{d}} d \bar{k} \int_{\mathbb{R}^{d}}d\eta~~\frac{i}{\hbar} \delta V\left(x,\frac{\hbar}{2}\eta,t\right)w\left(x, \bar{k}, t\right)exp\left(i\eta \cdot \left(k-\bar{k}\right)\right),
\end{equation}
In practice, Eq.(3) is usually approximated by numerical integration techniques.

In modeling the electron plasma in metal,  it is necessary to include the scattering processes of electrons with phonons quantum mechanically \cite{MGH}. The Levinson's formalism of interaction terms properly introduces the intracollisional field effect, while the transformation is entirely nontrivial\cite{GGK}. In real simulations, there are two classical approaches to formulating the scattering effect, namely, the relaxation time model and the Fokker-Planck model. 

The relaxation time model is expressed as
\begin{equation}
Q\left(w\right)=\frac{1}{\tau}\left(\frac{n}{n_{0}}w_{0}-w\right),\quad n\left(x,t\right)=\int_{\mathbb{R}^{d}}dk~~w\left(x,k,t\right), \quad n_{0}\left(x,t\right)=\int_{\mathbb{R}^{d}}dk~~w_{0}\left(x,k\right),
\end{equation}
which lumps all dissipation process into one macroscopic parameter: the relaxation time $\tau$.

The Fokker-Planck term model is given by
\begin{equation}
Q\left(w\right)=\frac{1}{\tau} div_{k} \left(\frac{mT_{0}}{\hbar^{2}}\nabla_{k}w+kw\right),
\end{equation}
where $T_{0}$ denote the lattice temperature.

In addition, it is reasonable to include the self-consistent electrostatic potential when simulating RTDs, which be achieved by coupling Eq.(2) with a Poisson equation
\begin{equation}
\Delta_{x}V^{self}\left(x,t\right)=\int_{\mathbb{R}^{d}}dk~~w\left(x,k,t\right)-D\left(x\right),
\end{equation} 
where $D\left(x\right)$ denotes the doping concentration.    

After changing the time scale (let $\frac{\hbar}{m}=1$), we arrive at the reduced collisional Wigner-Poisson equation, the quantum analogue to the Vlasov-Poisson model \cite{GRT}, 
\begin{equation}
\begin{split}
&\partial_{t}w+k \cdot \nabla_{x} w+ \theta\left[V\right] w=Q\left(w\right), \quad \theta\left[V\right]= V\left(x+\frac{1}{2i}\nabla_{k}\right)-V\left(x-\frac{1}{2i}\nabla_{k}\right),\\
&V=V^{ext}+V^{self},\quad \Delta_{x}V^{self}\left(x,t\right)=\int_{\mathbb{R}^{d}}dk~~w\left(x,k,t\right)-D\left(x\right), \\
&Q\left(w\right)=\frac{1}{\tau}\left(\frac{n}{n_{0}}w_{0}-w\right), \quad w\left(x,k,0\right)=w_{0}\left(x,k\right).
\end{split}
\end{equation}

The existence and uniqueness of a global classical solution of collisionless Wigner-Poisson equation (namely, ignoring $Q\left(\omega\right)$ term) is given by Brezzi and Markowich \cite{BM}, via the reformulation of the quantum transport problem as a system of countably many Schr\"{o}dinger equations coupled to a Poisson equation. For more details of Wigner function, one can refer to \cite{TV}. 

\section{Numerical scheme}
\hspace{0.5cm}In this section, we mainly discuss the numerical scheme of solving Eq.(7). It is observed that the second term (the diffusion term) is a simple hyperbolic operator, while the pseudo-differential operator is nonlocal and does not possess classical characteristics. Traditional numerical scheme can be roughly summarized as the following three steps:

(1) Transform Eq.(7) into a hyperbolic system through discretization in $k-$direction;

(2) Use the finite difference/element/volume method to tackle the hyperbolic operator $k \cdot \nabla_{x}$;

(3) Use an implicit-explicit ODE solver to integrate the resulting dynamical system.

The pseudo-differential operator is discretized by the spectral collocation method or the numerical integration formulas, and the resulting dynamical system is tackled by either implicit finite/element/volume method or spectral method\cite{SLC,LMN}. An implicit treatment is much more preferable, since the dynamical system is always a stiff problem. The resulting set of linear equations is solved by iterative Newton methods.

This approach, however, is very demanding for high dimensional problems, since the number of grid points increases dramatically and the coefficient matrix of the linear equations is extremely large. Therefore, the explicit methods are much more useful in solving the high dimensional problem owing to their lower computational complexity. It is also found that the Wigner function oscillates rapidly in phase space due to the quantum interference. The step size in $x-$direction should be sufficiently small, otherwise severe numerical errors will be observed. 
 
An alternative way of solving Eq.(7) is derived from its integrated form (or the mild solution). Under the spectral representation, the reduced hyperbolic system can be rewritten as an abstract ODE (or its mild solution). Assume that the integrand function is continuous with respect to $t$, then the integral can be approximated by extrapolation polynomials on the nodes of backward characteristics. Before discussing the explicit multistep scheme, we first turn to the spectral representation of the Wigner equation.    

\subsection*{A. Spectral collocation method}
\hspace{0.5cm}The spectral collocation method, proposed by Ringhofer \cite{RS, RC}, is based on the fact that the plane waves are the eigenfunctions of the pseudo-differential operator associated with the smooth Wigner potentials. It has been demonstrated the spectral collocation method is well-posed and convergent\cite{RS}, with the assumption that $w$ and $V\left(x\right)$ have sufficient regularities and $w\left(x,k,t\right)$ has a compact support. 

Following Ringhofer, assume that $w\left(x,k,t\right)$ has a compact support in $k \in \left[-\frac{\pi}{\alpha},\frac{\pi}{\alpha}\right]^{d}$, then we can approximate the Wigner function by trigonometric polynomial of the form in $L^{2}$ space,
 \begin{equation}
 w\approx w_{N}\left(x,k,t\right)=\sum_{n\in \mathbb{N}}c\left(x,n,t\right) \phi_{n}\left(k\right), ~~\mathbb{N}=\left\{-N, \cdots ,N\right\}^{d},~~N=2^{m}.
 \end{equation}
 
 The plane wave basis is given by
 \begin{equation}
 \phi_{n}\left(k\right)=\left(\frac{\alpha}{2\pi}\right)^{d/2}exp\left(i\alpha n \cdot k\right),
 \end{equation}
 which satisfies 
 \begin{equation}
 \int_{\left[-\frac{\pi}{\alpha},\frac{\pi}{\alpha}\right]}\phi_{m}^{*}\phi_{n}= \delta\left(m-n\right), \quad
 \left(\Delta k\right)^{d} \sum_{s\in \mathbb{N}} \phi_{m}\left(k_{s}\right)^{*}\phi_{n}\left(k_{s}\right)=\delta_{N}\left(m-n\right),
 \end{equation}
where $\delta_{\mathbb{N}}$ is a Kronecker $\delta$ with period $N$, $\Delta k=\displaystyle{\frac{\pi}{N\alpha}}$ and $\displaystyle{k_{s}=\frac{s  \pi}{N\alpha}}$.

Since
 \begin{equation}
 c\left(x,n,t\right)=\left(\Delta k\right)^{d} \sum_{s\in \mathbb{N}} \phi_{n}^{*}\left(k_{s}\right)w_{N}\left(x,k_{s},t\right),
 \end{equation} 
 $\theta\left[V\right]w$ can be approximated by
 \begin{equation}
B\left(x,k_{m},t\right)= \theta\left[V\right] w_{N}\left(x,k_{m},t\right)=i\frac{\left(\Delta k\right)^{d}}{\hbar}\sum_{n\in \mathbb{N}}\sum_{s\in \mathbb{N}} \delta V\left(x,\frac{\alpha \hbar}{2}n,t\right)w_{N}\left(x,k_{s},t\right)\phi_{n}^{*}\left(k_{s}\right)\phi_{n}\left(k_{m}\right),
 \end{equation}
 where $B$ is called a tensor matrix.
 
 When an explicit method is used, the computation of tensor matrix $B$ can be carried out by two successive FFTs, for
 \begin{equation}
B\left(x,k_{m},t\right)= \theta\left[V\right] w_{N}\left(x,k_{m},t\right)=i\frac{\left(\Delta k\right)^{d}}{\hbar}\sum_{n\in \mathbb{N}}\left[\sum_{s\in \mathbb{N}} w_{N}\left(x,k_{s},t\right)\phi_{n}^{*}\left(k_{s}\right)\right]\delta V\left(x,\frac{\alpha \hbar}{2}n,t\right)\phi_{n}\left(k_{m}\right).
 \end{equation} 
 
For simplicity, we assume $d=1$. Denote by
\begin{equation}
\begin{split}
&H_{n}=\sum_{s=-N}^{N}w\left(x,k_{s},t\right)e^{-in\cdot k_{s}}=\sum_{s=-N}^{N}w\left(x,k_{s},t\right)e^{-2\pi i \frac{ns}{2N}},~~n=-N+1,\cdots,N\\
&H_{-N}=\sum_{s=-N}^{N}w\left(x,k_{s},t\right)e^{s\pi i}.\\
\end{split}
\end{equation}

A simple calculation yields
\begin{equation}
H_{n}=\sum_{s=0}^{2N-1}w\left(x,k_{s},t\right)e^{-2\pi i \frac{ns}{2N}}+w\left(x,k_{-N},t\right)e^{in\pi},
\end{equation}
where $k_{s}=k_{s-2N}\left(s=N+1,\cdots 2N-1\right)$. Thus, the first term on the right-hand side can be calculated by standard FFT program.

It remains to calculate
 \begin{equation}
 \begin{split}
&B\left(x,k_{m},t\right)=\sum_{n=-N}^{N}\left[H_{n}\delta \psi\left(x,\alpha n,t\right)\right] e^{2\pi i \frac{mn}{2N}},~~m=-N+1,\cdots, N,\\
&B\left(x,k_{-N},t\right)=\sum_{n=-N}^{N}\left[H_{n}\delta \psi\left(x,\alpha n,t\right)\right] e^{-in\pi}
\end{split}
\end{equation}
 via inverse FFT program as
 \begin{equation}
 B\left(x,k_{m},t\right)=\sum_{n=0}^{2N-1}S_{n}e^{2\pi i \frac{mn}{2N}}+H_{-N}\delta\psi\left(x, \alpha n,t\right)e^{-im\pi},~~m=-N+1,\cdots,N,
 \end{equation}
where $S_{n}=H_{n}\delta \psi \left(x,\alpha n,t\right)$ and  $S_{n}= S_{n-2N}\left(n=N+1,\cdots,2N-1\right)$. 

Now the Wigner equation (7) is simply approximated by collocations at the appropriate equally space nodes, 
\begin{equation}
\partial_{t}w_{N}\left(x,k_{m},t\right)+k_{m} \cdot \nabla_{x} w_{N}\left(x,k_{m},t\right)+ B\left(x, k_{m},t\right)=Q\left(w_{N}\left(x,k_{m},t\right)\right),~~m=-N,\cdots,N.
\end{equation}

This section ends with several discussions about numerical methods for scattering term and discontinuous potential. In general, the relaxation time model is handled by numerical integration techniques, like composite Simpson rule.  The Fokker-Planck model is handled by either Monte Carlo method or deterministic numerical methods.

As the Wigner distribution is now approximated by a $L^{2}$-periodic function, the aliasing error induced by the interactions between the original function and its artificial images in $L^{2}-$space should be handled carefully. Sufficient smoothness of $w$ and $V$ is required so that the aliasing error will decay rapidly on the boundary of the computational domain\cite{RS,RCO}. However, the above requirement is not necessarily satisfied, as the potential $V\left(x\right)$ may have some gaps (for instance, in simulating the barriers in semiconductors)\cite{LMN}. This problem can be partially resolved by artificially splitting the potential into two parts, namely, $V=V^{barrier}+V^{self}$, where $V^{barrier}$ is a discontinuous barrier potential, and $V^{self}$ is the self-consistent electrostatic field.  Hence, the smooth part $V^{self}$ can be tackled by the spectral collocation method, while the non-smooth barrier potential  by numerical integration techniques. 

\subsection*{B. A multistep scheme for the hyperbolic system}
\hspace{0.5cm}The remaining part is to discuss a numerical solver for the hyperbolic system (18). The multistep scheme is derived by observing that the initial-value problem (18) can be represented as an abstract ODE, in the light of the operator semigroup theory.

For a fixed $k_{m}$, denote by $A=-k_{m}\cdot \nabla_{x}$ and $T\left(t\right)=e^{tA}$ the operator semigroup generated by $A$ in the Banach space $\mathbb{X} $. Since $A$ is a symmetric operator, $T\left(t\right)$ is a $C_{0}$-semigroup \cite{RRC}.

Now we seek a solution $w_{N}\in C^{1}\left(\mathbb{X}\times \mathbb{K}, \left[0, T\right]\right)$.  Rewrite Eq.(18) in its integrated form,
\begin{equation}
w_{N}\left(x,k_{m},t\right)=T\left(t\right)w_{N}\left(x,k_{m},0\right)-\int_{0}^{t}ds ~~ T\left(t-s\right)\left[B\left(x,k_{m},s\right)-Q\left(w_{N}\left(x,k_{m},s\right)\right)\right].
\end{equation}
Since $T\left(t\right) x=x-k_{m}t$, it yields
\begin{equation}
w_{N}\left(x,k_{m},t+\Delta t\right)=w_{N}\left(x-k_{m}\Delta t, k_{m}, t\right)-\int_{t}^{t+\Delta t} ds \left[B-Qw_{N}\right]\left(x-k_{m}\left(t+\Delta t-s\right),k_{m},s\right).
\end{equation}

To derive a numerical scheme for Eq.(20), a direct choice is to use interpolation or extrapolation polynomials to estimate the integrand functions, using the same idea as the Adams multistep methods in numerical ODEs\cite{HNW}. 

Denote by $g_{m}\left(x,s; t_{n+1}\right)=T\left(t_{n+1}-s\right)\left[B-Qw_{N}\right]\left(x,k_{m},s\right)$, where the subindex $m$ of $g_{m}\left(x,s;t_{n+1}\right)$ indicates that $T\left(t\right)$ is generated by the operator $-k_{m} \cdot \nabla_{x}$.  Assume that $g_{m}\left(x,s; t_{n+1}\right)\in C\left(\mathbb{X}\times \mathbb{K} ,\left[0,t_{n+1}\right]\right)$, then $g_{m}\left(x,t_{n+1};t_{n+1}\right)$ can be approximated by an extrapolation polynomial $p\left(x, k_{m},t\right)$ on nodes $(x-k_{m}\Delta t, k_{m}, t_{n})$, $(x-2k_{m}\Delta t, k_{m}, t_{n-1})$, $\cdots$ $(x-(p+1)k_{m}\Delta t, k_{m}, t_{n-p})$, which is expressed in terms of backward differences,
\begin{equation}
\nabla^{0}g_{m}\left(x,t_{n};t_{n+1}\right)=g_{m}\left(x,t_{n};t_{n+1}\right),\quad \nabla^{j+1}g_{m}\left(x,t_{n};t_{n+1}\right)=\nabla^{j}g_{m}\left(x,t_{n};t_{n+1}\right)-\nabla^{j} g_{m}\left(x,t_{n-1};t_{n+1}\right),
\end{equation}
as follows:
\begin{equation}
p\left(x,k_{m},t\right)=p\left(x,k_{m},t_{n}+s\Delta t\right)=\sum_{j=0}^{k-1} \left(-1\right)^{j}\begin{pmatrix}-s\\j\end{pmatrix} \nabla^{j} g_{m}\left(x,t_{n};t_{n+1}\right).
\end{equation}

Inserting Eq.(22) into Eq.(20), we arrive at the generalized Adams methods for solving the hyperbolic equations (18). We denote $\tilde{w}_{N}$ the numerical solution of $w$ and $\tilde{B}=\theta\left[V\right] \tilde{w}_{N}$.
\begin{algorithm}Explicit Adams methods
\begin{eqnarray}
\begin{split}
p=0:~~&\tilde{w}_{N}\left(x,k_{m},t_{n+1}\right)=\tilde{w}_{N}\left(x-k_{m}\Delta t,k_{m},t_{n}\right)
-\Delta t \left[\tilde{B}-Q\tilde{w}_{N}\right]\left(x-k_{m}\Delta t,k_{m},t_{n}\right),\\
p=1:~~&\tilde{w}_{N}\left(x,k_{m},t_{n+1}\right)=\tilde{w}_{N}\left(x-k_{m}\Delta t,k_{m},t_{n}\right)
-\frac{3}{2}\Delta t \left[\tilde{B}-Q\tilde{w}_{N}\right]\left(x-k_{m}\Delta t,k_{m},t_{n}\right)\\
&+\frac{1}{2}\Delta t \left[\tilde{B}-Q\tilde{w}_{N}\right]\left(x-2k_{m}\Delta t,k_{m},t_{n-1}\right),\\
p=2:~~&\tilde{w}_{N}\left(x,k_{m},t_{n+1}\right)=\tilde{w}_{N}\left(x-k_{m}\Delta t,k_{m},t_{n}\right)
-\frac{23}{12}\Delta t \left[\tilde{B}-Q\tilde{w}_{N}\right]\left(x-k_{m}\Delta t,k_{m},t_{n}\right)\\
&+\frac{16}{12}\Delta t \left[\tilde{B}-Q\tilde{w}_{N}\right]\left(x-2k_{m}\Delta t,k_{m},t_{n-1}\right)-\frac{5}{12}\Delta t \left[\tilde{B}-Q\tilde{w}_{N}\right]\left(x-3k_{m}\Delta t,k_{m},t_{n-2}\right).
\end{split}
\end{eqnarray}
\end{algorithm}
Similarly, the integrands can be approximated by interpolation polynomials, yielding
\begin{algorithm}Implicit Adams methods
\begin{eqnarray}
\begin{split}
p=0:~~&\tilde{w}_{N}\left(x,k_{m},t_{n+1}\right)=\tilde{w}_{N}\left(x-k_{m}\Delta t,k_{m},t_{n}\right)
-\frac{1}{2}\Delta t \left[\tilde{B}-Q\tilde{w}_{N}\right]\left(x,k_{m},t_{n+1}\right)\\
&-\frac{1}{2}\Delta t \left[\tilde{B}-Q\tilde{w}_{N}\right]\left(x-k_{m}\Delta t,k_{m},t_{n}\right),\\
p=1:~~&\tilde{w}_{N}\left(x,k_{m},t_{n+1}\right)=\tilde{w}_{N}\left(x-k_{m}\Delta t,k_{m},t_{n}\right)
-\frac{5}{12}\Delta t \left[\tilde{B}-Q\tilde{w}_{N}\right]\left(x,k_{m},t_{n+1}\right)\\
&-\frac{8}{12}\Delta t \left[\tilde{B}-Q\tilde{w}_{N}\right]\left(x-k_{m}\Delta t,k_{m},t_{n}\right)+\frac{1}{12}\Delta t \left[\tilde{B}-Q\tilde{w}_{N}\right]\left(x-2k_{m}\Delta t,k_{m},t_{n-1}\right).
\end{split}
\end{eqnarray}
\end{algorithm}

Implicit Adams methods are not so practical in solving the hyperbolic systems (18) directly, but they can be used to correct the predicted value of $w$ through explicit methods, known as the predictor-corrector scheme. Numerical methods of higher order can be derived in a similar way. To guarantee the consistency and stability of the numerical scheme, the coefficients should satisfy the root condition and certain algebraic relations\cite{HNW}. 

The generalized Adams methods are devised for an initial-value problem, without a prior assumption of boundary conditions. Therefore, one can have more freedom to choose an appropriate formulation of the boundary condition, or simply make a nullification outside the computational domain.  In addition, the  Adams methods can be easily adapted in arbitrary dimension, owing to the way of approximating an integral with respect to time variable.  

Another remarkable feature of multistep methods is that they track the Lagrangian advection in $x-$direction, resulting from the operator $T\left(t\right)$. Therefore the above methods, which make use of the backwards characteristics to construct extrapolation (or interpolation) functions, are just extensions of semi-Lagrangian scheme. The multistep methods are expected to be free from the restriction of Courant number and allow a longer time step, since they exploit the exact Lagrangian advection.      

The price to pay is to reconstruct a regular grid using cubic spline interpolation. For the Wigner problem, it is relatively easy since the characteristic of $w_{t}+k\cdot \nabla_{x} w=0$ can be solved exactly. It is recommended to sample grid points along the characteristic line, so that the grid mesh obtains as many shifted grid points as possible. In general, we can choose $\Delta x=N_{x} \left(\Delta k\cdot \Delta t\right)$, where $N_{x}$ is an integer that indicates the numerical resolution.  When the characteristic end is not lying on the grid mesh, it can be computed by cubic spline interpolation. (As illustrated in \cite{SRBG}, the linear interpolation is too dissipative to be used, also shown in Section 4.) We call the numerical resolution is sufficiently high if $N_{x}=1$, as the grid mesh contains all the shifted points except those deviating from the computational domain. 

When explicit methods or predictor-corrector methods are employed, all the calculations (including interpolation function, tensor matrix $B$ and collision term $Q$) can be carried out independently, thereby allowing an easy strategy for parallel and distributed computing. Thus, it is expected that multistep method are much more advantageous in high dimensional problem and High Performance Computing (HPC) environment, just like the semi-Lagrangian methods.

The multistep scheme requires more initial values to start up, which can be obtained from one-step methods, like FDMs and operator splitting scheme, with a smaller time step. However, the implementation of one-step methods usually requires some information of boundary conditions. An alternative way is introduced to overcome this problem, by using the explicit backward Euler method (the first formula in Algorithm 3.1) for prediction and the implicit mid-point Euler method (the first formula in Algorithm 3.2) for correction, with a smaller time step. 

\subsection*{C. Boundary conditions}
\hspace{0.5cm}So far we have not discussed the boundary conditions yet. No prior formulation of boundary conditions is necessary in the multistep methods, since they are devised to tackle a Cauchy problem. Nevertheless, the computational domain $\Omega$ cannot be infinitely large and needs a reasonable truncation.  For the multistep scheme, we choose a simple nullification outside the computational domain, with a Dirichlet boundary condition $w=0$ on $\partial \Omega$. This approach eliminates both inflow and outflow in $x-$direction. 

Several formulations of boundary conditions in open quantum systems and corresponding mathematical concepts have been illustrated in \cite{AAN,FWR}. We only review the well-known inflow boundary conditions, proposed by Frensley,  
\begin{equation}
\begin{split}
&w\left(x_{L},k,t\right)=w_{L}\left(x_{L},k\right),~~k>0,\\
&w\left(x_{R},k,t\right)=w_{R}\left(x_{R},k\right),~~k<0,\\
\end{split}
\end{equation}
where $w_{L}\left(k,t\right)$ and $w_{R}\left(k,t\right)$ can be approximated by the Fermi-Dirac distribution.
 
Solving the Wigner-Poisson equation is much more complicated due to the coupling self-consistent Poisson equation. In principle, the multistep scheme can tackle the nonlinear problems straightforwardly, like its counterpart in numerical ODEs. However, a boundary condition, which asymptotically conserves the charge neutrality in the quantum system, is necessary for solving the Poisson equation\cite{FWR}. In previous papers, a time-dependent Dirichlet boundary condition was introduced\cite{KKF}, although its validation was not illustrated.  Therefore we only focus on the linear Wigner equation in the next section. The self-consistent field will be discussed in subsequent papers.

\section{Numerical results}
\subsection*{A. Test problems}
\hspace{0.5cm}The numerical results are presented by simulating the motion of a Gaussian wave packet (GWP) in several barrier potentials, which have been studied in \cite{KKF,BBA,SLC}. To facilitate a comparison, the author uses the same physical units and quantities as in \cite{SLC,BBA}, listed in Table 1. 

\begin{table}
 \centering
 \caption{\label{1} Units and parameters.}
 \begin{tabular}{llc}
  \toprule
  \toprule
  Physical quantity & Unit  & Value\\
  \midrule
  Time & $fs$ & -  \\
  Length & $nm$ & -  \\
  Energy & $eV$ & -\\
  Temperature & $K$ & -\\
  Electron mass $m_{e}$& $eV\cdot fs^{2}\cdot nm^{-2}$ & $5.68562966$\\
  Planck constant $\hbar$ & $eV\cdot fs$ & $0.658211899$\\
  Boltzmann constant $k_{B}$ & $eV\cdot K^{-1}$ & $8.61734279\times 10^{5}$\\
  \bottomrule
  \bottomrule
  \end{tabular}
\end{table}

The purpose of numerical tests is twofold. Firstly, we test the accuracy and convergence of multistep methods. The performance metric is based on either the exact solutions or numerical solutions with high resolution ($N_{x}=1, \Delta t=0.05$). Secondly, we investigate both the quantum tunneling effect and the scattering effect. The scattering process is modeled by the relaxation time model,
which effectively removes the correlation and introduces irreversibility\cite{KKF}.

The rescaled collisionless Wigner equation in one dimension is 
\begin{equation}
\frac{\partial w}{\partial t}+\frac{\hbar k}{m} \frac{\partial w}{\partial x}+\frac{1}{2\pi \hbar} \theta\left[V\right]w=0.
\end{equation}
When including the scattering effect, it yields the collisional Wigner equation
\begin{equation}
\frac{\partial w}{\partial t}+\frac{\hbar k}{m} \frac{\partial w}{\partial x}+\frac{1}{2\pi \hbar} \theta\left[V\right]w=\frac{1}{\tau}\left(\frac{n}{n_{0}}w_{0}-w\right).
\end{equation}

The wave function of a GWP is expressed as
\begin{equation}
\psi\left(x,t\right)=\left[\frac{1}{2\pi a^{2} \left(1+i\beta t\right)^{2}}\right]^{\frac{1}{4}}e^{i\left(k_{0}x-\omega_{0}t\right)}exp\left[-\frac{\left(x-v_{0}t\right)^{2}}{4 \alpha^{2}\left(1+i\beta t\right)}\right],
\end{equation}
where $v_{0}$ is the average velocity, $\alpha$ is the minimum position spread, and
\begin{equation}
\beta=\frac{\hbar}{2m\alpha^{2}},~~v_{0}=\frac{\hbar k_{0}}{m}=\frac{2\omega_{0}}{k_{0}}.
\end{equation}

The Wigner-function description of Eq.(28) is
\begin{equation}
w\left(x,k,t\right)=2exp\left\{-\frac{\left(x-x_{0}-v_{0}t\right)^{2}}{2a^{2}\left(1+\beta^{2}t^{2}\right)}\right\} exp\left\{-2\alpha^{2}\left(1+\beta^{2}t^{2}\right)^{2}\left[\left(k-k_{0}\right)-\frac{\beta t\left(x-x_{0}-v_{0}t\right)}{2\alpha^{2}\left(1+\beta^{2}t^{2}\right)}\right]^{2}\right\},
\end{equation}
which is the exact solution of Eq.(26) when $V=0$.

The initial condition for the GWP simulation is
\begin{equation}
w\left(x,k,0\right)=2exp\left[-\frac{\left(x-x_{0}\right)^{2}}{2\alpha^{2}}\right]exp\left[-2\alpha^{2}\left(k-k_{0}\right)^{2}\right].
\end{equation}

The quantum tunneling effect is investigated by simulating a GWP hitting a Gaussian barrier with three different heights. The Gaussian barrier with a width $\omega$ is given
\begin{equation}
V\left(x\right)=H exp^{-\frac{x^{2}}{2\omega^{2}}},
\end{equation} 
where the self-consistent electrostatic potential is not included. In subsequent simulations, the heights $H$ of $V\left(x\right)$ are chosen to be 0.3, 1.3 and 2.3, respectively, with $\omega=1$. 

If not specified, the coefficients in simulations are chosen as $\alpha=2.825$, $m=0.0665m_{e}$, $x_{0}=-30$ and $k_{0}=1.4$ so that the kinetic energy of GWP $E_{0}\approx 1.12$. The computational domain $\Omega= \mathbb{X} \times \mathbb{K}$ is  $\left[-\frac{125\pi \hbar}{16m},\frac{125\pi\hbar}{16m}\right] \times \left[-2\pi,2\pi \right]$, with $\Delta x=\frac{\pi\hbar}{64m}$, $\Delta k=\frac{\pi}{64}$ (1001 grid points in $x-$direction and 257 grid points in $k-$direction). Different time steps $\Delta t$ are investigated, from 0.05 to 0.2 (the maximum Courant number is $\sigma=\frac{\hbar \Delta t}{m \Delta x}=4.08$).

For the multistep methods, the one-step predictor-corrector method is used to obtain the missing starting points, with time step $\Delta t=0.05$. The shifted grid points are interpolated by cubic spline interpolation if not lying on the grid mesh. And the boundary condition is chosen as the Dirichlet type, $w=0$ on $\partial \Omega$, with a nullification for the shifted grid points outside $\Omega$. While the inflow boundary condition is employed for the upwind finite difference method,
\begin{equation}
\begin{split}
&w\left(x_{L},k,t\right)=w\left(x_{L},k,0\right),~~k>0,\\
&w\left(x_{R},k,t\right)=w\left(x_{R},k,0\right),~~k<0.\\
\end{split}
\end{equation}
The numerical error induced by the inflow boundary conditions is negligible, as the computational domain $\Omega$ is chosen large enough. In fact, it has $w\left(x_{L},k,0\right)\approx 0$ and $w\left(x_{R},k,0\right)\approx 0$.

\subsection*{B. Numerical results}
\hspace{0.5cm}In order to test the accuracy of multistep methods, a comparison is made between the multistep scheme and the upwind FDM by monitoring the error when simulating the time evolution of the GWP in the free space $\left(V=0\right)$, without the collision term. The performance metrics are  $L^{2}$ error ($\epsilon_{2}$) and $L^{\infty}$ error ($\epsilon_{\infty}$)\cite{SLC}.
\begin{equation}
\begin{split}
&\epsilon_{2}\left(t\right)=\left[\int_{\mathbb{X} \times \mathbb{K}}\left(\Delta w\left(x,k,t\right)\right)^{2}dxdk\right]^{\frac{1}{2}},\\
&\epsilon_{\infty}\left(t\right)=\displaystyle{max\left\{\Delta w\left(x,k,t\right)\right\},~~\left(x,k\right) \in \mathbb{X} \times \mathbb{K}},\\
\end{split}
\end{equation}
where $\Delta w\left(x,k,t\right)=\left|w^{reference}\left(x,k,t\right)-w^{num}\left(x,k,t\right)\right|$. In practice, $\epsilon_{2}$ is approximated by
\begin{equation}
\bar{\epsilon}_{2}\left(t\right)=\left[\sum_{\mathbb{X} \times \mathbb{K}}\left(\Delta w\left(x,k,t\right)\right)^{2}\Delta x \Delta k\right]^{\frac{1}{2}}.
\end{equation}

The evolution of a GWP in the free space $\left(V=0\right)$ is simulated by both methods with the same time step $\Delta t=0.05$. In this case, the multistep scheme reduces to 
\begin{equation}
\tilde{w}_{N}\left(x,k_{m},t_{n+1}\right)=\tilde{w}_{N}\left(x-\frac{\hbar}{m}k_{m}\Delta t,k_{m},t_{n}\right),
\end{equation}
which is the exact solution of $\partial_{t}w+\frac{\hbar}{m}k_{m} \nabla_{x} w=0$. Numerical results are listed as follows. 

\begin{figure}[h]
    \centering
    \subfigure[$L^{2}$ error for the collisionless case $\left(V=0\right)$]{
    \label{fig:subfig:a}
    \includegraphics[width=2.4in,height=1.8in]{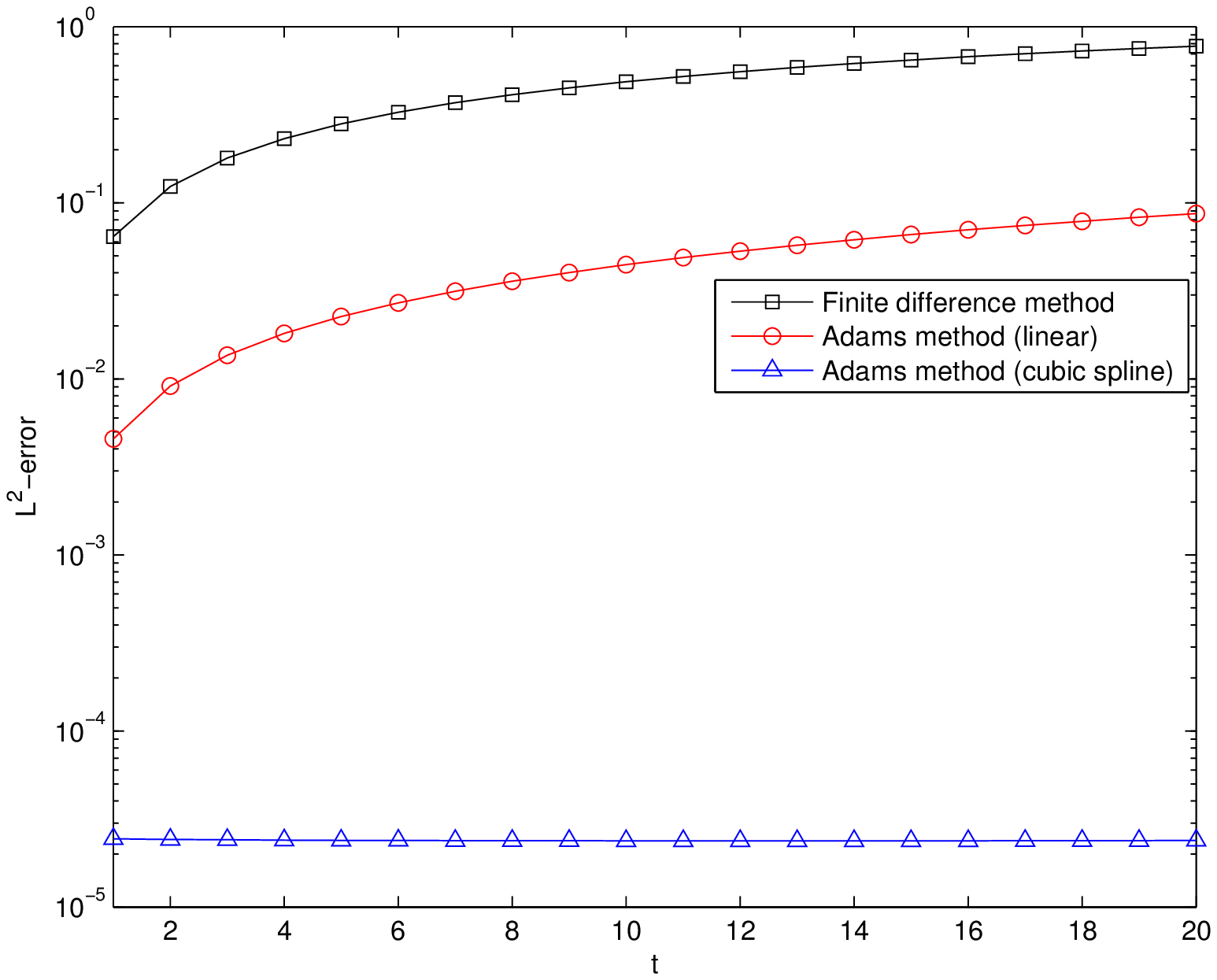}}
    \hspace{0.2in}
    \subfigure[$L^{\infty}$ error for the collisionless case $\left(V=0\right)$]{
    \label{fig:subfig:b}
    \includegraphics[width=2.4in,height=1.8in]{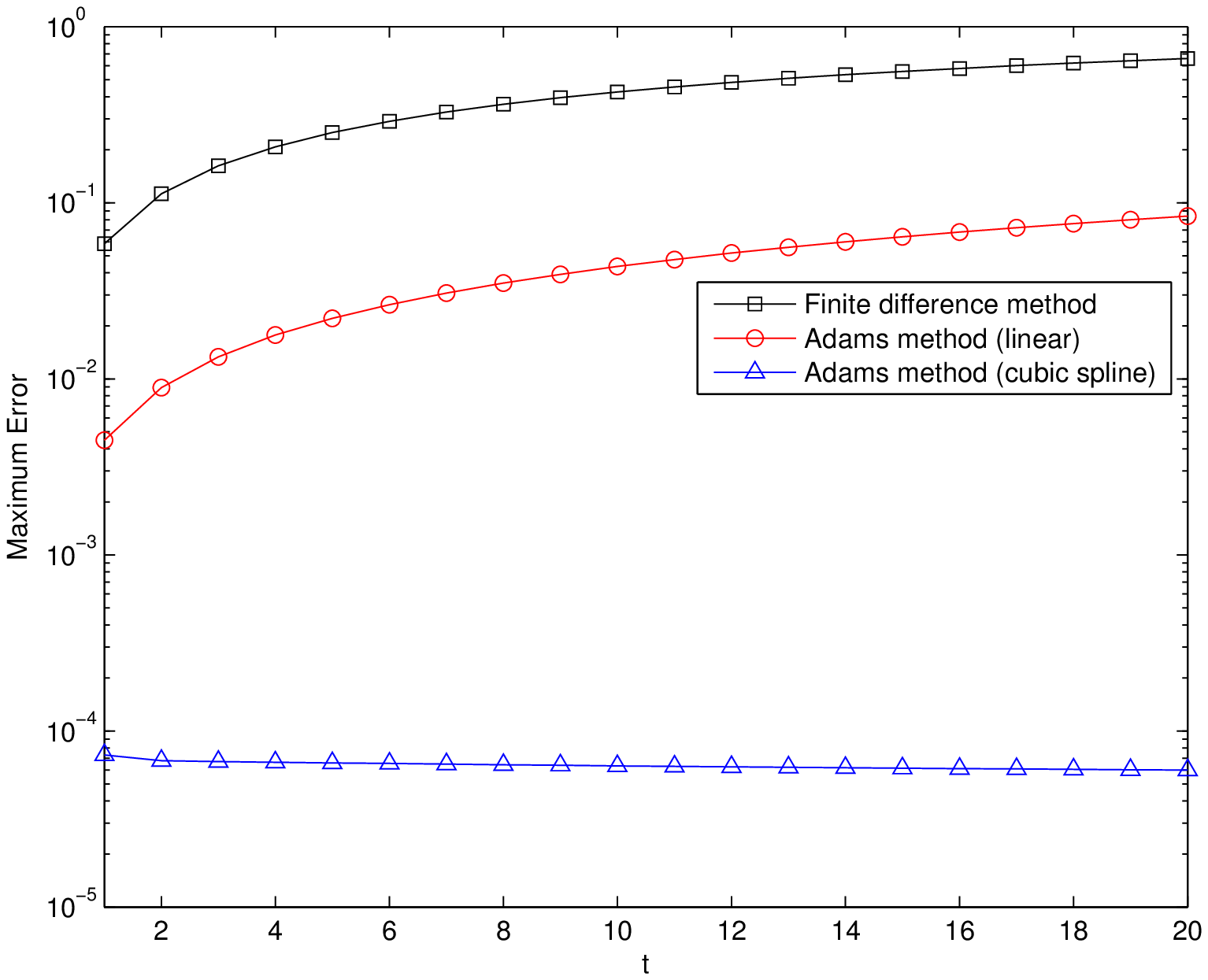}}
     \caption{The time evolution of numerical errors of the first-order upwind finite difference method, linear interpolation method and cubic spline method, under the flat potential $\left(V=0\right)$}  
\end{figure} 
  
We make a comparison between the first-order upwind finite difference method and the explicit Adams method, where both linear interpolation and cubic spline interpolation are tested. The evolution of numerical errors is demonstrated in Figure 1. It is shown that the cubic spline interpolation method yields the most accurate results. The linear interpolation method is also more accurate than the FDM, although the accumulation of global error is still observed.

For the simple test problem $\partial_{t}w+\frac{\hbar}{m}k_{m} \nabla_{x} w=0$, the initial value and the exact Wigner function at $t=20$ are plotted in Figure 2a and 2b. Figure 2c shows the distribution of the absolute error $\Delta w$ at $t=20$ through cubic spline interpolation, indicating that the numerical result is very accurate (the maximum $\Delta w$ is less than $6 \times 10^{-5}$). In Figure 2d we make a comparison of numerical waveforms at $k=1.3744, t=20$. The cubic spline interpolation method gives a precise waveform, while both the linear interpolation method and FDM suffer from the numerical dissipation. This accords with the observation in \cite{SRBG}. Therefore, in the following simulations, we only employ cubic spline interpolation to compute the shifted grid points.  

\begin{figure}[h]
    \centering
    \subfigure[The initial Wigner function at t=0]{
    \label{fig:subfig:a}
    \includegraphics[width=2.4in,height=1.8in]{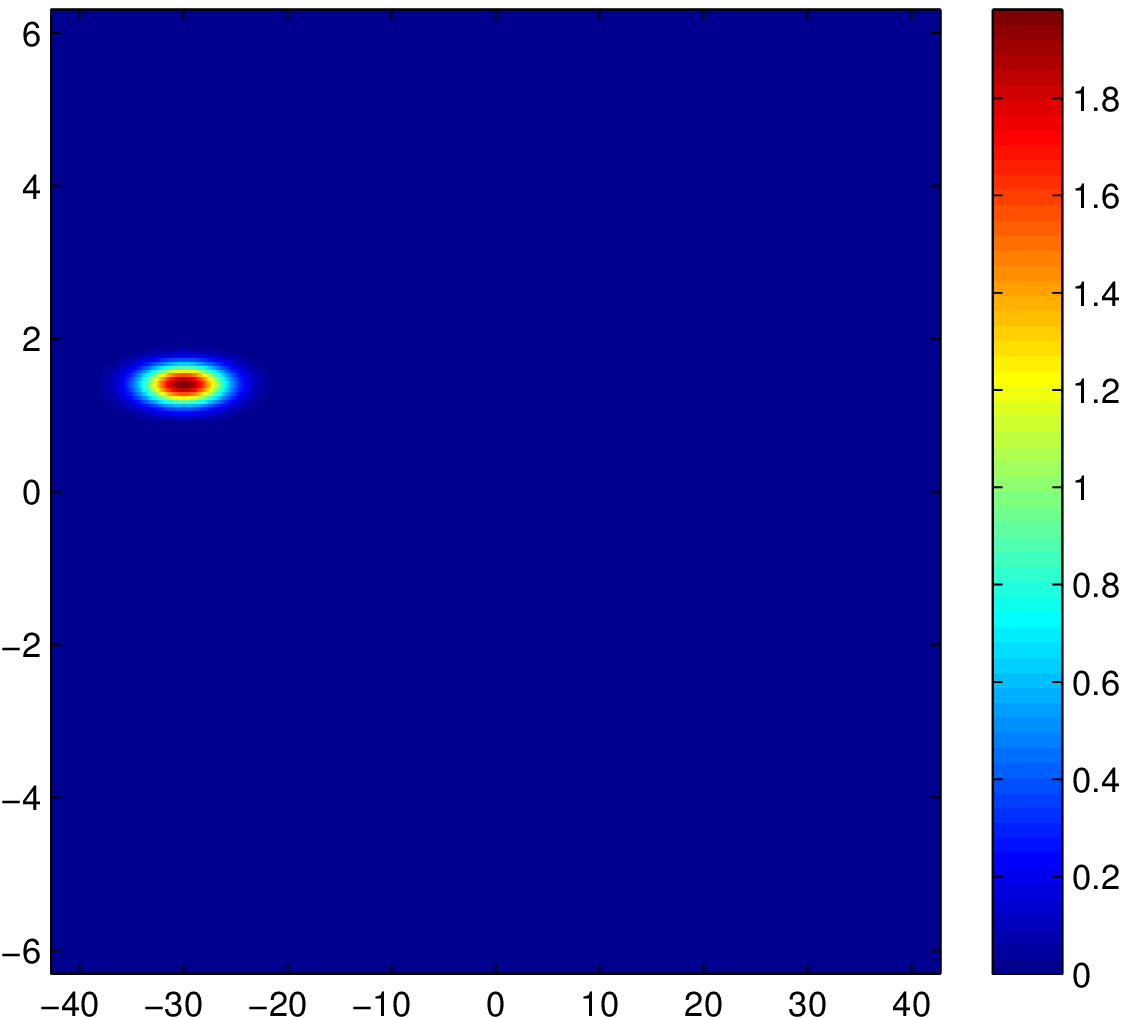}}
    \hspace{0.2in}
    \subfigure[The exact Wigner function at t=20]{
    \label{fig:subfig:b}
    \includegraphics[width=2.4in,height=1.8in]{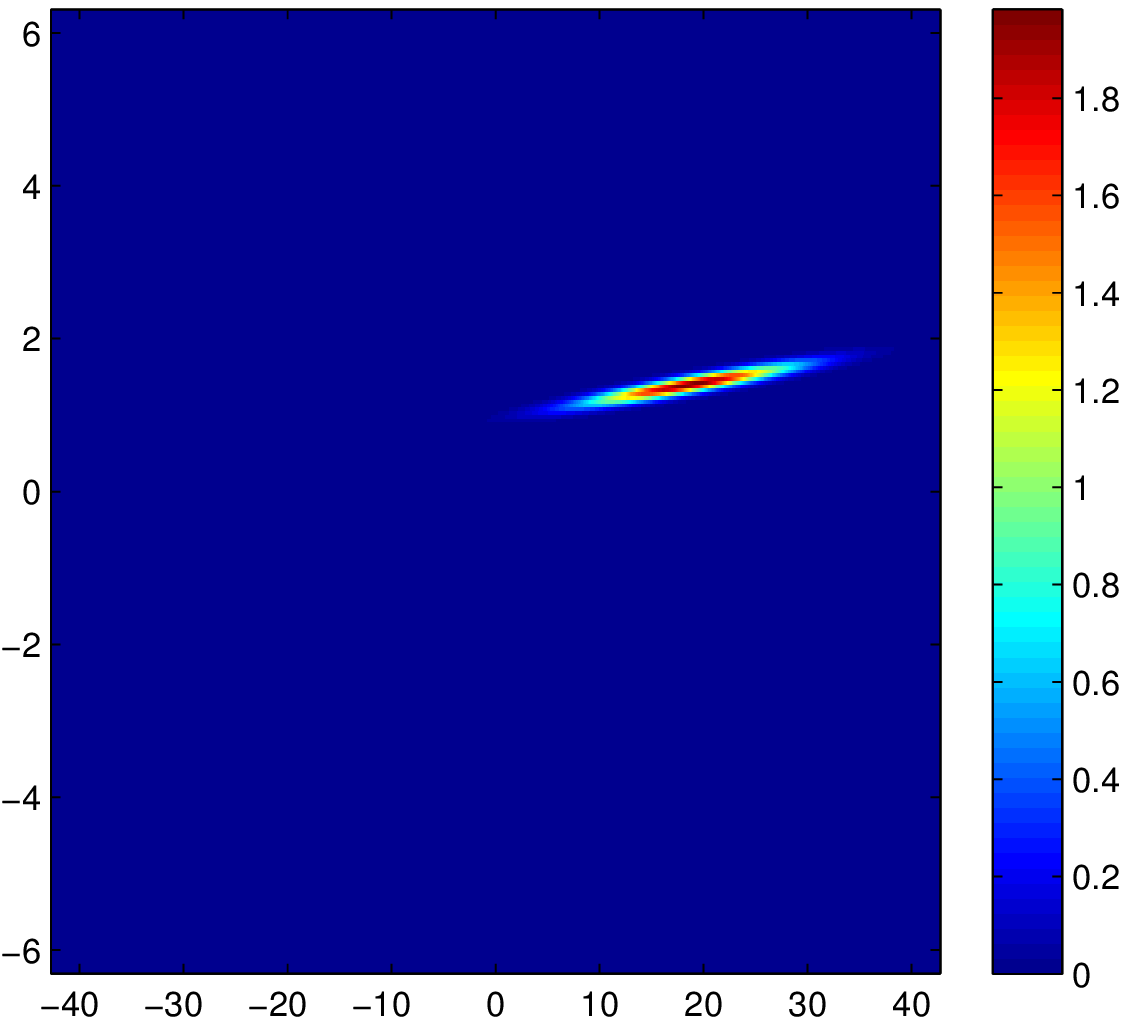}}
    \\
    \centering
    \subfigure[$\Delta w$ at t=20, by cubic spline interpolation]{
    \label{fig:subfig:c}
    \includegraphics[width=2.4in,height=1.8in]{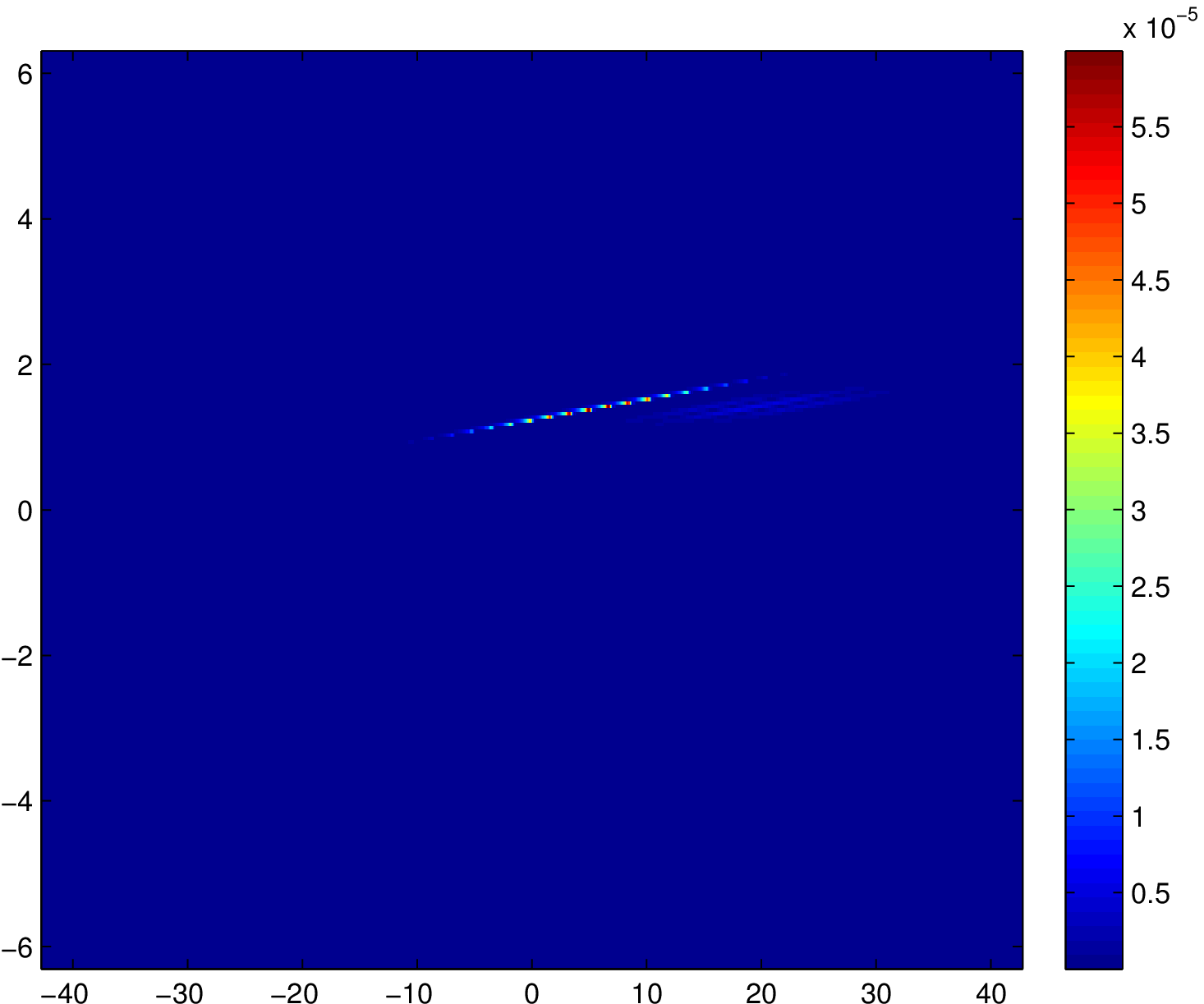}}
    \hspace{0.2in}
    \subfigure[The numerical waveforms at k=1.3744, t=20]{
    \label{fig:subfig:d}
    \includegraphics[width=2.4in,height=1.8in]{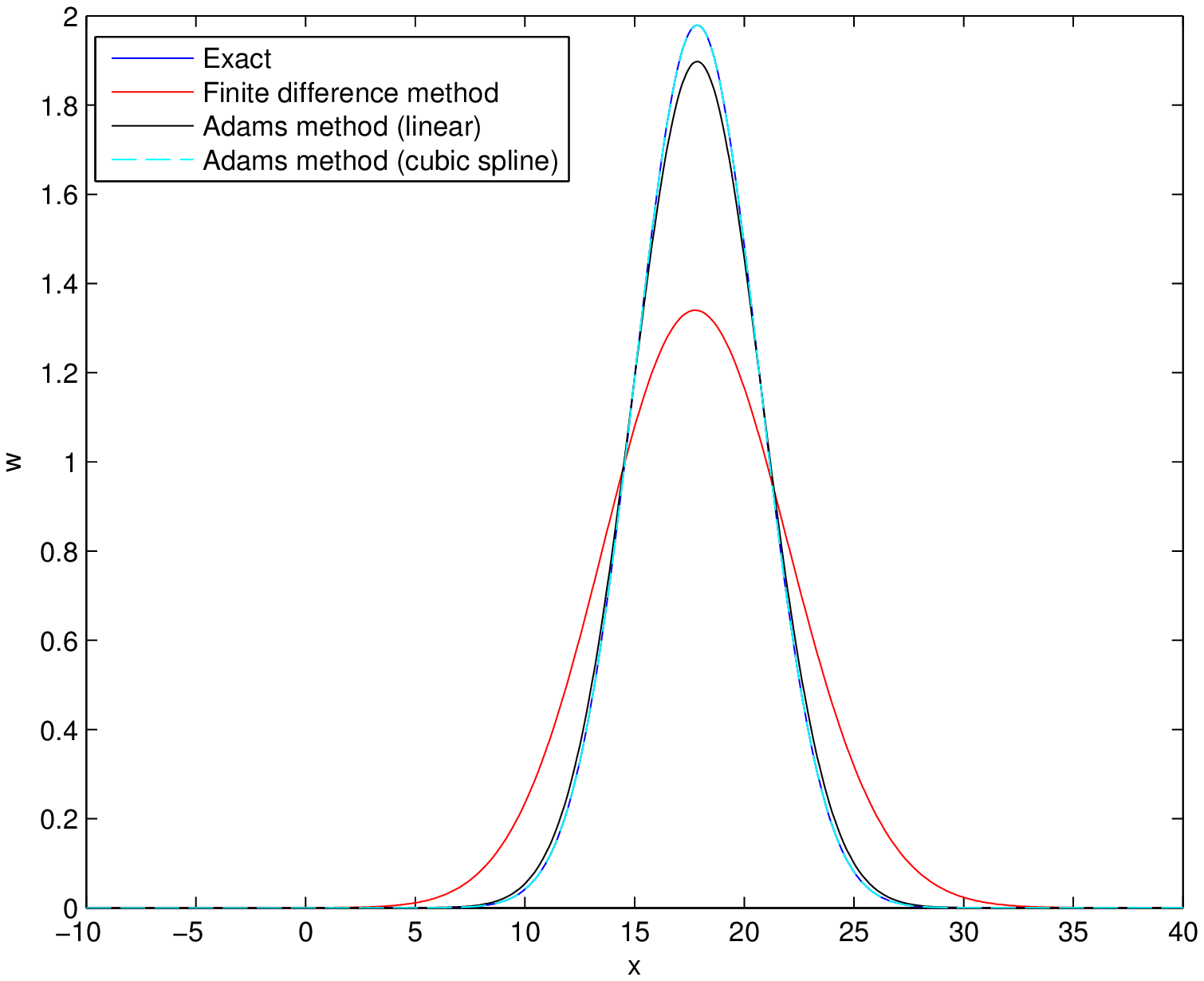}}
     \caption{The time evolution of a GWP in the free space $\left(V=0\right)$}  
\end{figure}

Now we let a GWP with kinetic energy $E_{0}\approx 1.12$ to hit several Gaussian barriers. The explicit three-step method is employed, combining with the spectral collocation method to discretize the pseudo-differential operator. The choice of Gaussian potential guarantees the consistency and convergence of the spectral collocation method. 

To show the convergence of multistep methods, we choose $H$ to be $1.3$ and monitor the $L^{2}$ and $L^{\infty}$ error under three different time steps, with the same uniform grid mesh. Since the exact solution of Eq. (26) is not trivial, we choose the numerical solution with $\Delta t=0.05$ and high resolution $N_{x}=1$ (20001 grid points in $x-$direction and $257$ grid points in $k-$direction, with $\Delta x=\frac{\pi\hbar}{1280m}$, $\Delta k=\frac{\pi}{64}$) as the reference. Numerical errors are significantly small in a short time, since free advection is dominant. Afterwards quantum interference becomes important and larger numerical errors are observed. As shown in Figure 3, numerical errors are reduced dramatically when the time step becomes smaller. In addition, when $\Delta t=0.05$, the numerical errors resulting from interpolations are less than $10^{-3}$, which demonstrates the accuracy of the cubic spline interpolation.

\begin{figure}[h]
    \centering
    \subfigure[$L^{2}$ error for the collisionless case $\left(H=1.3\right)$]{
    \label{fig:subfig:a}
    \includegraphics[width=2.4in,height=1.8in]{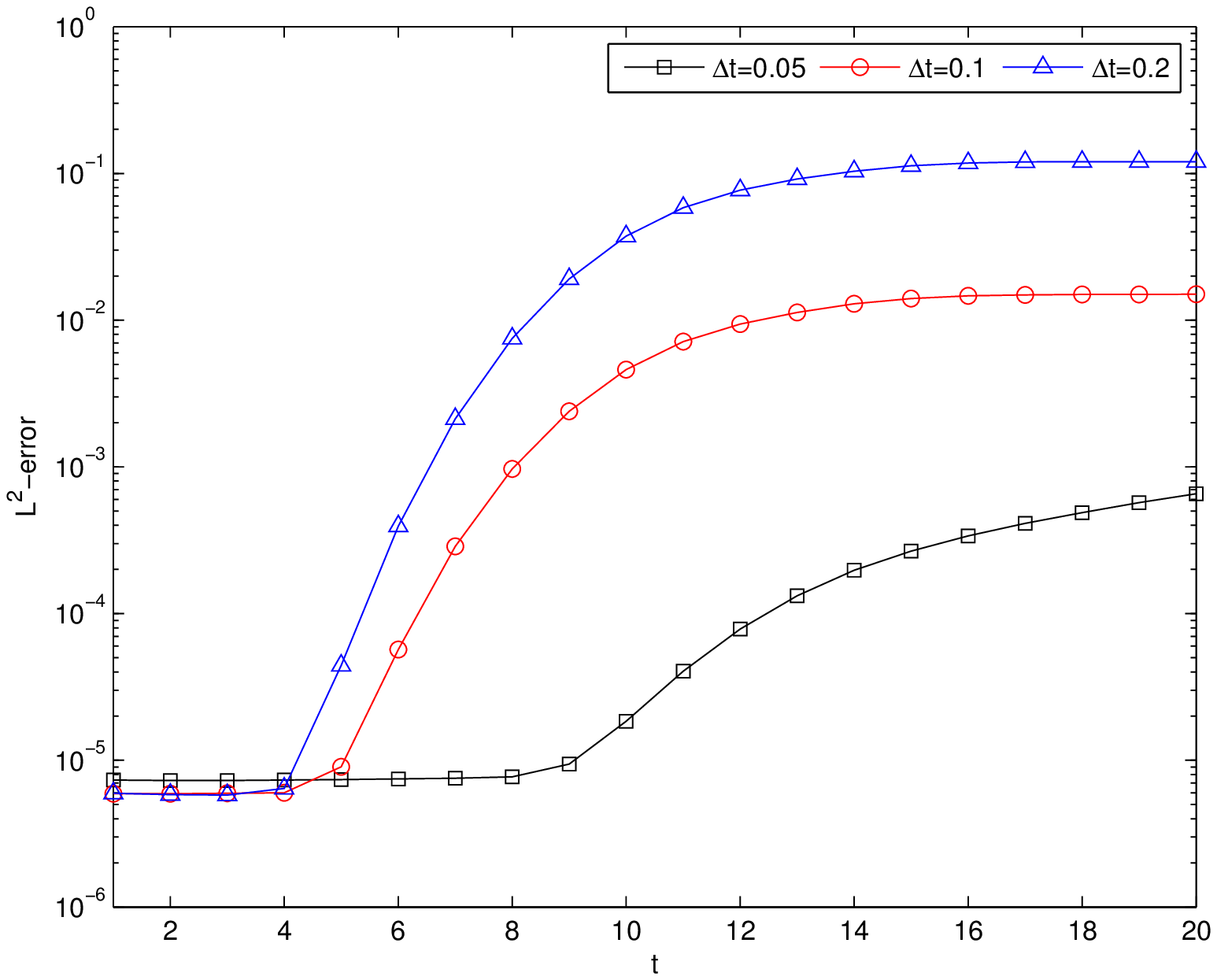}}
    \hspace{0.2in}
    \subfigure[$L^{\infty}$ error for the collisionless case $\left(H=1.3\right)$]{
    \label{fig:subfig:b}
    \includegraphics[width=2.4in,height=1.8in]{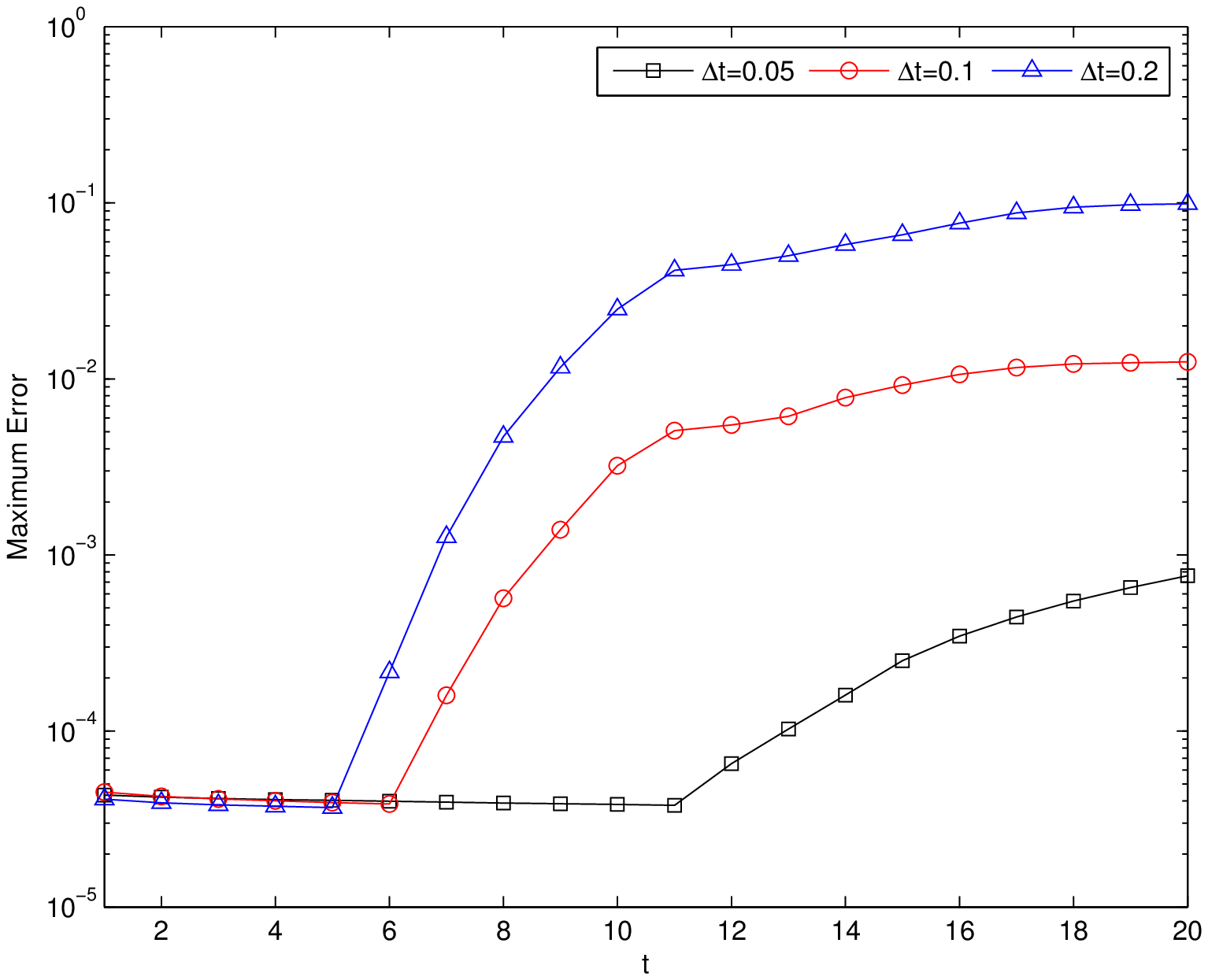}}
     \caption{The convergence history of the explicit three-step method with respect to $\Delta t$ ($H=1.3$)  }  
\end{figure} 

Next we begin to investigate the quantum interference with different Gaussian barriers. We choose the same grid mesh ($N_{x}=20$) and use explicit three-step Adams method, with $\Delta t=0.05$ and cubic spline interpolation. 

Figure 4 shows the Wigner function for the GWP interacting with the Gaussian barrier $V\left(x\right)=0.3e^{-\frac{x^{2}}{2}}$. The kinetic energy $E_{0}$ of GWP is much greater than the barrier height. Therefore, the GWP travels across the barrier easily.

\begin{figure}[h]
    \centering
    \subfigure[t=5]{
    \label{fig:subfig:a}
    \includegraphics[width=2.4in,height=1.8in]{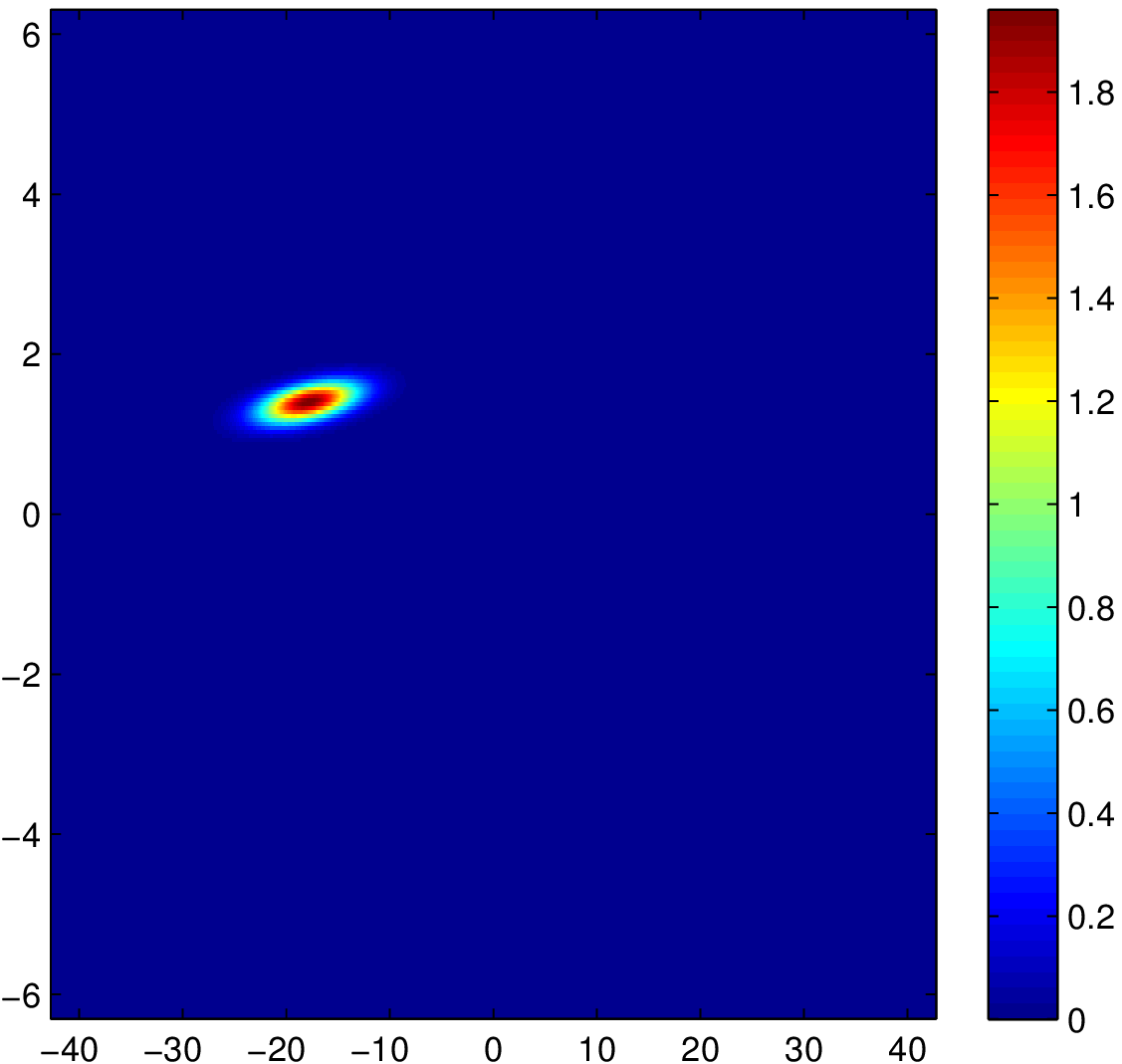}}
    \hspace{0.2in}
    \subfigure[t=10]{
    \label{fig:subfig:b}
    \includegraphics[width=2.4in,height=1.8in]{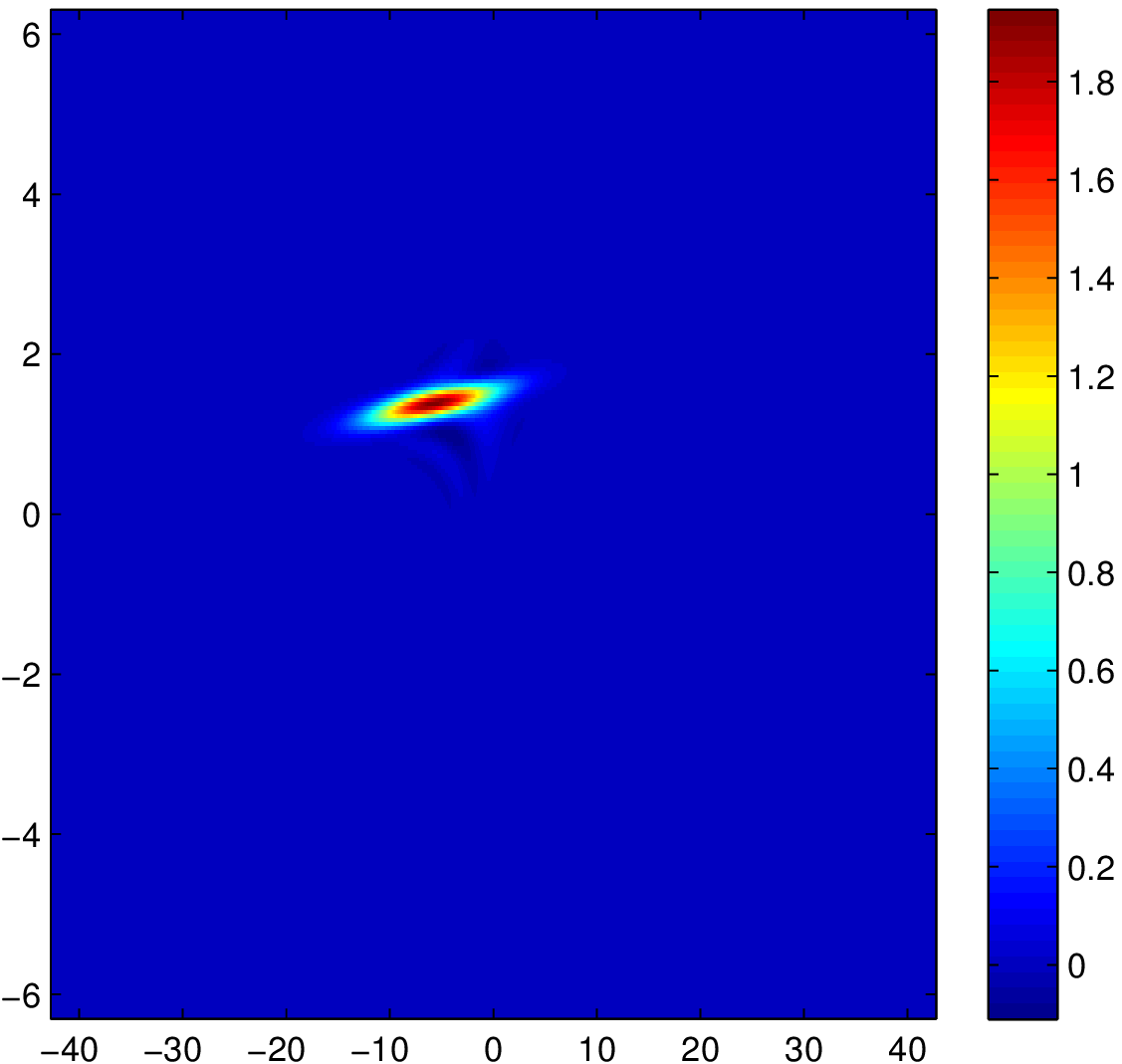}}
    \\
    \centering
    \subfigure[t=15]{
    \label{fig:subfig:c}
    \includegraphics[width=2.4in,height=1.8in]{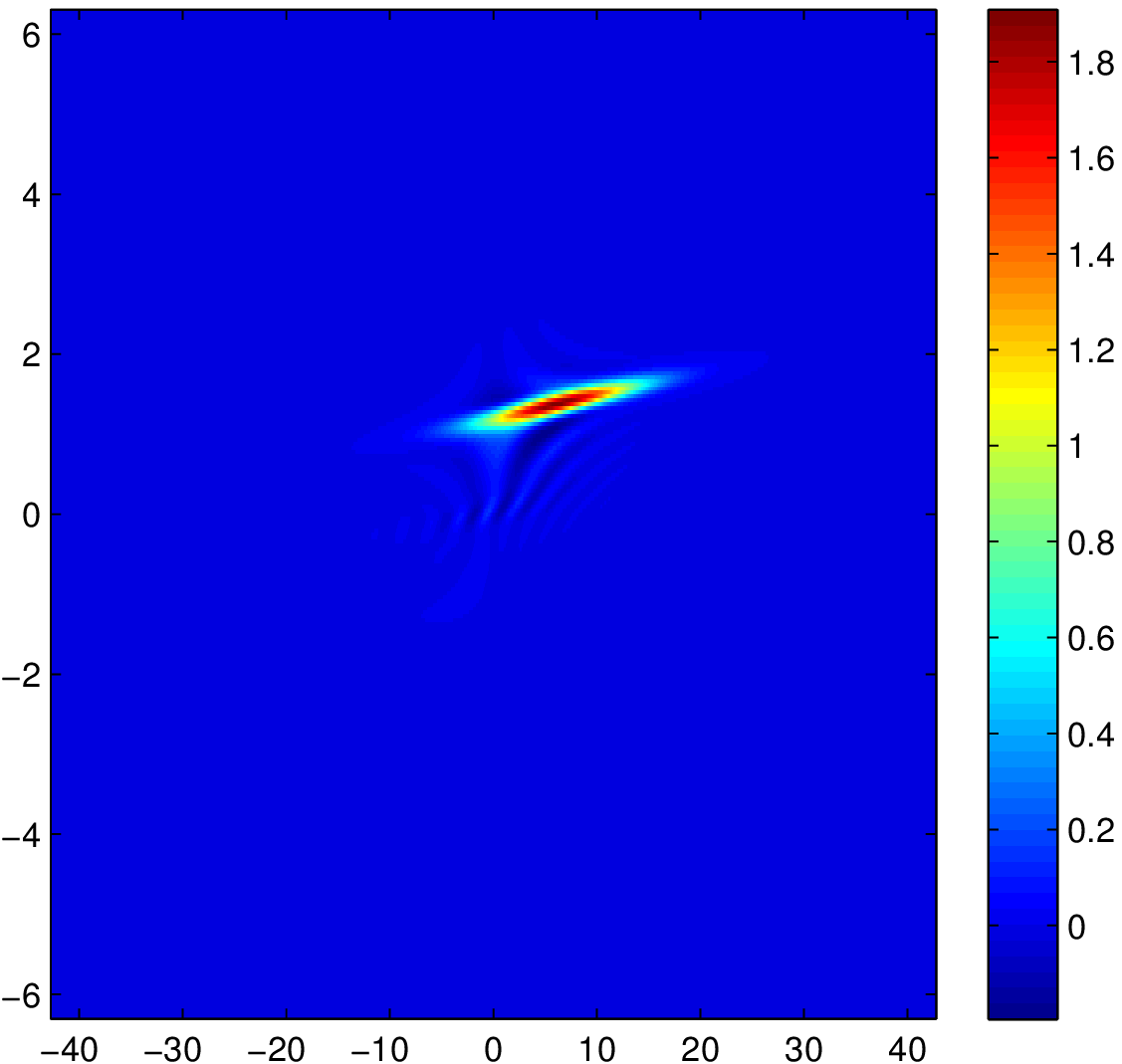}}
    \hspace{0.2in}
    \subfigure[t=20]{
    \label{fig:subfig:d}
    \includegraphics[width=2.4in,height=1.8in]{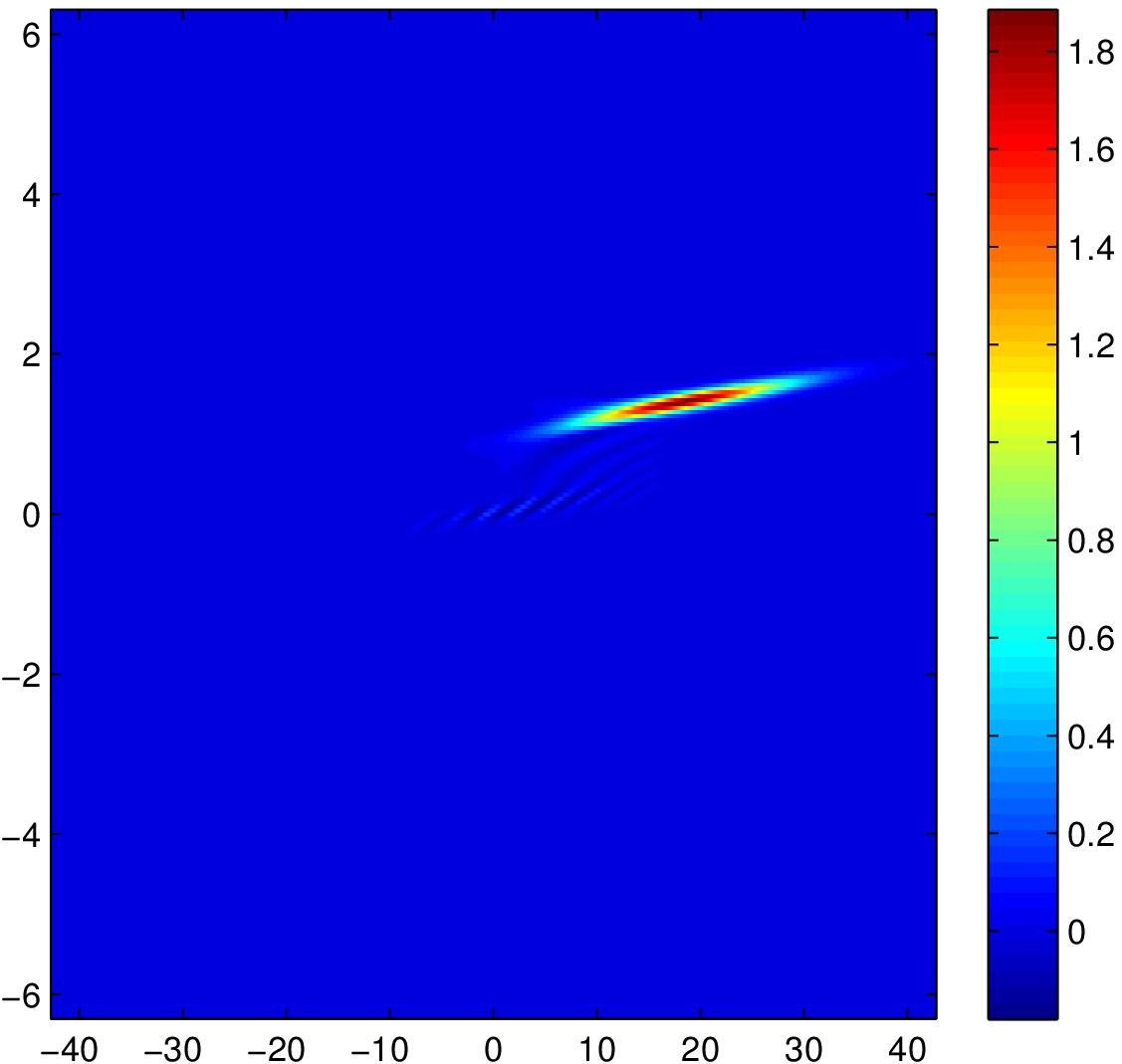}}
     \caption{The time evolution of a GWP interacting with a Gaussian barrier $\left(H=0.3\right)$}  
\end{figure}

If the height of barrier is comparable to $E_{0}$, the GWP is separated into two wave packets due to the quantum interference with the Gaussian barrier. Although the kinetic energy is smaller than the barrier height, a part of wave packet is still able to travel across the barrier, while another is reflected back, seen in Figure 5. When the height of potential barrier grows even larger, like $H=2.3$, the GWP is almost completely reflected back, presented in Figure 6. Besides, in both cases, the sign of $w$ changes rapidly around $k=0$, which indicates an oscillation of the Wigner distribution in phase space.

\begin{figure}[!h]
    \centering
    \subfigure[t=7.5]{
    \label{fig:subfig:a}
    \includegraphics[width=2.4in,height=1.8in]{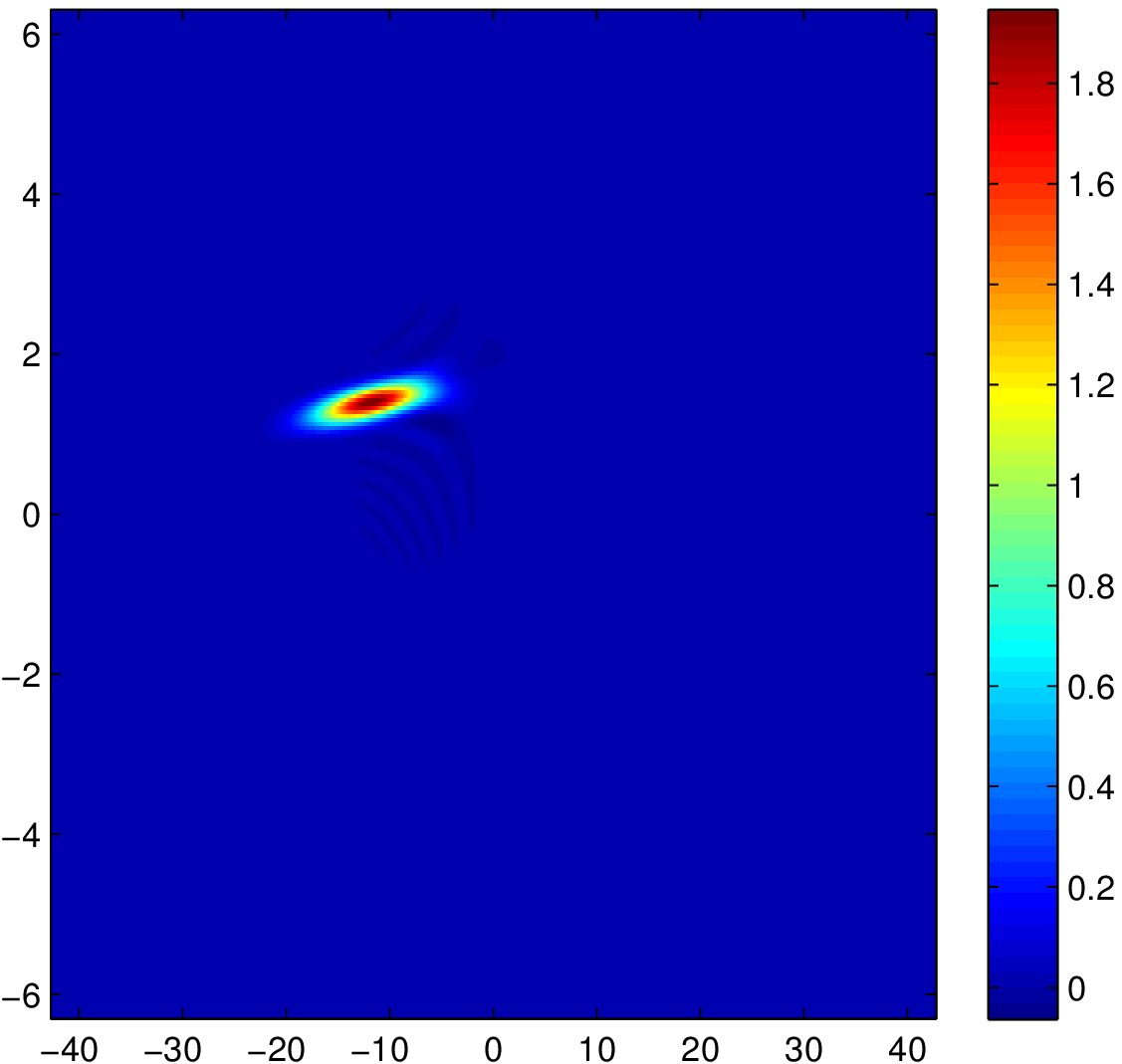}}
    \hspace{0.2in}
    \subfigure[t=10]{
    \label{fig:subfig:b}
    \includegraphics[width=2.4in,height=1.8in]{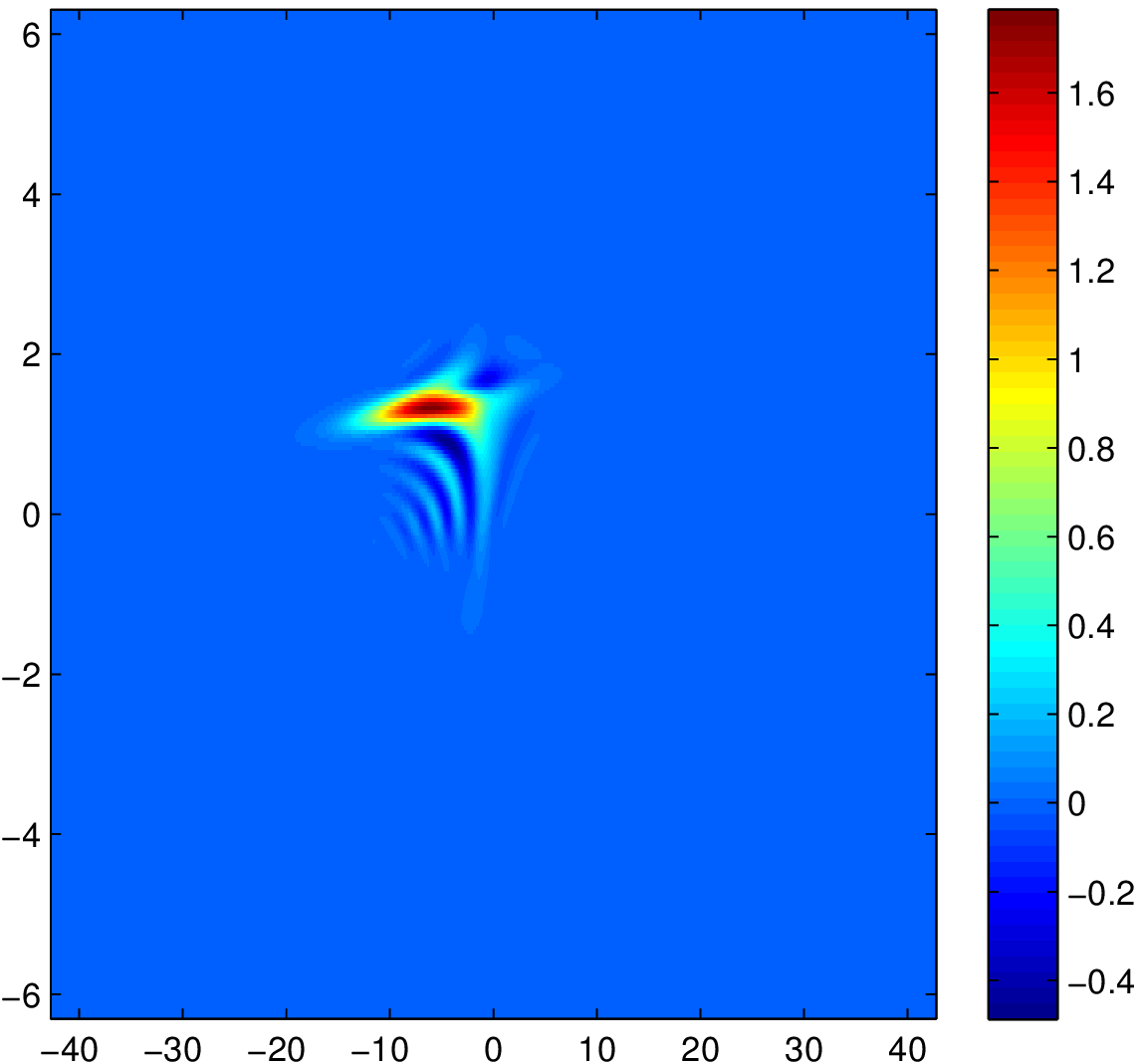}}
    \\
    \centering
    \subfigure[t=12.5]{
    \label{fig:subfig:c}
    \includegraphics[width=2.4in,height=1.8in]{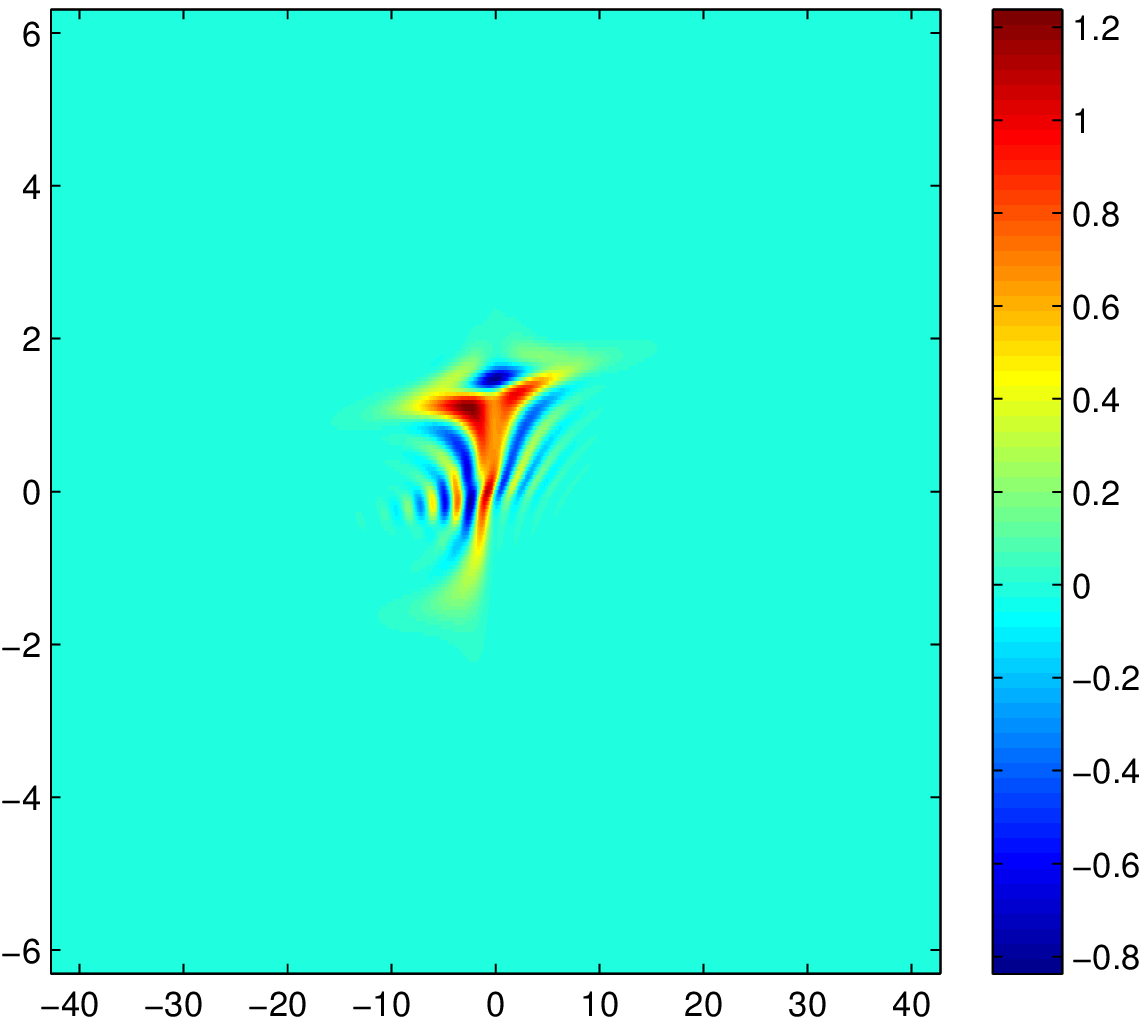}}
    \hspace{0.2in}
    \subfigure[t=15]{
    \label{fig:subfig:d}
    \includegraphics[width=2.4in,height=1.8in]{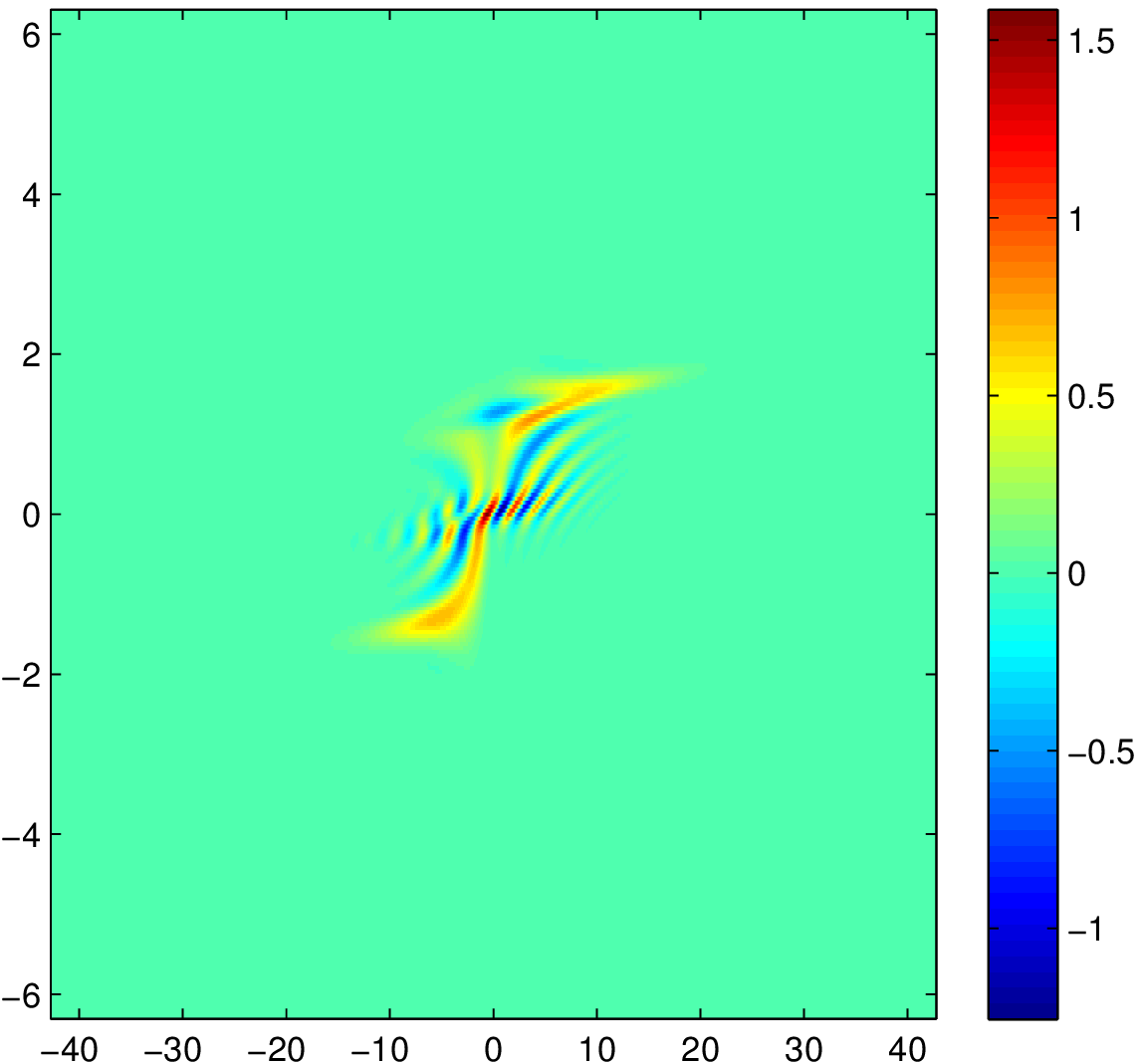}}
    \\
    \centering
    \subfigure[t=17.5]{
    \label{fig:subfig:e}
    \includegraphics[width=2.4in,height=1.8in]{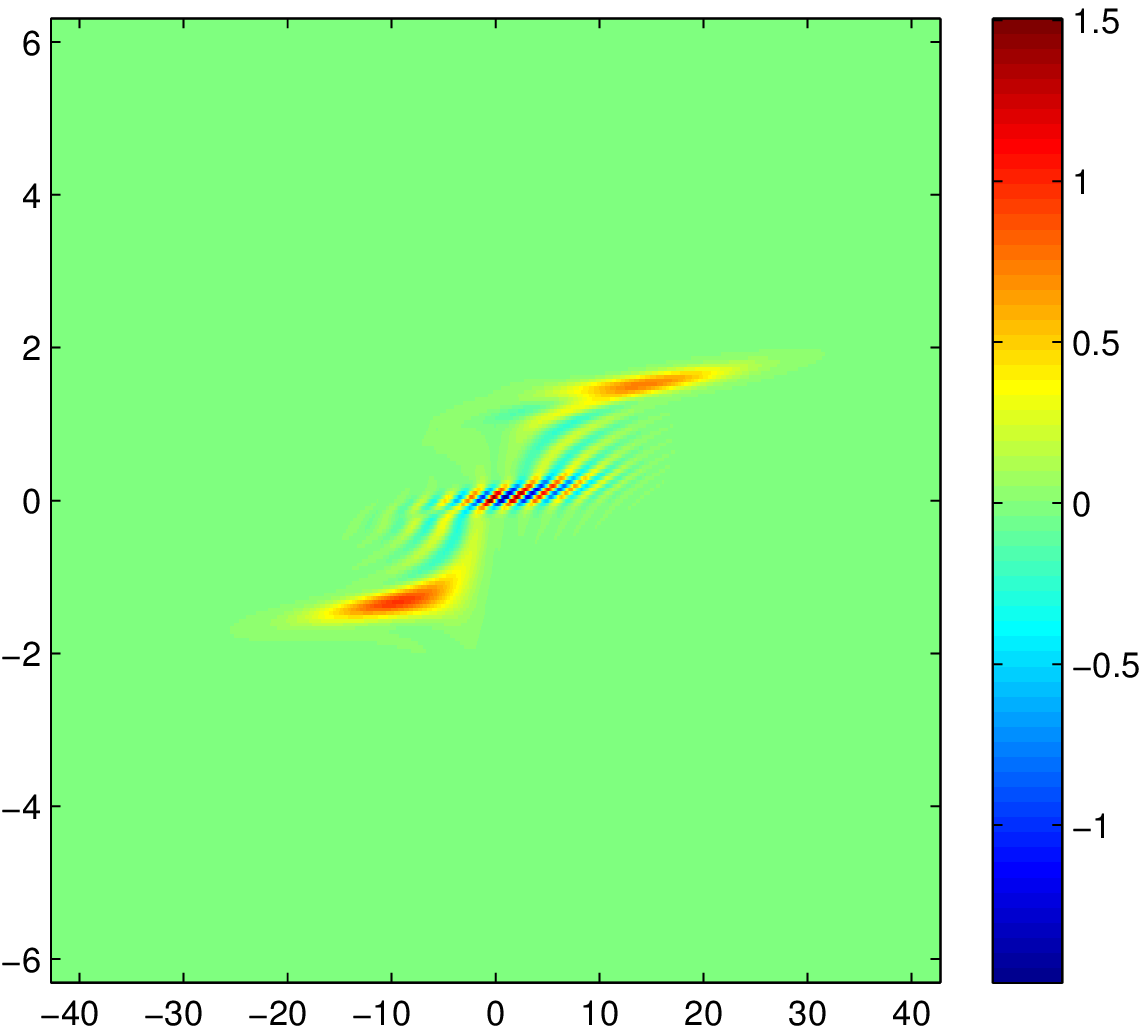}}
    \hspace{0.2in}
    \subfigure[t=20]{
    \label{fig:subfig:f}
    \includegraphics[width=2.4in,height=1.8in]{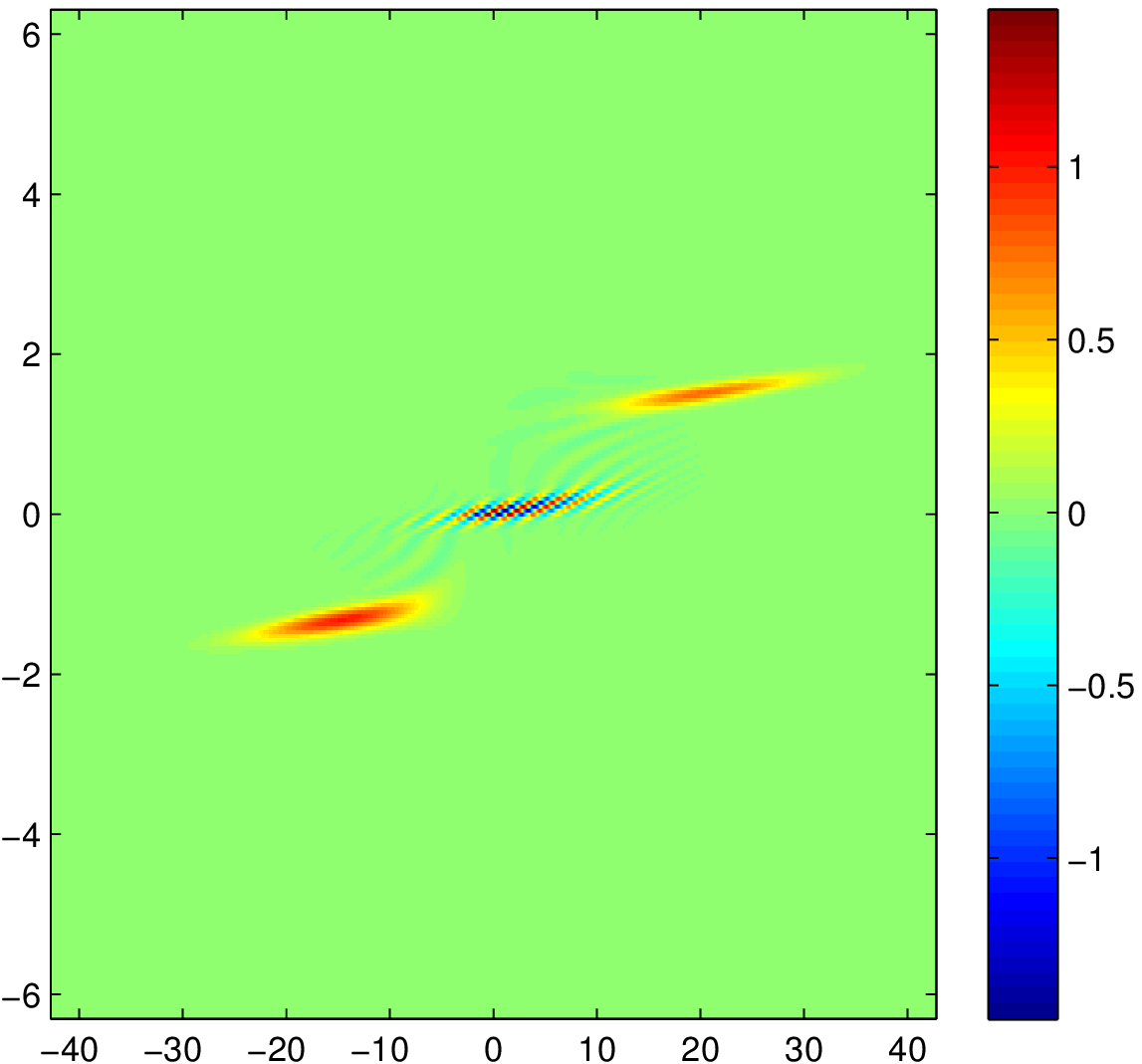}}
     \caption{The time evolution of a GWP interacting with a Gaussian barrier $\left(H=1.3\right)$}  
\end{figure}

\begin{figure}[!h]
    \centering
    \subfigure[t=7.5]{
    \label{fig:subfig:a}
    \includegraphics[width=2.4in,height=1.8in]{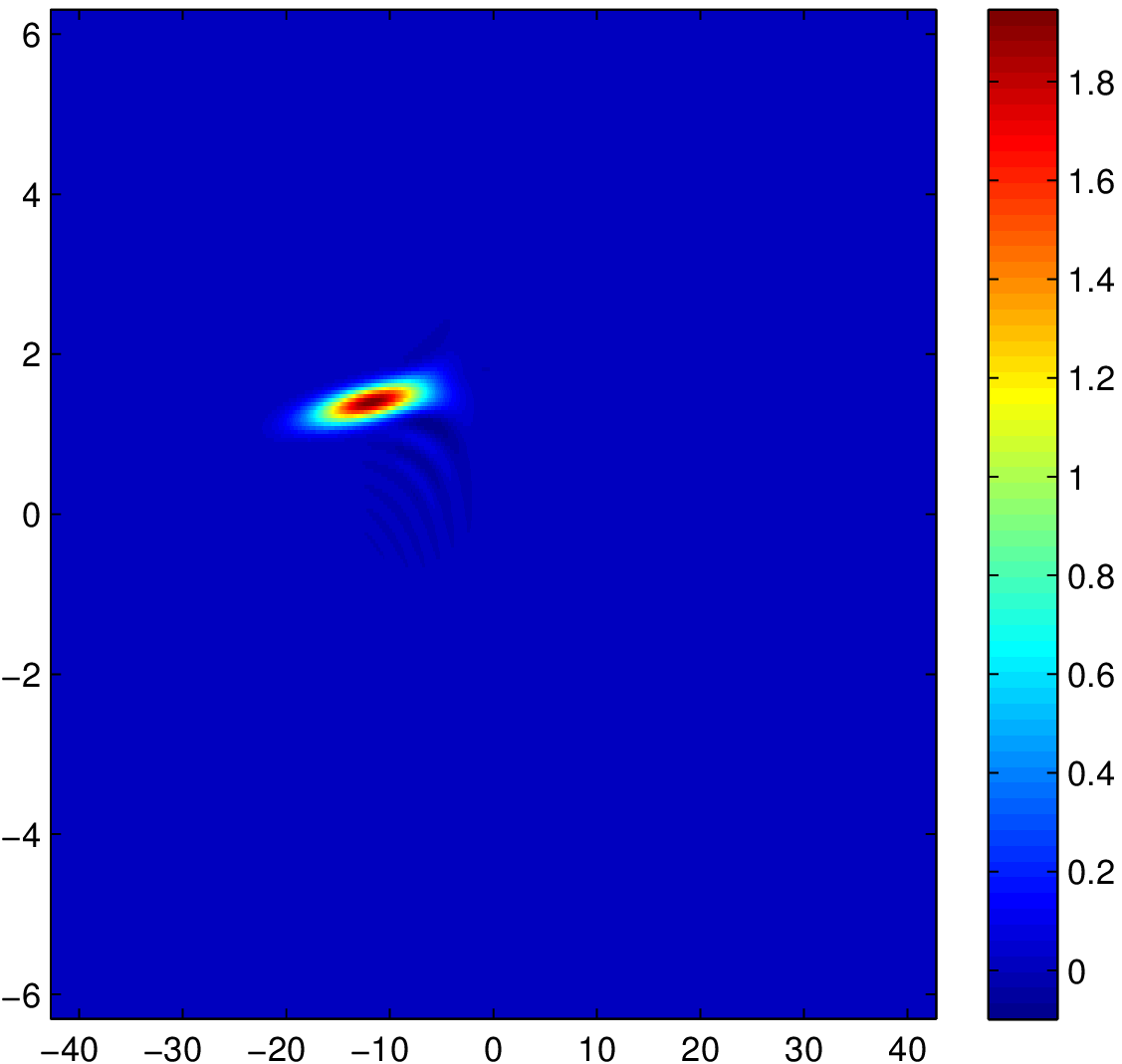}}
    \hspace{0.2in}
    \subfigure[t=10]{
    \label{fig:subfig:b}
    \includegraphics[width=2.4in,height=1.8in]{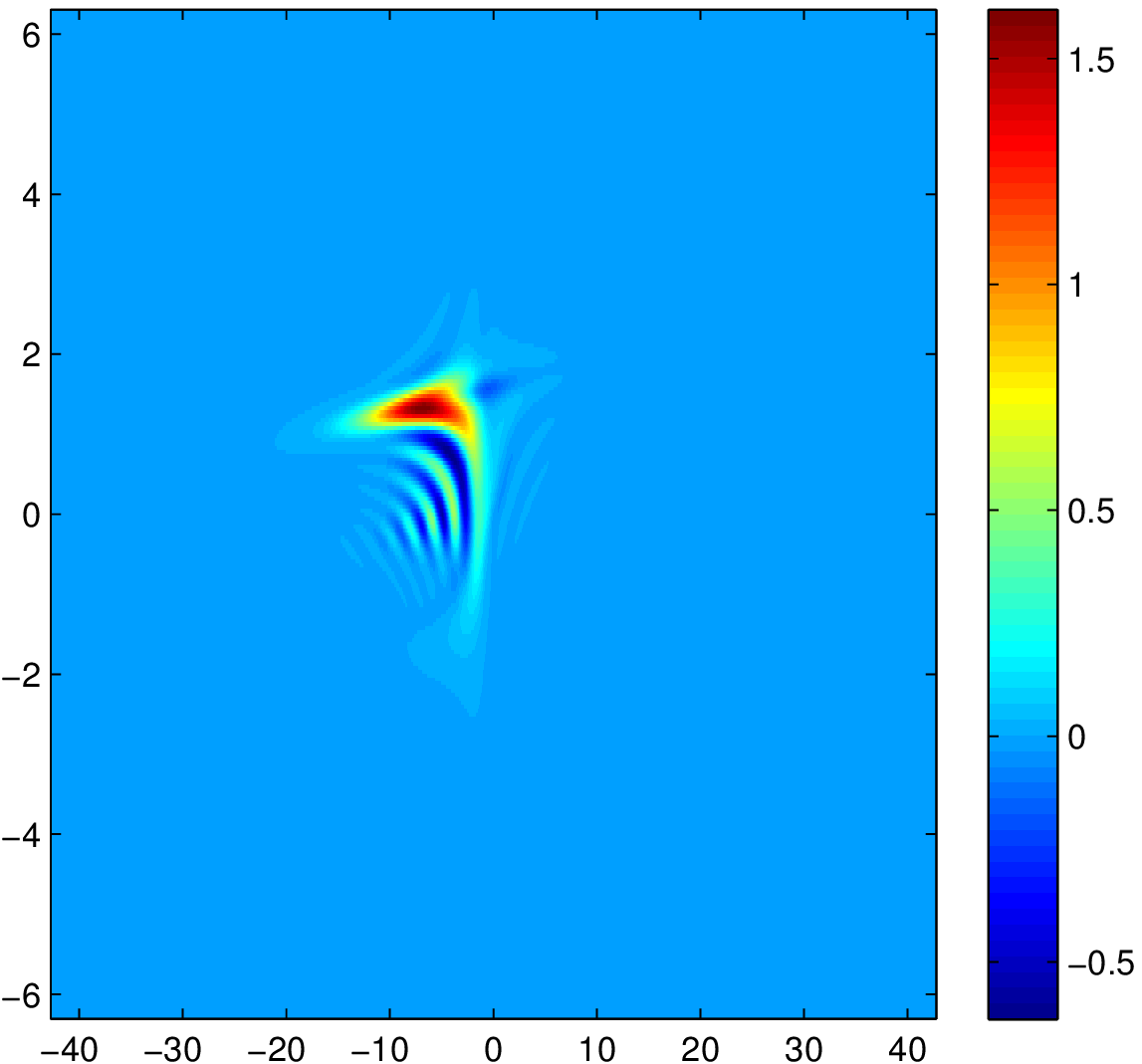}}
    \\
    \centering
    \subfigure[t=12.5]{
    \label{fig:subfig:c}
    \includegraphics[width=2.4in,height=1.8in]{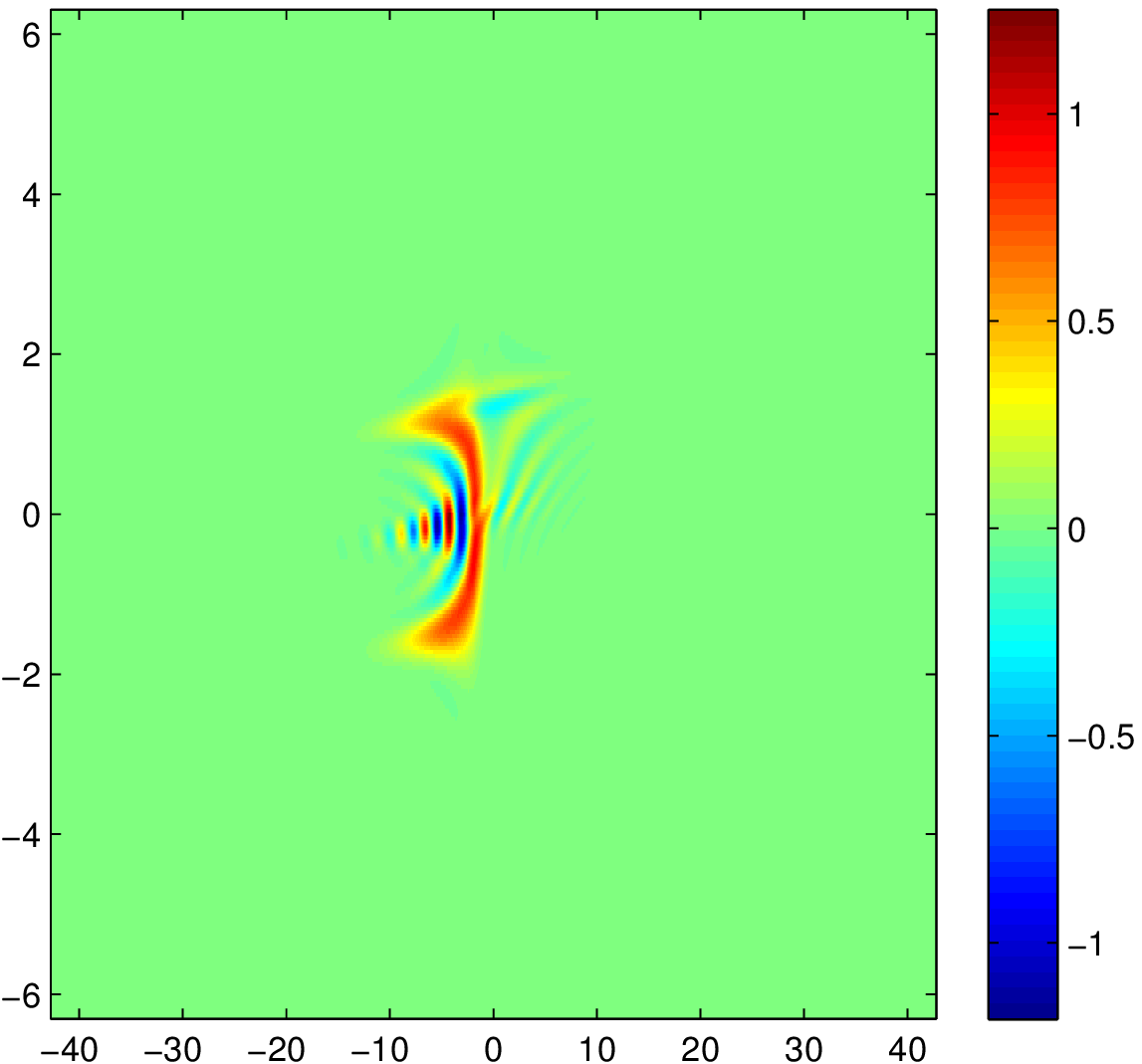}}
    \hspace{0.2in}
    \subfigure[t=15]{
    \label{fig:subfig:d}
    \includegraphics[width=2.4in,height=1.8in]{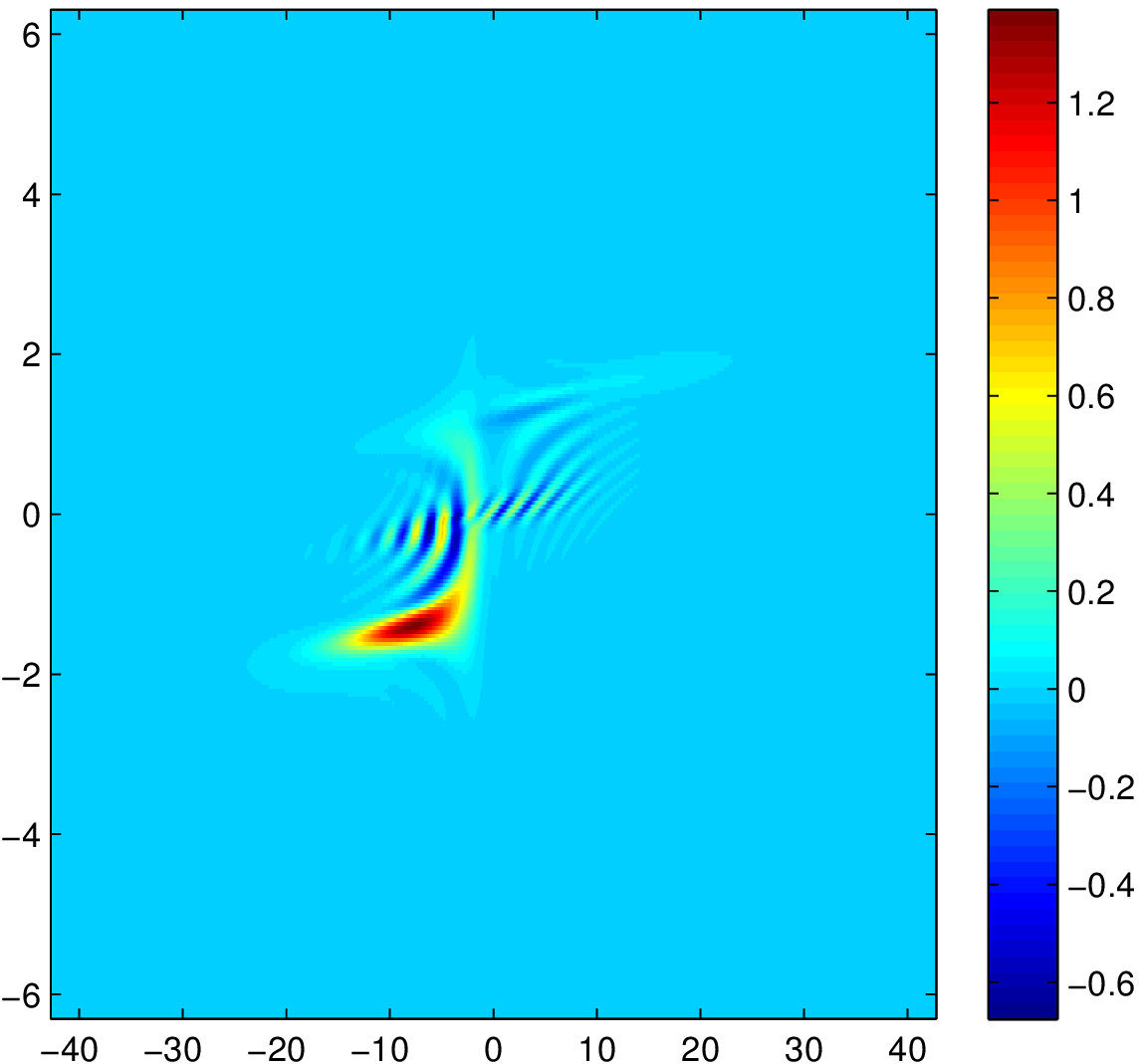}}
    \\
    \centering
    \subfigure[t=17.5]{
    \label{fig:subfig:e}
    \includegraphics[width=2.4in,height=1.8in]{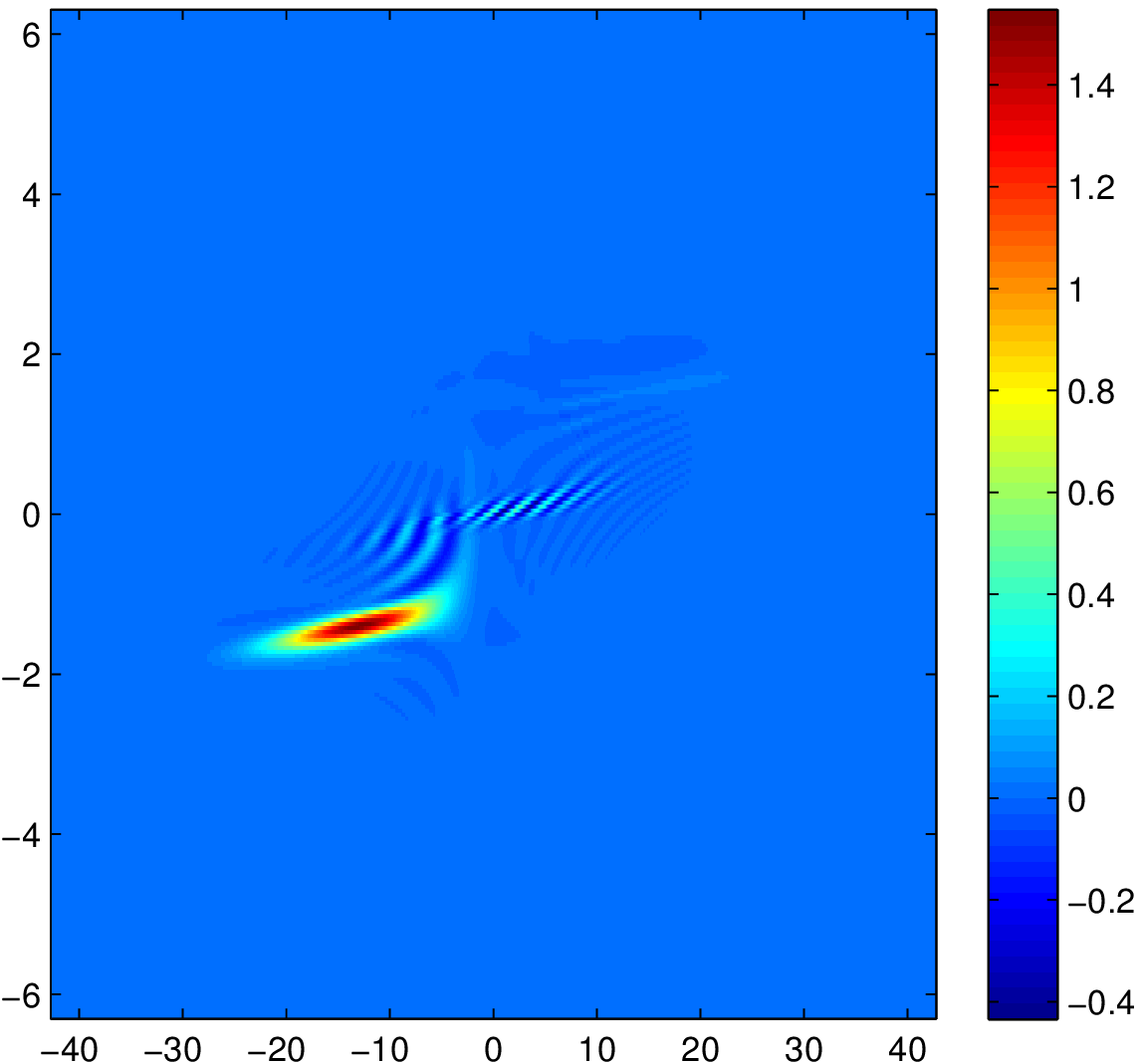}}
    \hspace{0.2in}
    \subfigure[t=20]{
    \label{fig:subfig:f}
    \includegraphics[width=2.4in,height=1.8in]{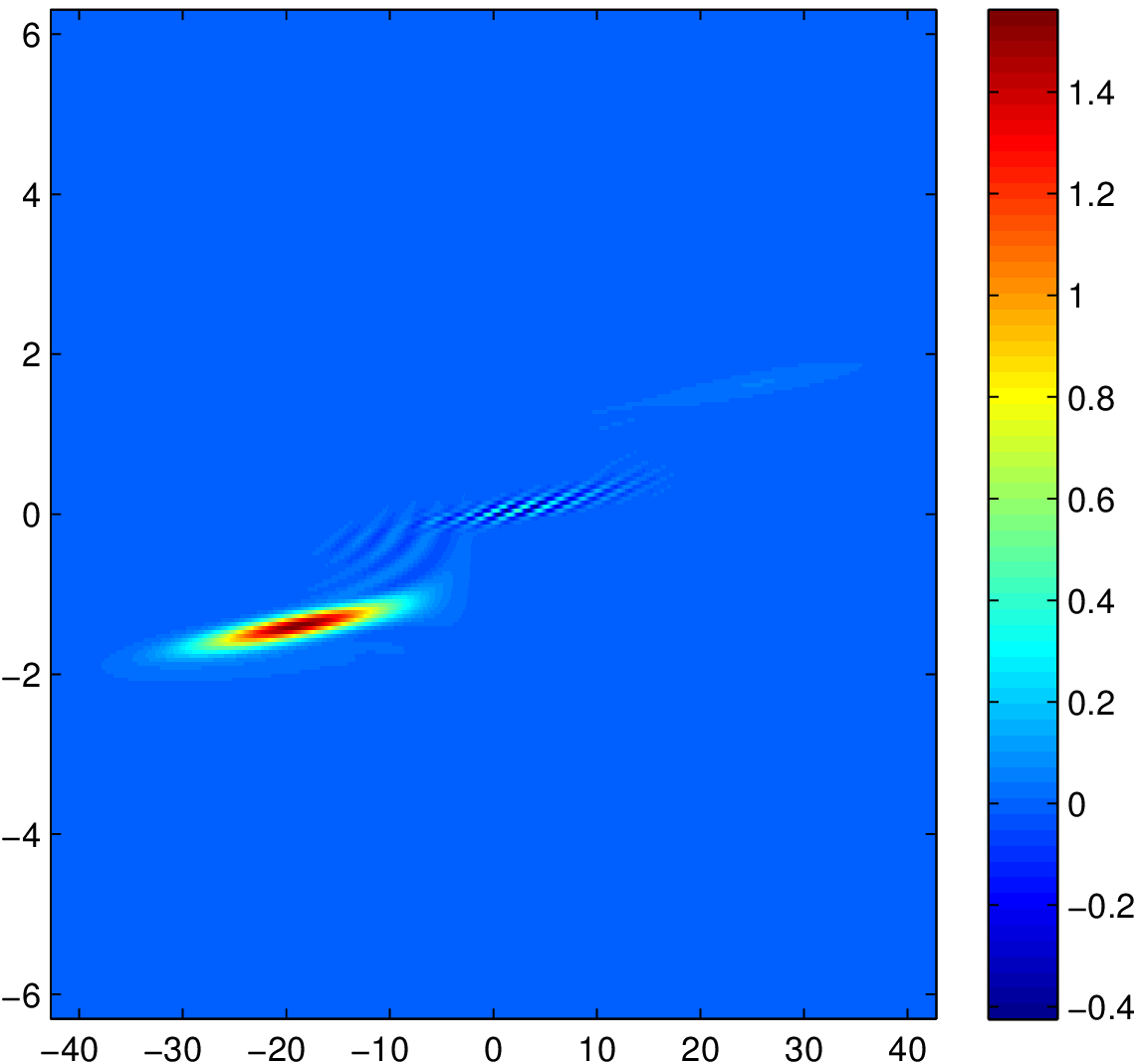}}
     \caption{The time evolution of a GWP interacting with a Gaussian barrier $\left(H=2.3\right)$}  
\end{figure}

We turn to the collisional Wigner equation.  The exact solution of Eq.(27) is also not trivial, even with a flat potential. For a special case, the analytical solution can be solved by separation of variables. Assume that the wave (the Wigner distribution) moves in a uniform velocity $v_{0}=\frac{\hbar k_{0}}{m}$ under a flat potential $V=0$, then the corresponding Wigner equation is
\begin{equation}
\frac{\partial w}{\partial t}+\frac{\hbar k_{0}}{m} \frac{\partial w}{\partial x}=\frac{1}{\tau}\left(\frac{n}{n_{0}}w_{0}-w\right).
\end{equation}
It is easy to verify that the exact solution of Eq.(37) is
\begin{equation}
w\left(x,k,t\right)=2exp\left[-\frac{\left(x-x_{0}-v_{0}t\right)^{2}}{2\alpha^{2}}\right]exp\left[-2\alpha^{2}\left(k-k_{0}\right)^{2}\right].
\end{equation}

The collision term involves an integral $n\left(x,t\right)=\int w\left(x,k,t\right) dk$, which can be approximated by the composite Simpson rule at the nodes $k_{-N}$ to $k_{N}$,
\begin{equation} 
n\left(x,t\right)=\frac{\Delta k}{3}\left(f\left(x, k_{-N},t\right)+f\left(x,k_{N},t\right)+4\sum_{i=1}^{N} f\left(x, k_{2i-1-N},t\right)+ 2\sum_{i=1}^{N-1}  f\left(x, k_{2i-N},t\right)\right).
\end{equation}
In this case, the accuracy of explicit three-step method is demonstrated in Figure 7, with time step $\Delta t=0.05$, $N_{x}=20$ and the relaxation time $\tau=1$.
\begin{figure}[!h]
    \centering
    \subfigure[The exact Wigner function at $t=20$]{
    \label{fig:subfig:a}
    \includegraphics[width=2.4in,height=1.8in]{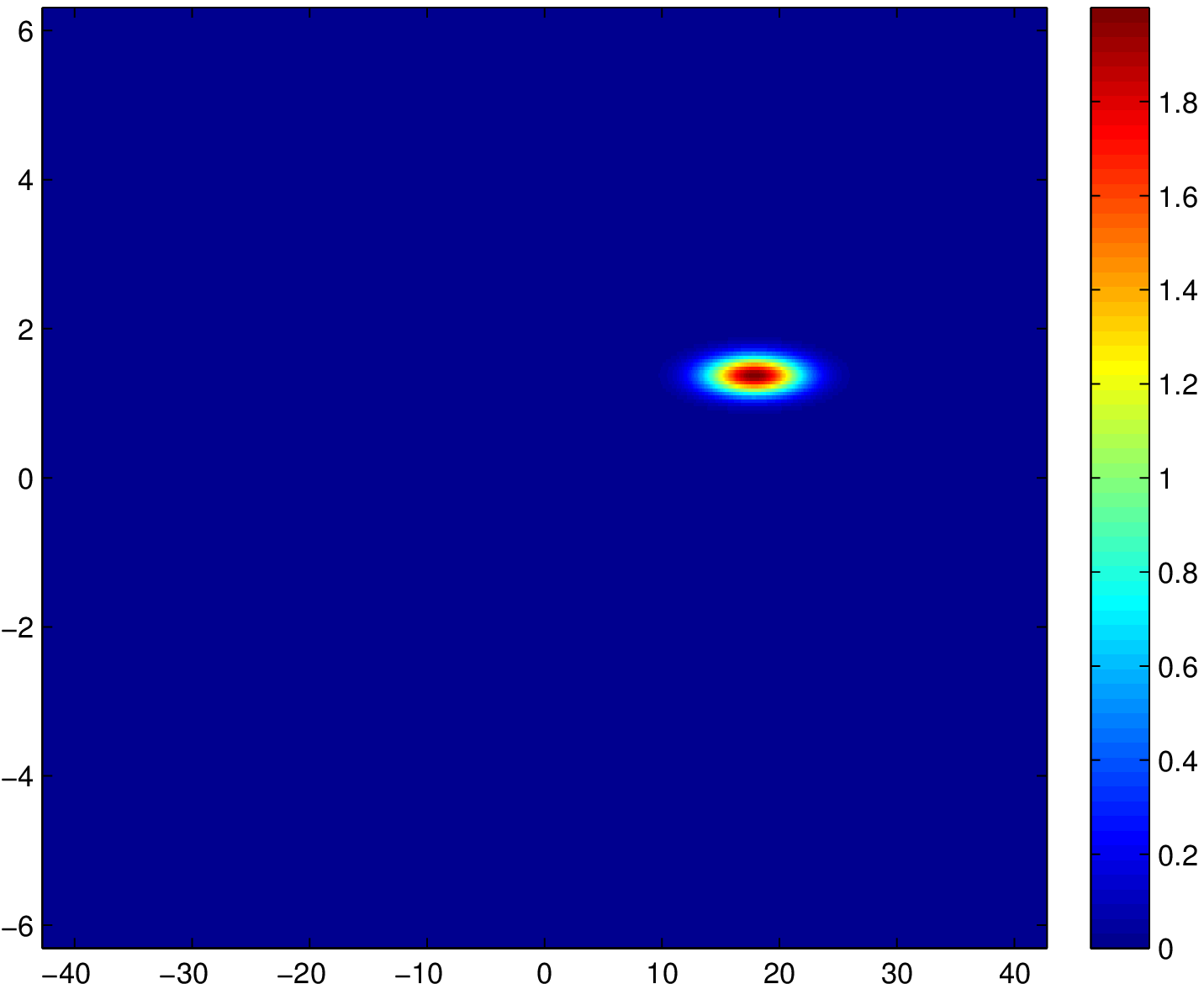}}
    \hspace{0.2in}
    \subfigure[$\Delta w$ at $t=20$, by explicit three-step method]{
    \label{fig:subfig:b}
    \includegraphics[width=2.4in,height=1.8in]{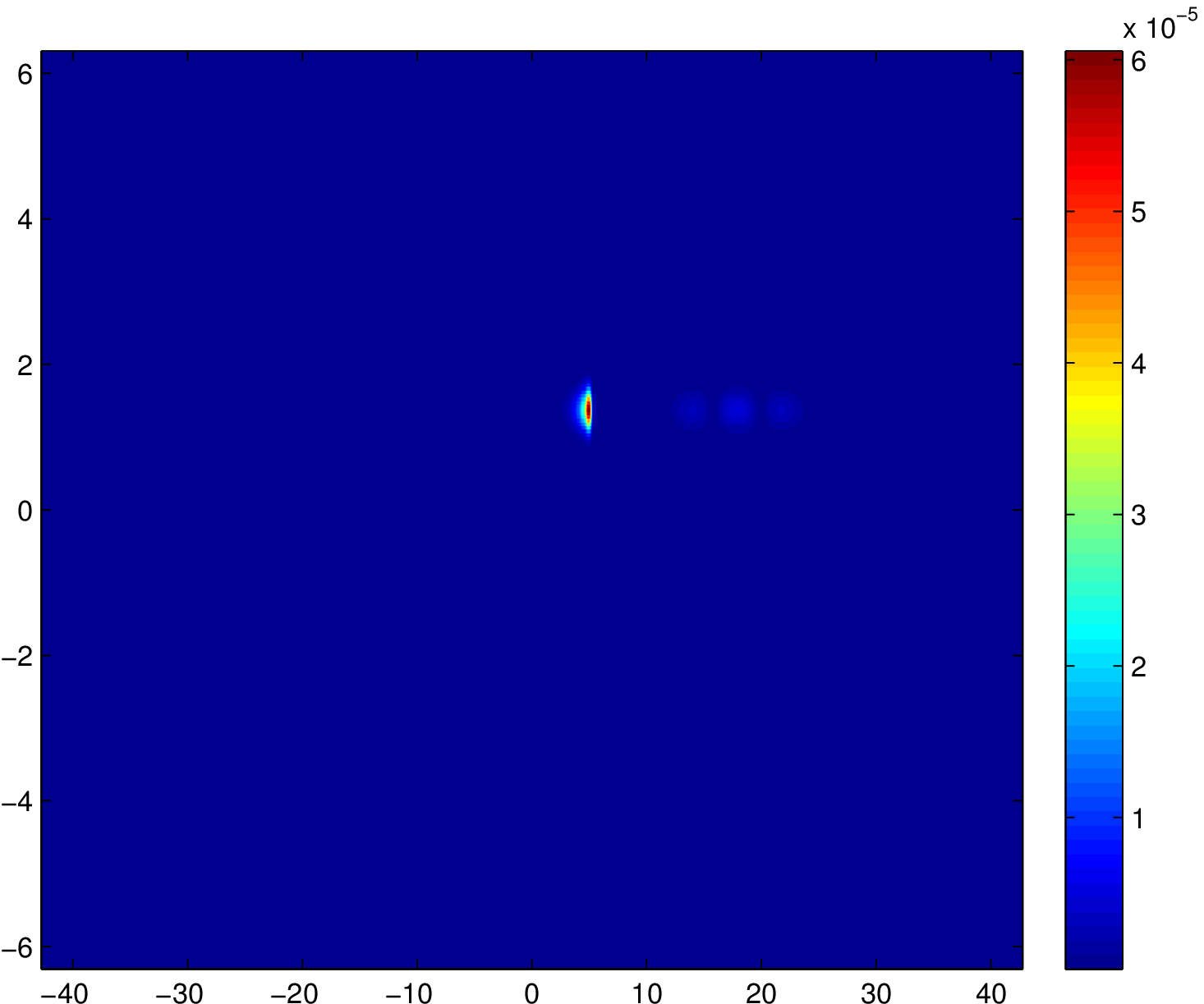}}
     \caption{Numerical error of the Wigner function in the collisional case $\left(\tau=1, V=0\right)$}  
\end{figure}

To make a further investigation of the convergence and robustness of Adams multistep scheme, a comparison is made between numerical results with different time steps or different numerical resolutions. The relaxation time $\tau$ is chosen $1$ and the wave velocity $v_{0}=1.3744$. We choose the same performance metrics, $L^{2}$ error and $L^{\infty}$ error, to monitor the numerical errors.  

As shown in Figure 8, the Adams method gives accurate numerical results, while the time step $\Delta t=0.1$ seems to yield the best numerical result. It is because the numerical errors come from both cubic spline interpolation and numerical integration.  A larger time step may increase the global error in time evolution, but it needs less interpolated grid points, thereby reducing the possible numerical error resulting from interpolations simultaneously. Therefore, an appropriate time step should be chosen to strike a balance between accuracy and efficiency. 

The relation between numerical errors and numerical resolution $N_{x}$ is plotted in Figure 9, where the time step is chosen to be $0.1$. The numerical solution is accurate and the convergence is clear. In this simple case, the lowest resolution $N_{x}=10$ gives the best numerical results, since the numerical error mainly comes from the discrete approximation of collisional term. Hence, this test makes us confident that a relatively low numerical resolution can be chosen to achieve the efficiency, without too much loss of accuracy. 
\begin{figure}[!h]
    \centering
    \subfigure[$L^{2}$ error for the Wigner functions $\left(\tau=1, V=0\right)$]{
    \label{fig:subfig:a}
    \includegraphics[width=2.4in,height=1.8in]{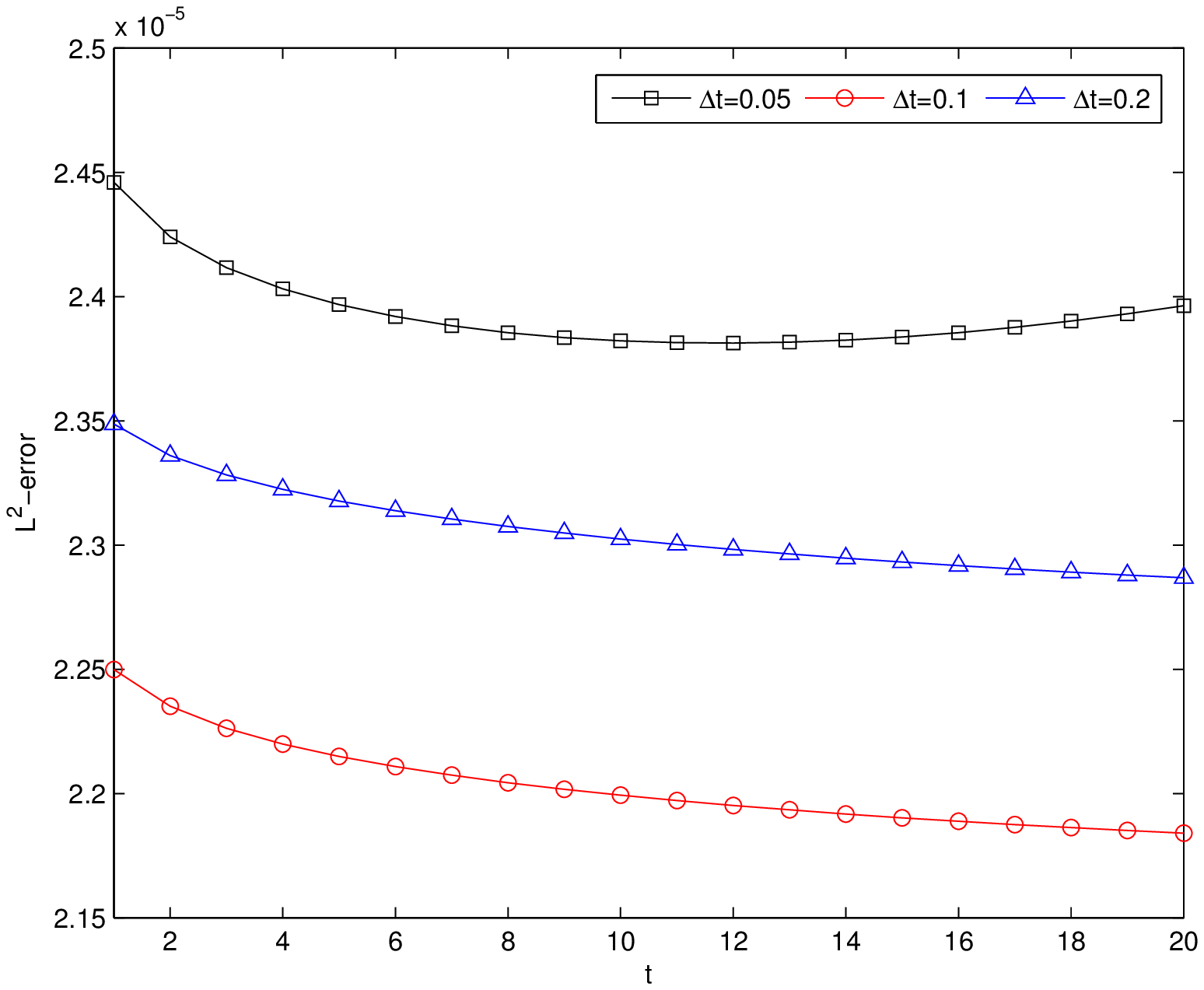}}
    \hspace{0.2in}
    \subfigure[$L^{\infty}$ error for the Wigner functions $\left(\tau=1, V=0\right)$]{
    \label{fig:subfig:b}
    \includegraphics[width=2.4in,height=1.8in]{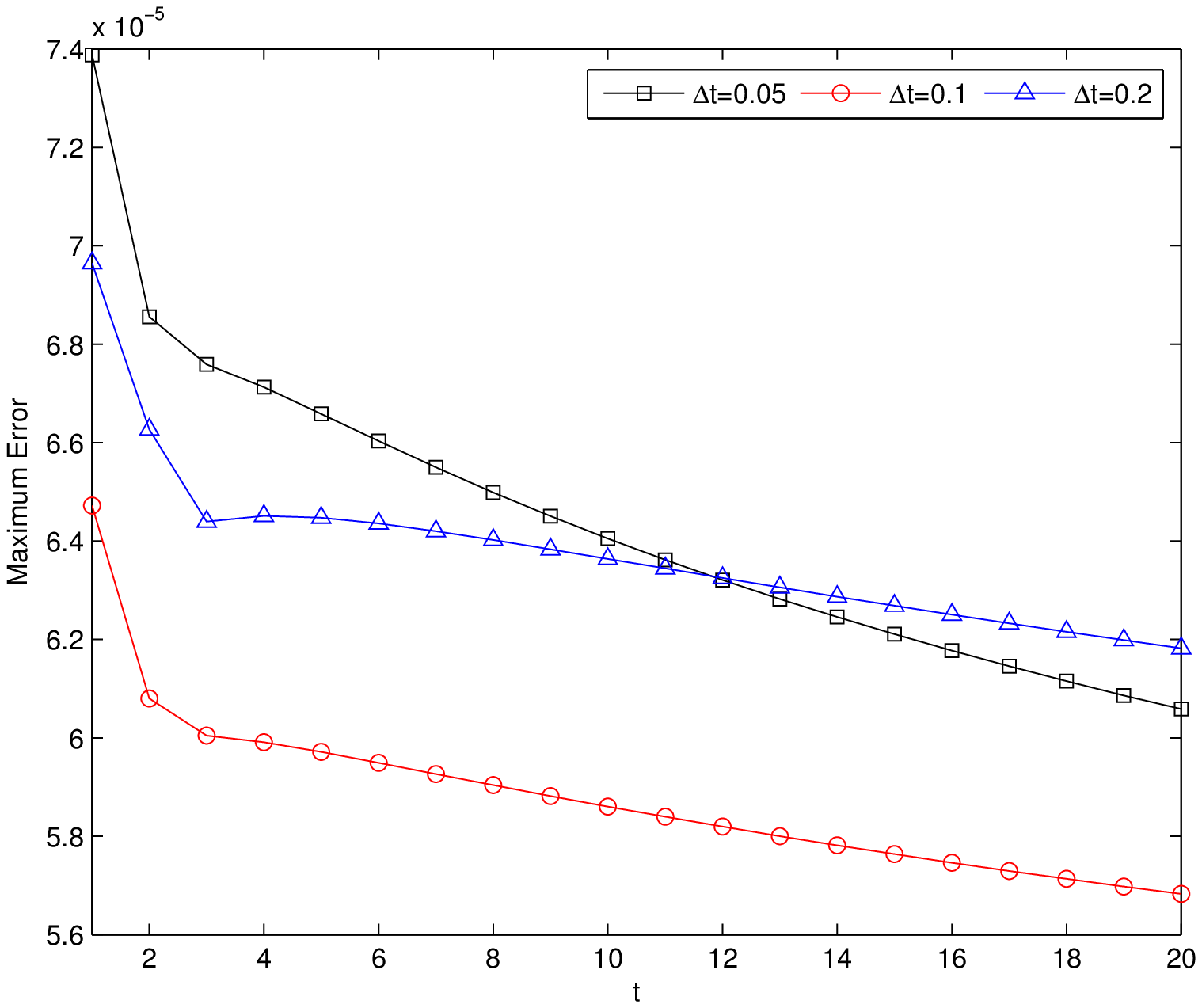}}
     \caption{Numerical errors for the test problem Eq.(37) with respect to time step $\Delta t$}      
\end{figure} 
     
\begin{figure}[!h]
    \centering
    \subfigure[$L^{2}$ error for the Wigner functions $\left(\tau=1, V=0\right)$]{
    \label{fig:subfig:a}
    \includegraphics[width=2.4in,height=1.8in]{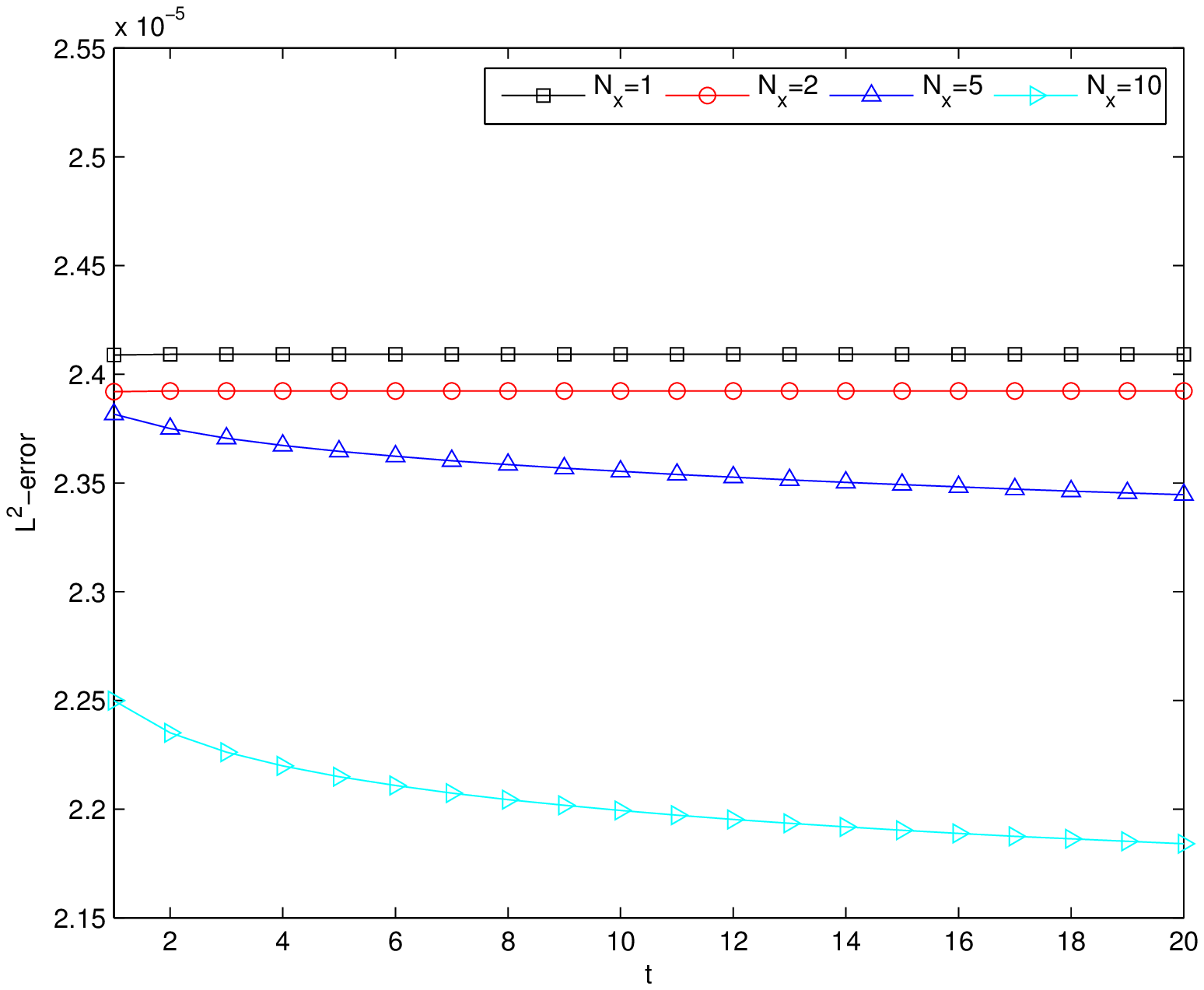}}
    \hspace{0.2in}
    \subfigure[$L^{\infty}$ error for the Wigner functions $\left(\tau=1, V=0\right)$]{
    \label{fig:subfig:b}
    \includegraphics[width=2.4in,height=1.8in]{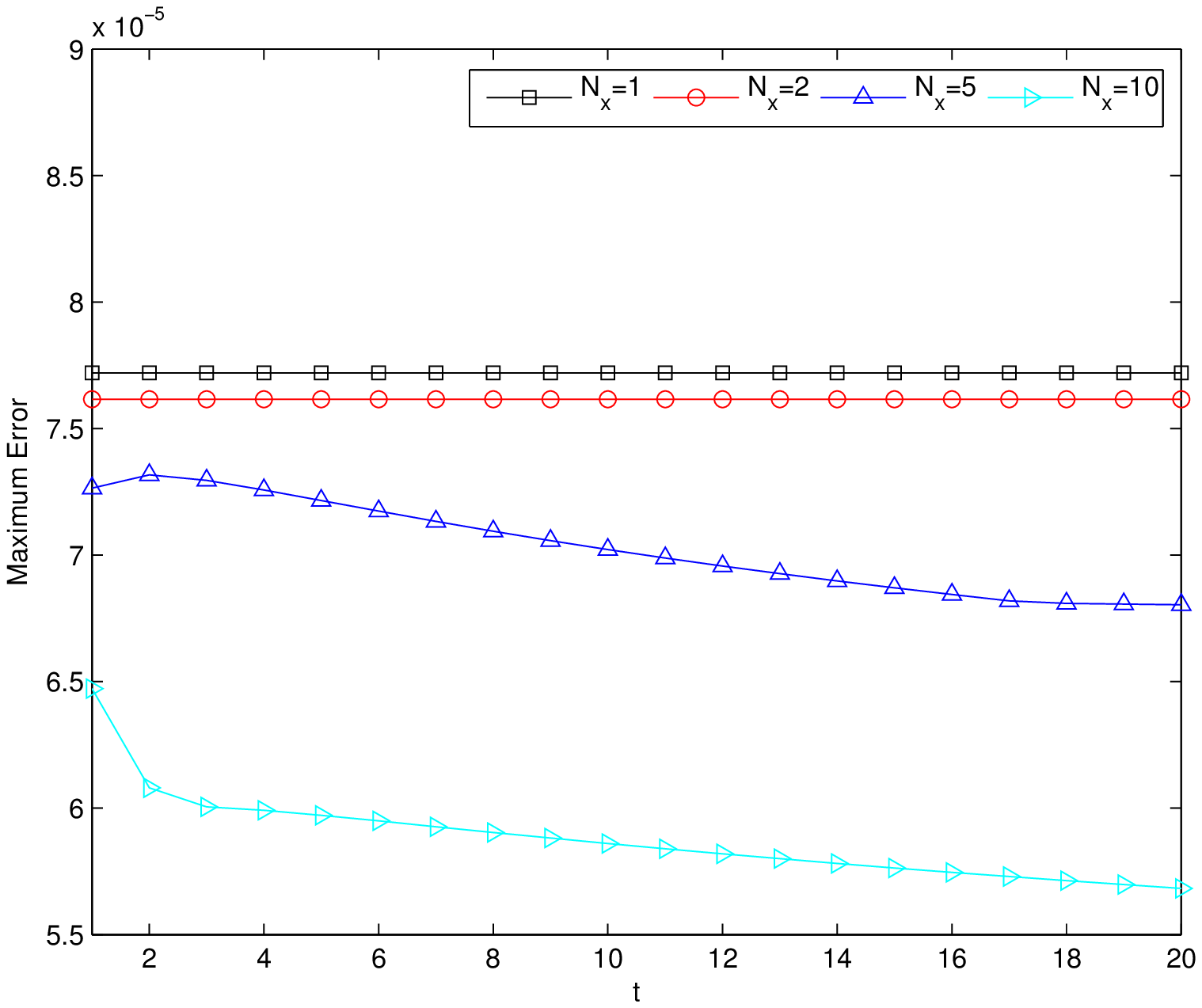}}
     \caption{Numerical errors for the test problem Eq.(37) with respect to numerical resolution $N_{x}$}  
\end{figure}
 
Now we include the quantum interference in the model and discuss how the scattering process influences the quantum tunneling effect, by solving Eq.(27) numerically. We let a GWP hit a Gaussian barrier $V\left(x\right)=2.3e^{-\frac{x^{2}}{2}}$ and investigate the time evolution of the Wigner function under different relaxation time $\tau$. 

Before our discussion, it needs to investigate the convergence of Adam multistep scheme in the collisional case. The explicit three-step method is used, with different time step $\Delta t=0.05, 0.1, 0.2$, respectively. The relaxation time $\tau$ is chosen as $1$. A uniform grid mesh is used, with $\Delta x=\frac{\pi\hbar}{64m}$, $\Delta k=\frac{\pi}{64}$.  The numerical solutions with $\Delta t=0.05$ and high resolution $N_{x}=1$ are chosen as the reference. We plot the convergence history in Figure 10. When the time step goes smaller, an obvious error reduction is observed, which validates the convergence of the explicit three-step Adams method. The error induced by cubic spline interpolation is also negligible (in fact, both $L^{2}$ error and $L^{\infty}$ error are less than $1\times 10^{-4}$).

\begin{figure}[!h]
    \centering
    \subfigure[$L^{2}$ error for the collisional case $\left(\tau=1, H=2.3\right)$]{
    \label{fig:subfig:a}
    \includegraphics[width=2.4in,height=1.8in]{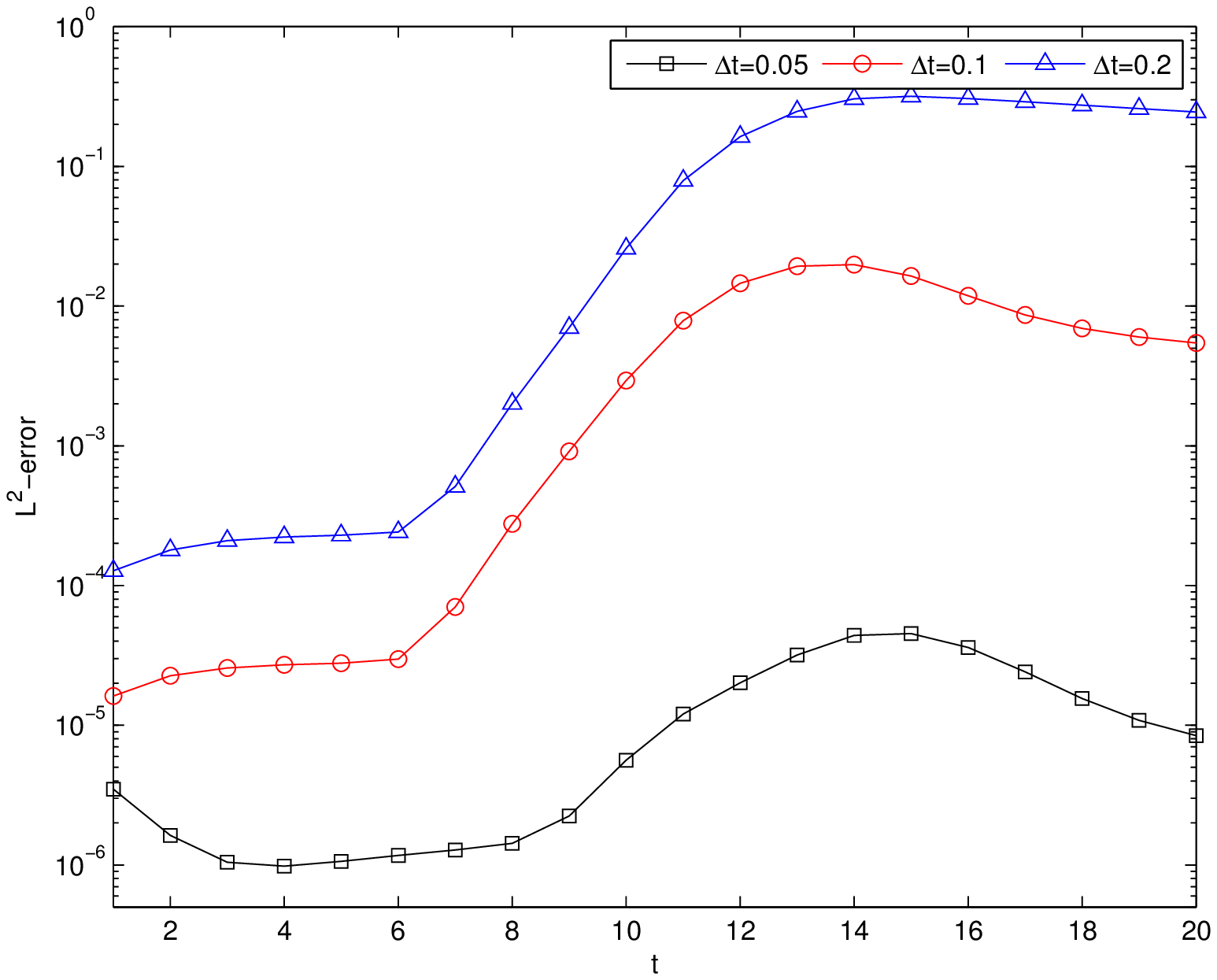}}
    \hspace{0.2in}
    \subfigure[$L^{\infty}$ error for the collisional case $\left(\tau=1, H=2.3\right)$]{
    \label{fig:subfig:b}
    \includegraphics[width=2.4in,height=1.8in]{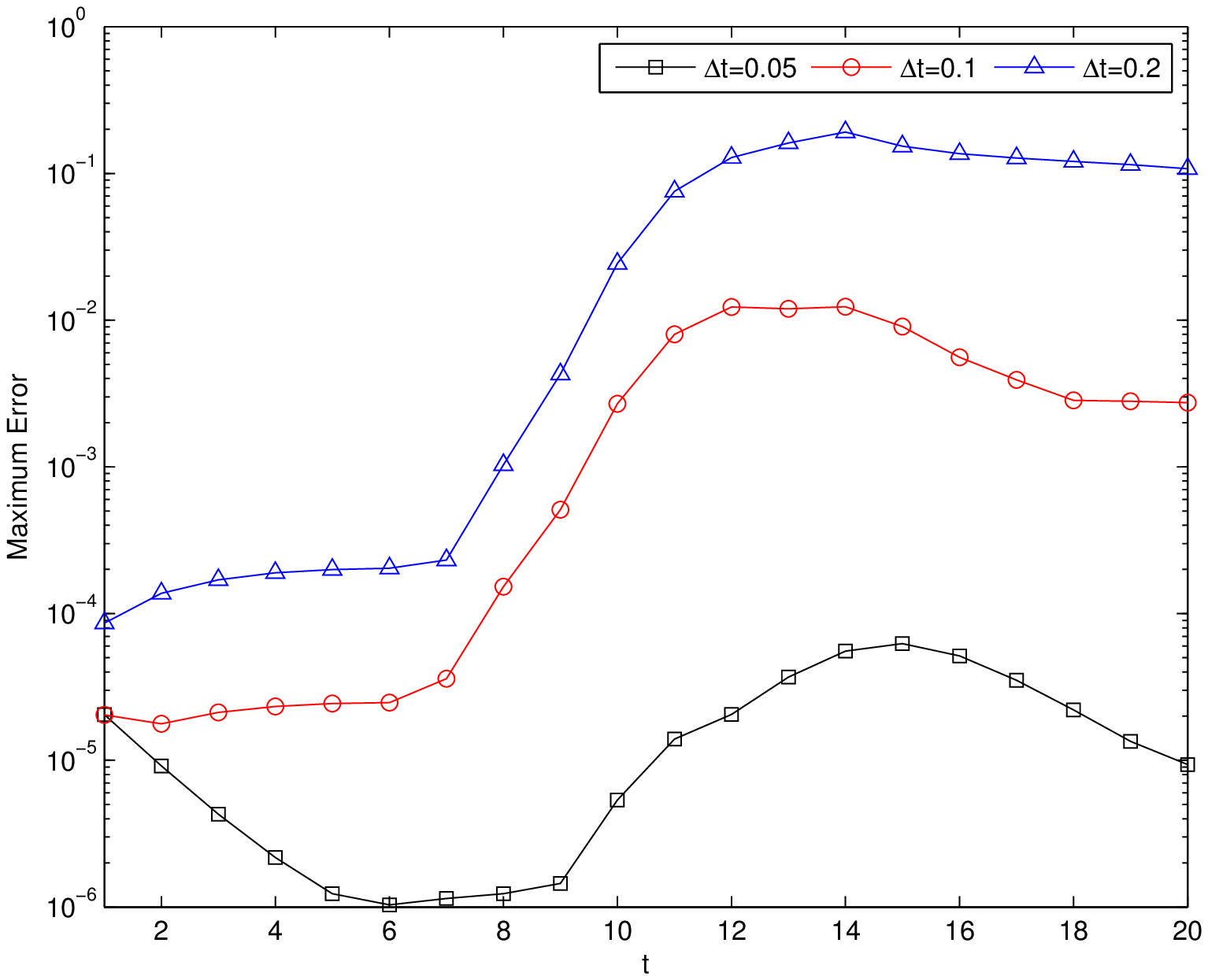}}
     \caption{The convergence history of the explicit three-step method with respect to $\Delta t$ ($H = 2.3$, $\tau=1$)}  
\end{figure}

Finally, we study the scattering effect through the relaxation time model. If the relaxation time $\tau$ is sufficiently large, a weak dissipation is expected. On the other hand, when the dissipation is strong enough (for a small $\tau$), it's easy for the perturbed Wigner function to return into its equilibrium state. Now we choose a small $\tau=1$. It is observed in Figure 11 that the GWP travels across the Gaussian potential easily, although the height of Gaussian barrier is sufficiently large. This observation is quite different from the collisionless case. In addition, the waveform has a change after a quantum mechanical interaction with a high barrier potential. It is noticed that the oscillatory structure of the Wigner distribution is still observed around $k=0$.
\begin{figure}[!h]
    \centering
    \subfigure[t=7.5]{
    \label{fig:subfig:a}
    \includegraphics[width=2.4in,height=1.8in]{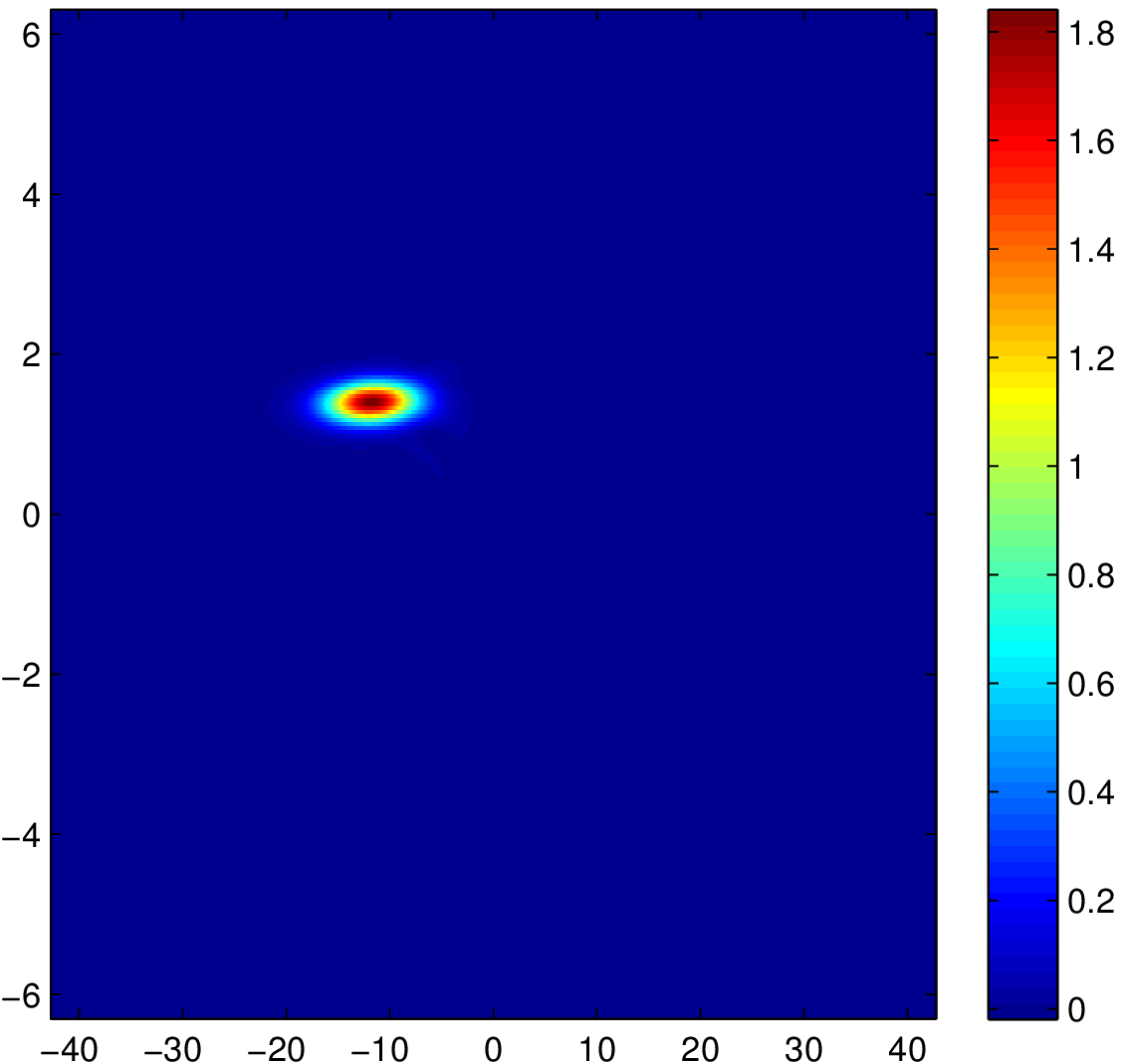}}
    \hspace{0.2in}
    \subfigure[t=10]{
    \label{fig:subfig:b}
    \includegraphics[width=2.4in,height=1.8in]{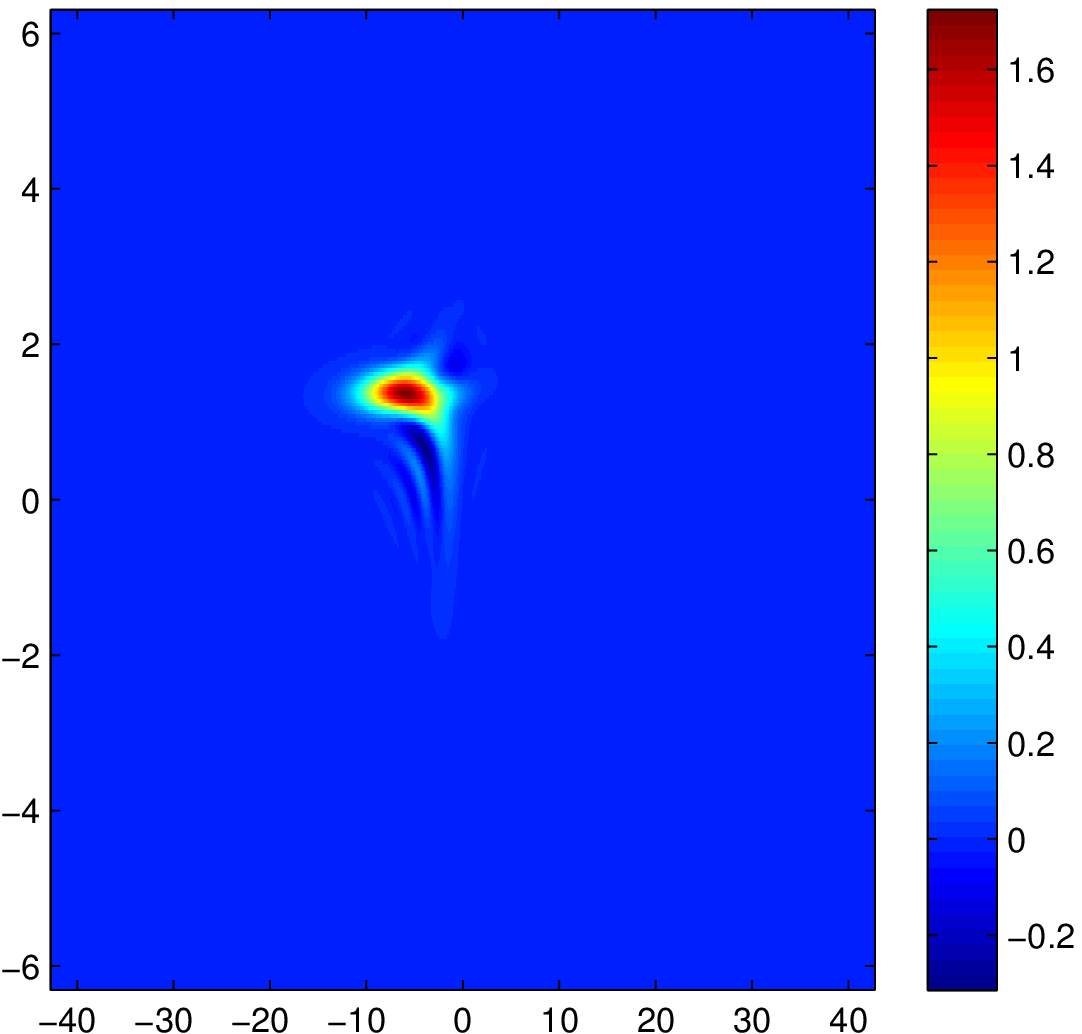}}
    \\
    \centering
    \subfigure[t=12.5]{
    \label{fig:subfig:c}
    \includegraphics[width=2.4in,height=1.8in]{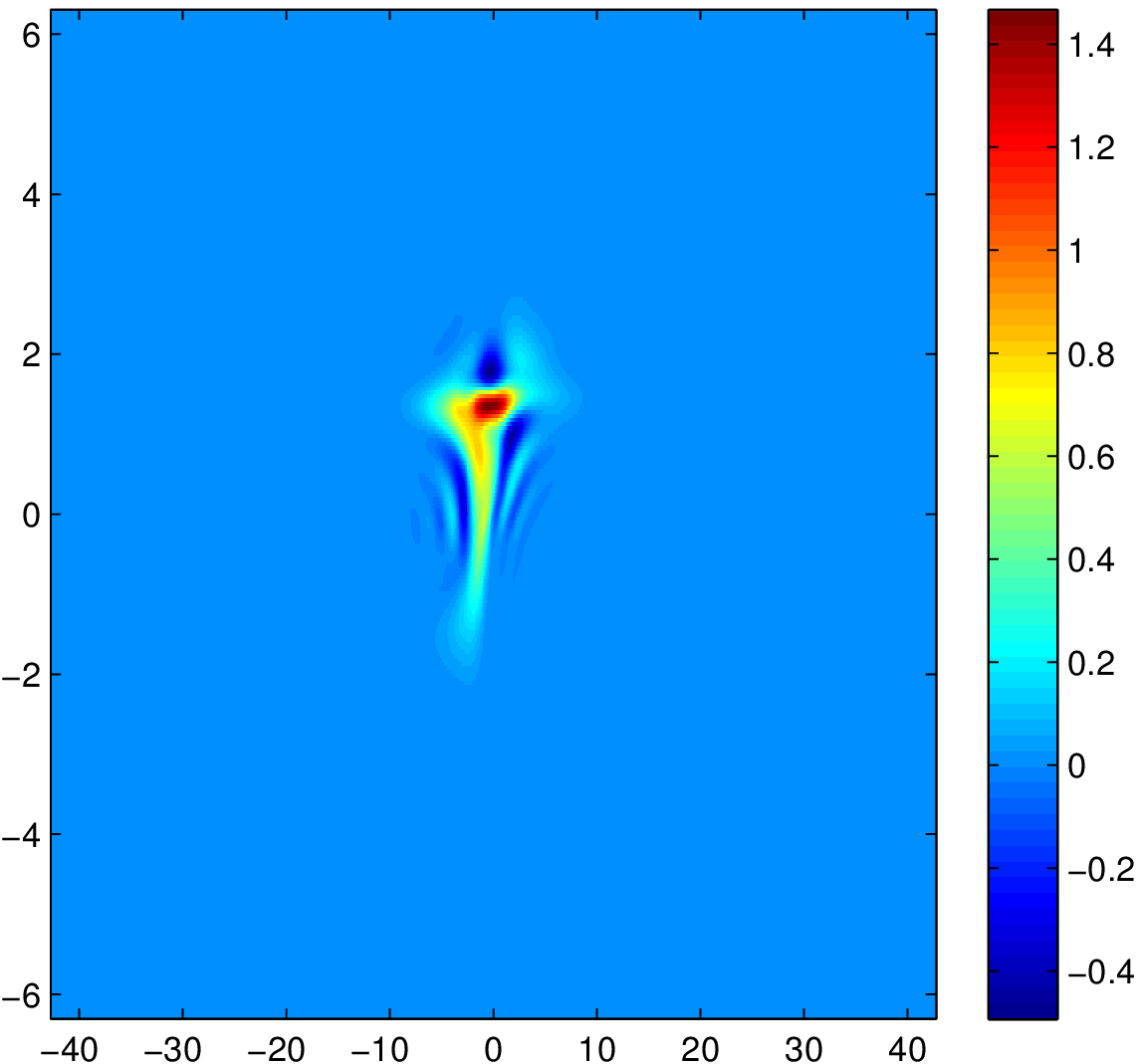}}
    \hspace{0.2in}
    \subfigure[t=15]{
    \label{fig:subfig:d}
    \includegraphics[width=2.4in,height=1.8in]{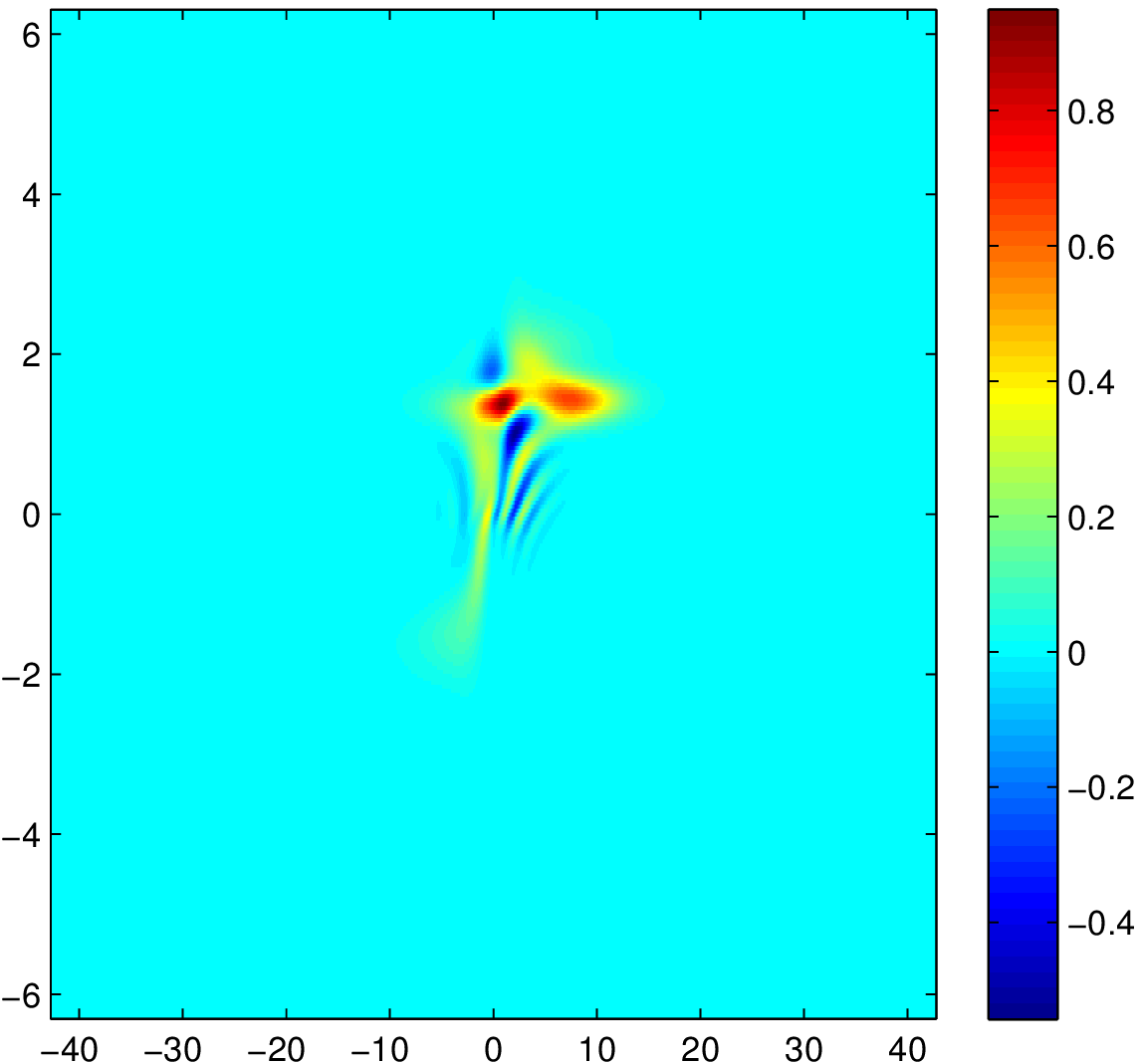}}
    \\
    \centering
    \subfigure[t=17.5]{
    \label{fig:subfig:e}
    \includegraphics[width=2.4in,height=1.8in]{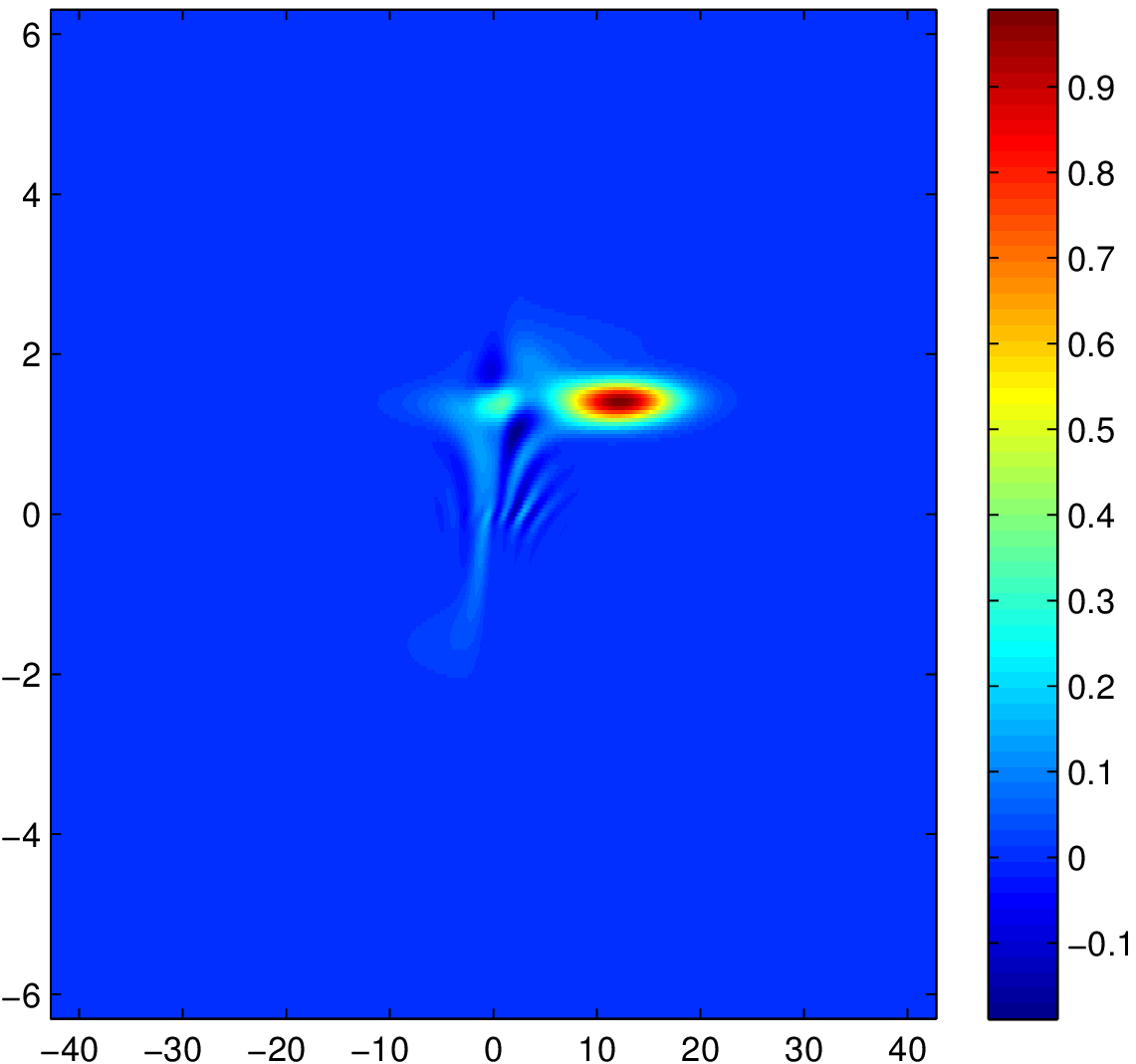}}
    \hspace{0.2in}
    \subfigure[t=20]{
    \label{fig:subfig:f}
    \includegraphics[width=2.4in,height=1.8in]{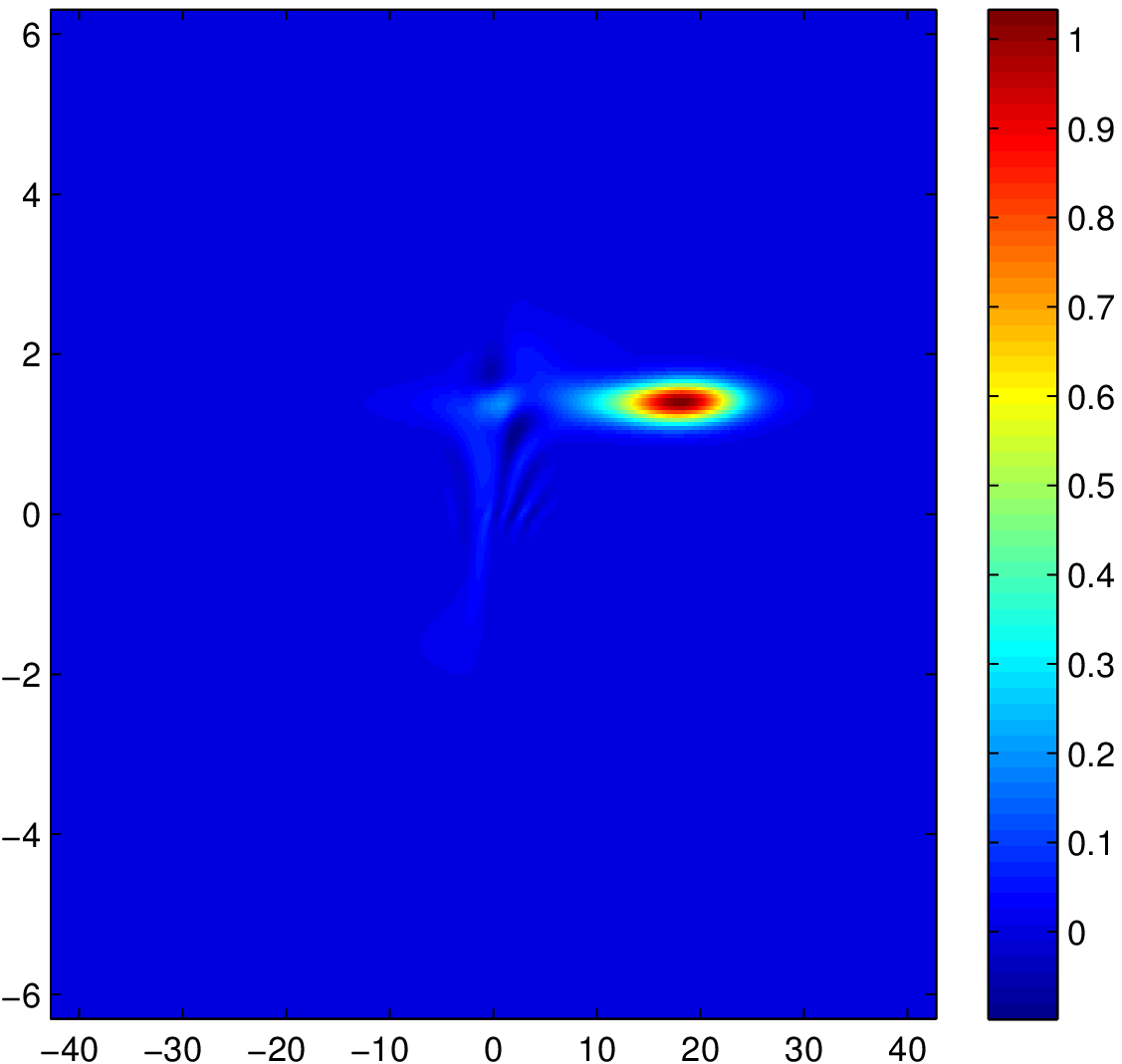}}
     \caption{The time evolution of a GWP interacting with a Gaussian barrier $\left(H=2.3, \tau=1\right)$}  
\end{figure}

As the relaxation time $\tau$ grows larger, the quantum effect becomes more obvious. Figure 12 shows that under a weaker dissipation $(\tau=3)$, the wave is still able to travel across a high barrier potential partly, while another part is either reflected away or transported back. Since the relaxation time model forces the wave to return into its equilibrium, the reflected wave is separated into two streams.

\begin{figure}[!h]
    \centering
    \subfigure[t=7.5]{
    \label{fig:subfig:a}
    \includegraphics[width=2.4in,height=1.8in]{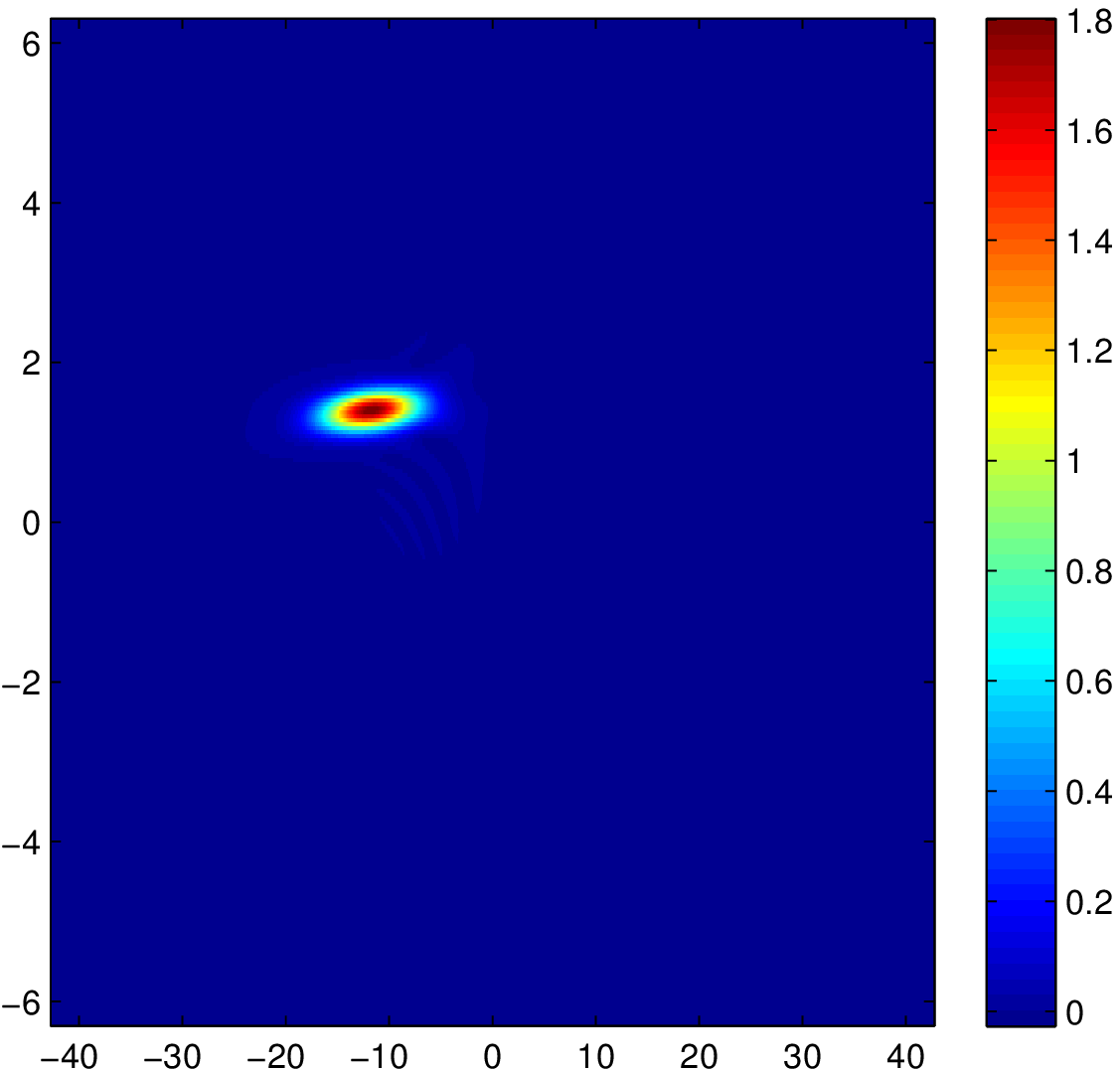}}
    \hspace{0.2in}
    \subfigure[t=10]{
    \label{fig:subfig:b}
    \includegraphics[width=2.4in,height=1.8in]{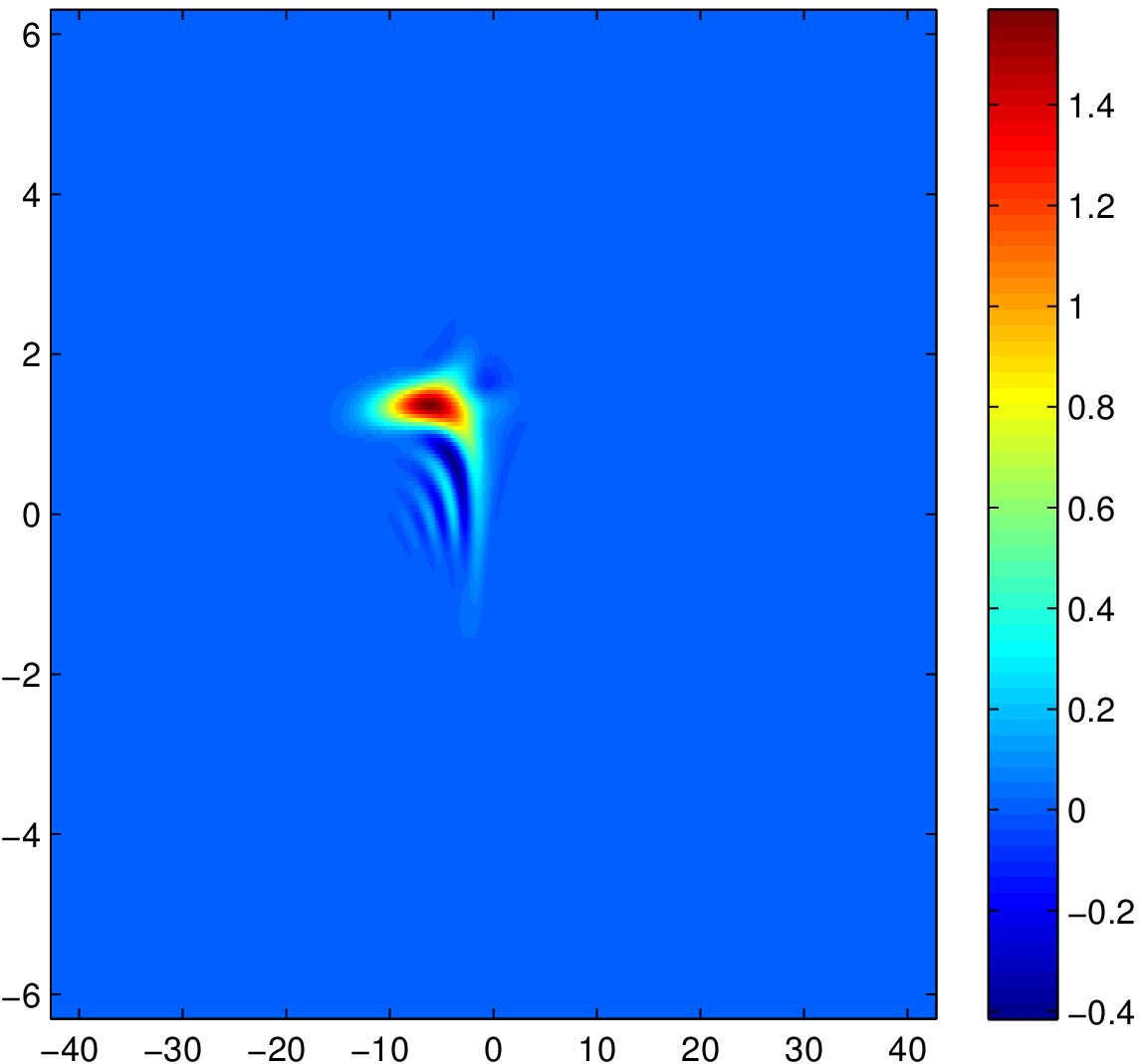}}
    \\
    \centering
    \subfigure[t=12.5]{
    \label{fig:subfig:c}
    \includegraphics[width=2.4in,height=1.8in]{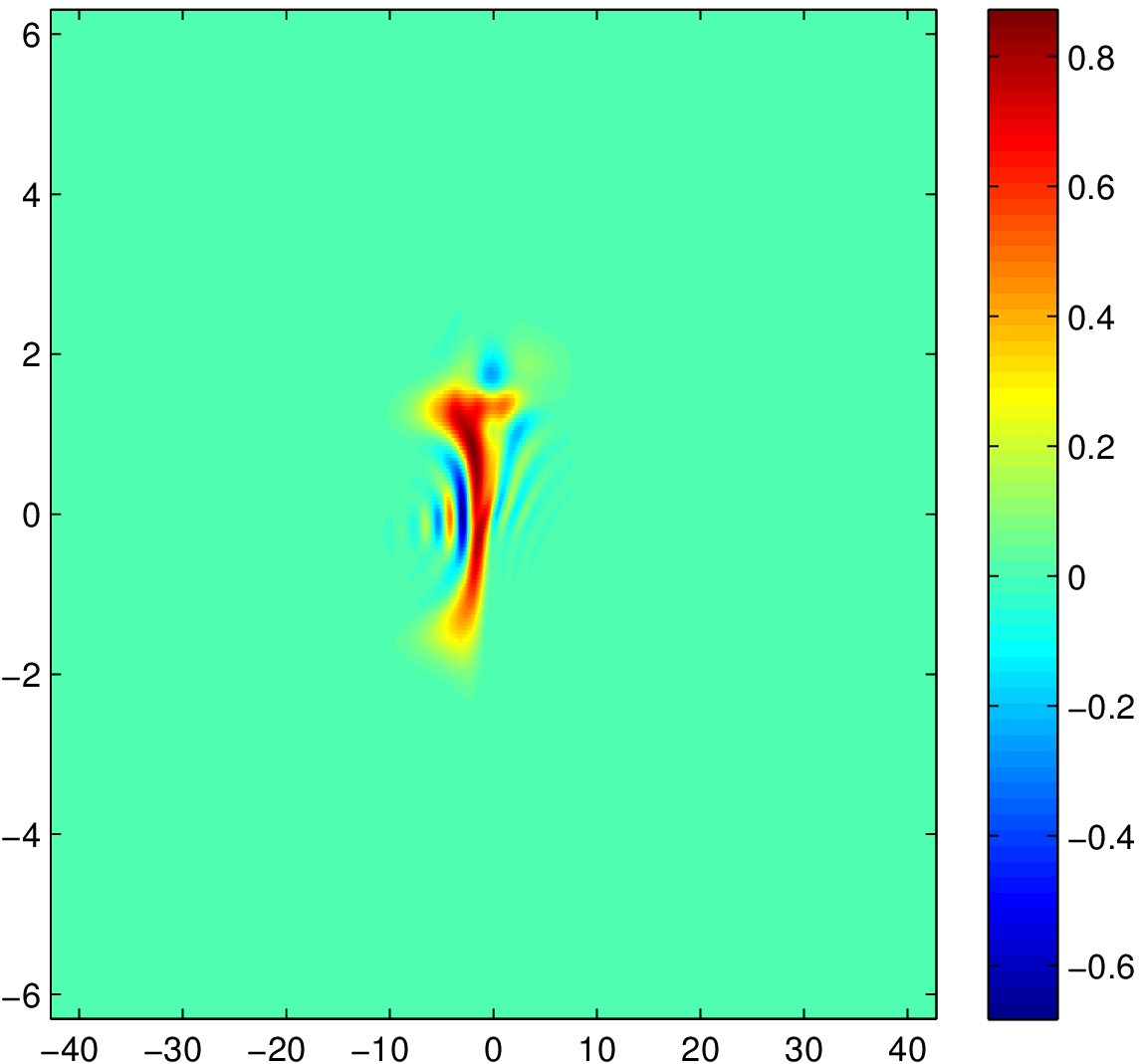}}
    \hspace{0.2in}
    \subfigure[t=15]{
    \label{fig:subfig:d}
    \includegraphics[width=2.4in,height=1.8in]{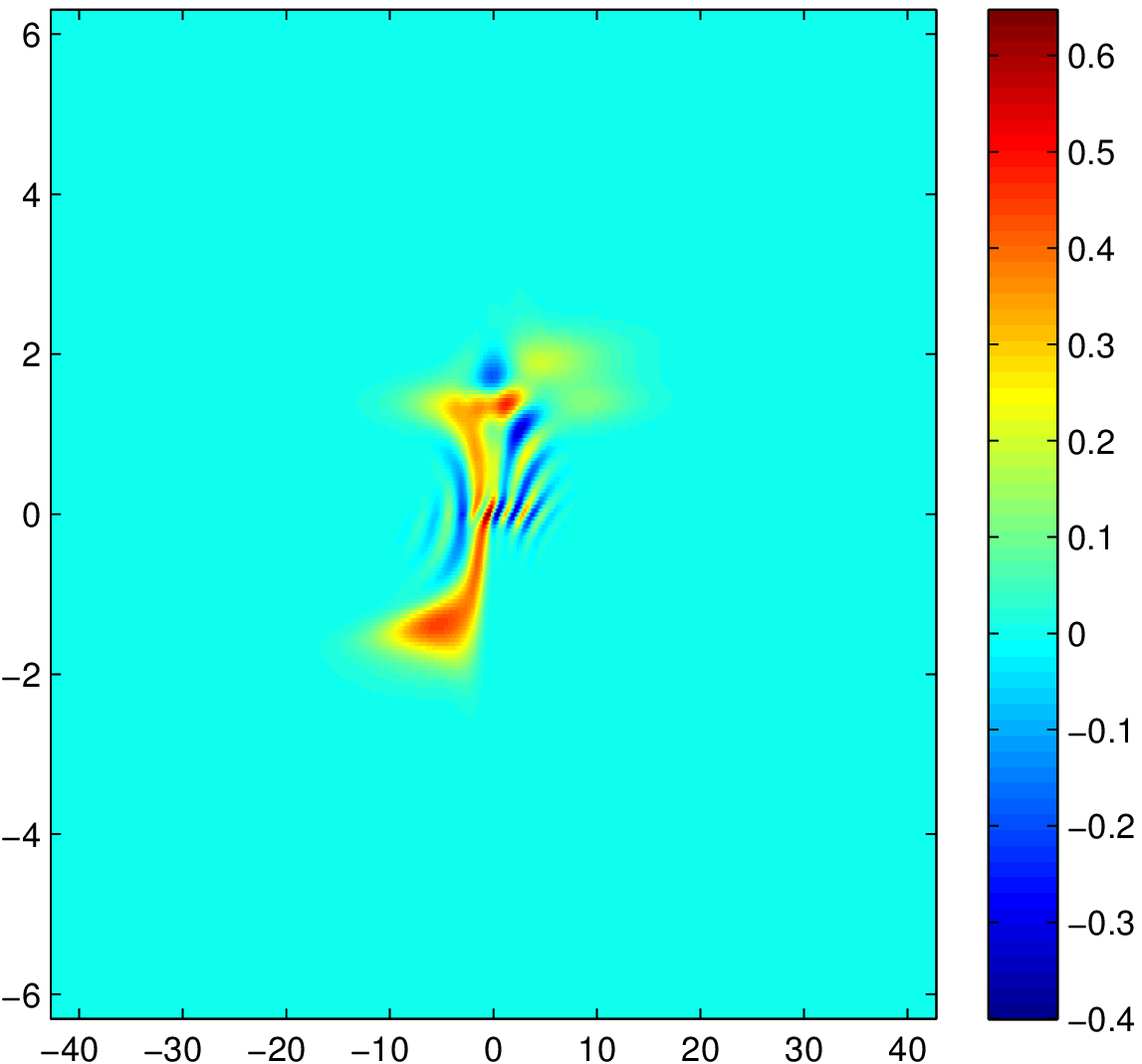}}
    \\
    \centering
    \subfigure[t=17.5]{
    \label{fig:subfig:e}
    \includegraphics[width=2.4in,height=1.8in]{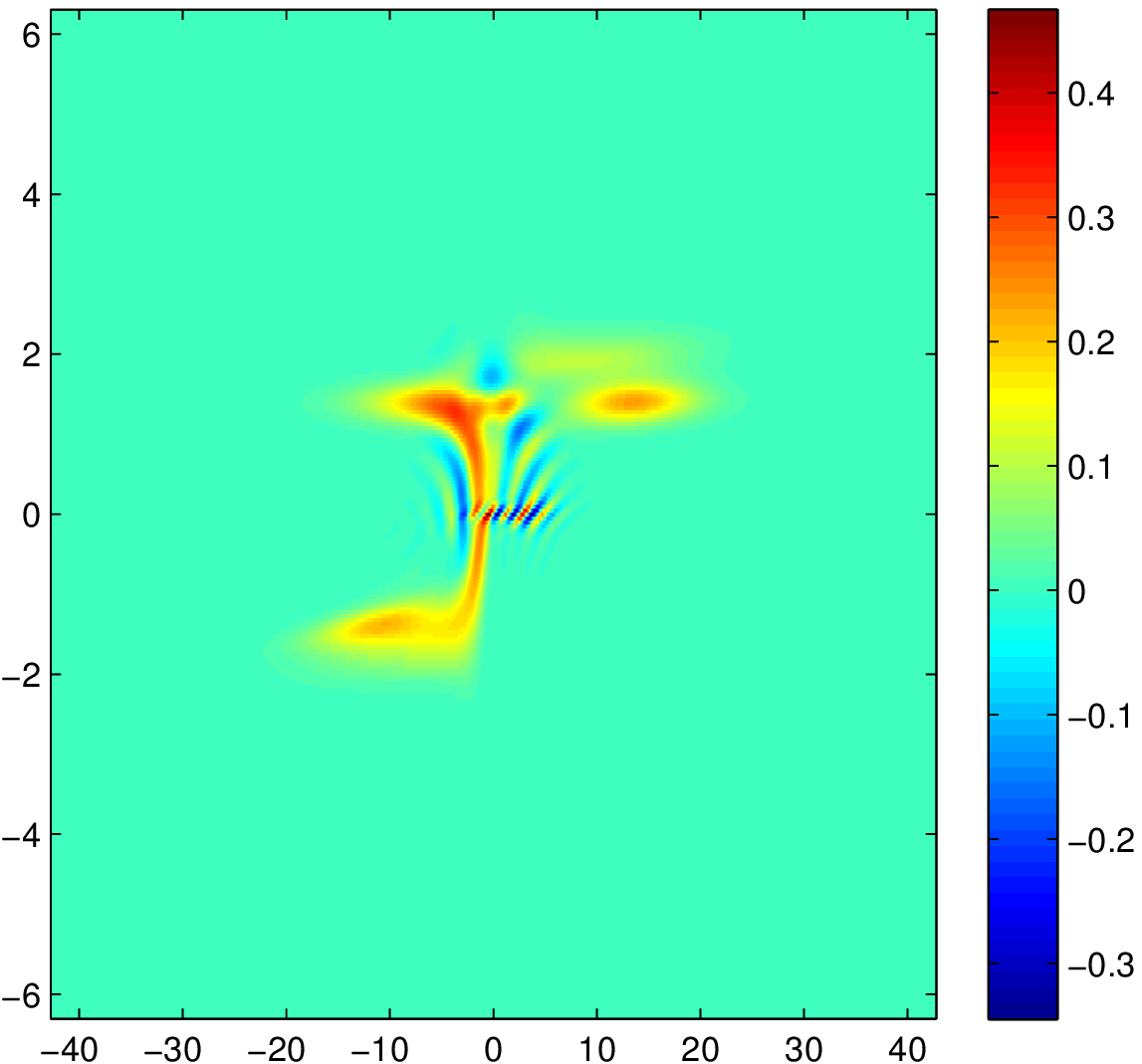}}
    \hspace{0.2in}
    \subfigure[t=20]{
    \label{fig:subfig:f}
    \includegraphics[width=2.4in,height=1.8in]{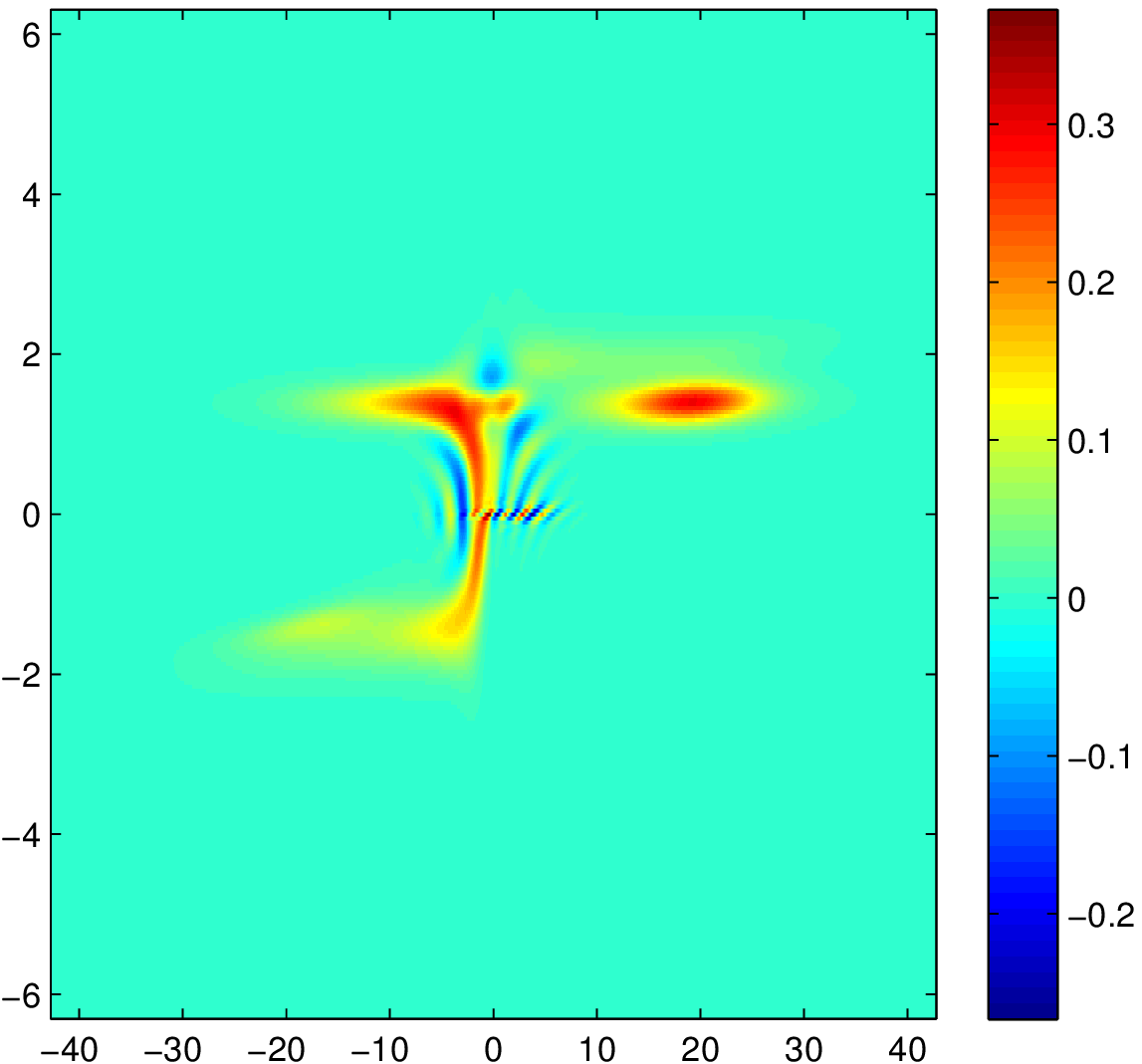}}
     \caption{The time evolution of a GWP interacting with a Gaussian barrier $\left(H=2.3, \tau=3\right)$}  
\end{figure}

When the relaxation time $\tau$ is chosen as $10$, the quantum effect is dominant and the wave packet is expected to be reflected back in a similar way as the collisionless case.  While the separation of wave packet is still observed in Figure 13, due to the mixing of quantum effect and scattering effect. 

\begin{figure}[!h]
    \centering
    \subfigure[t=7.5]{
    \label{fig:subfig:a}
    \includegraphics[width=2.4in,height=1.8in]{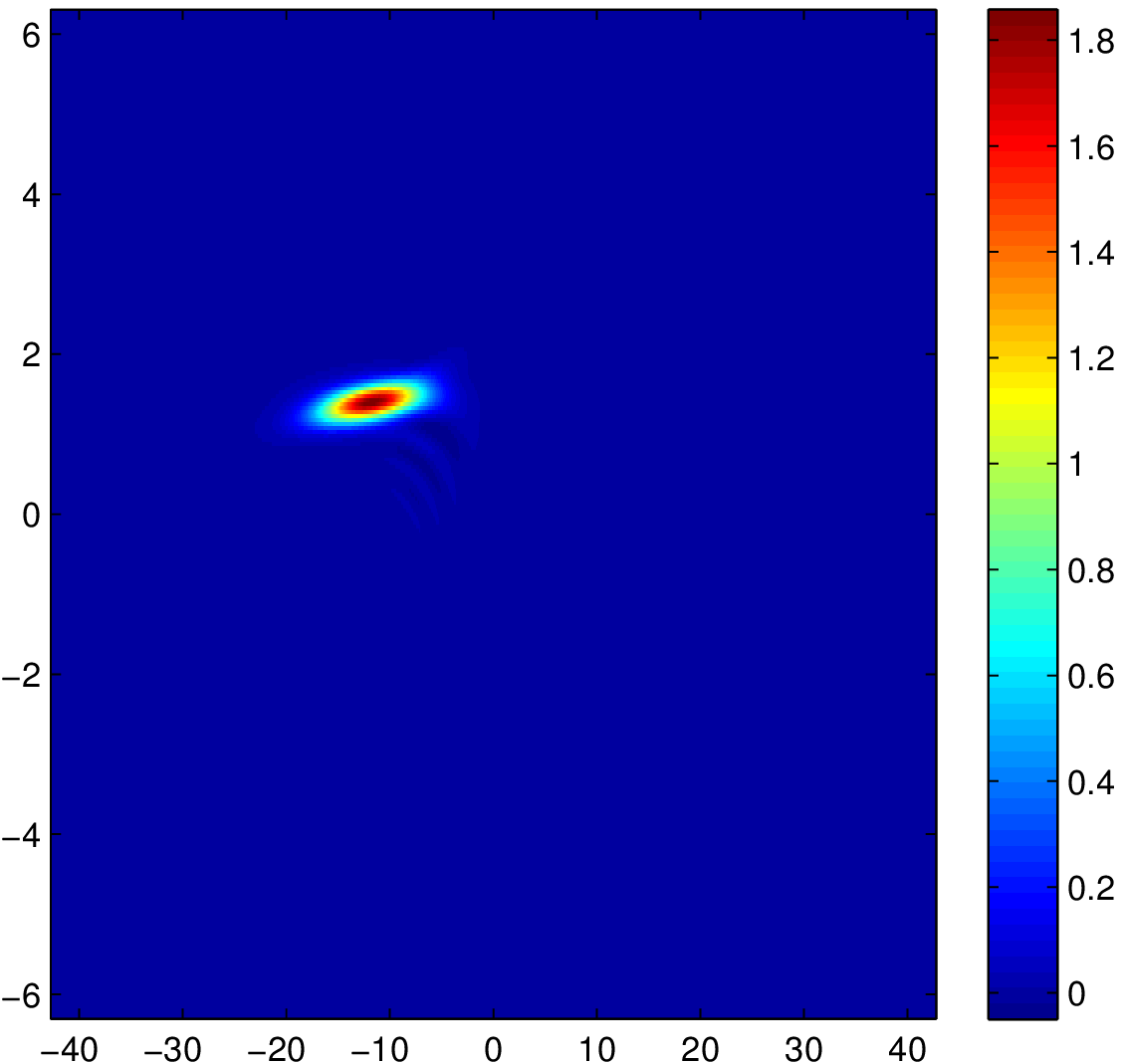}}
    \hspace{0.2in}
    \subfigure[t=10]{
    \label{fig:subfig:b}
    \includegraphics[width=2.4in,height=1.8in]{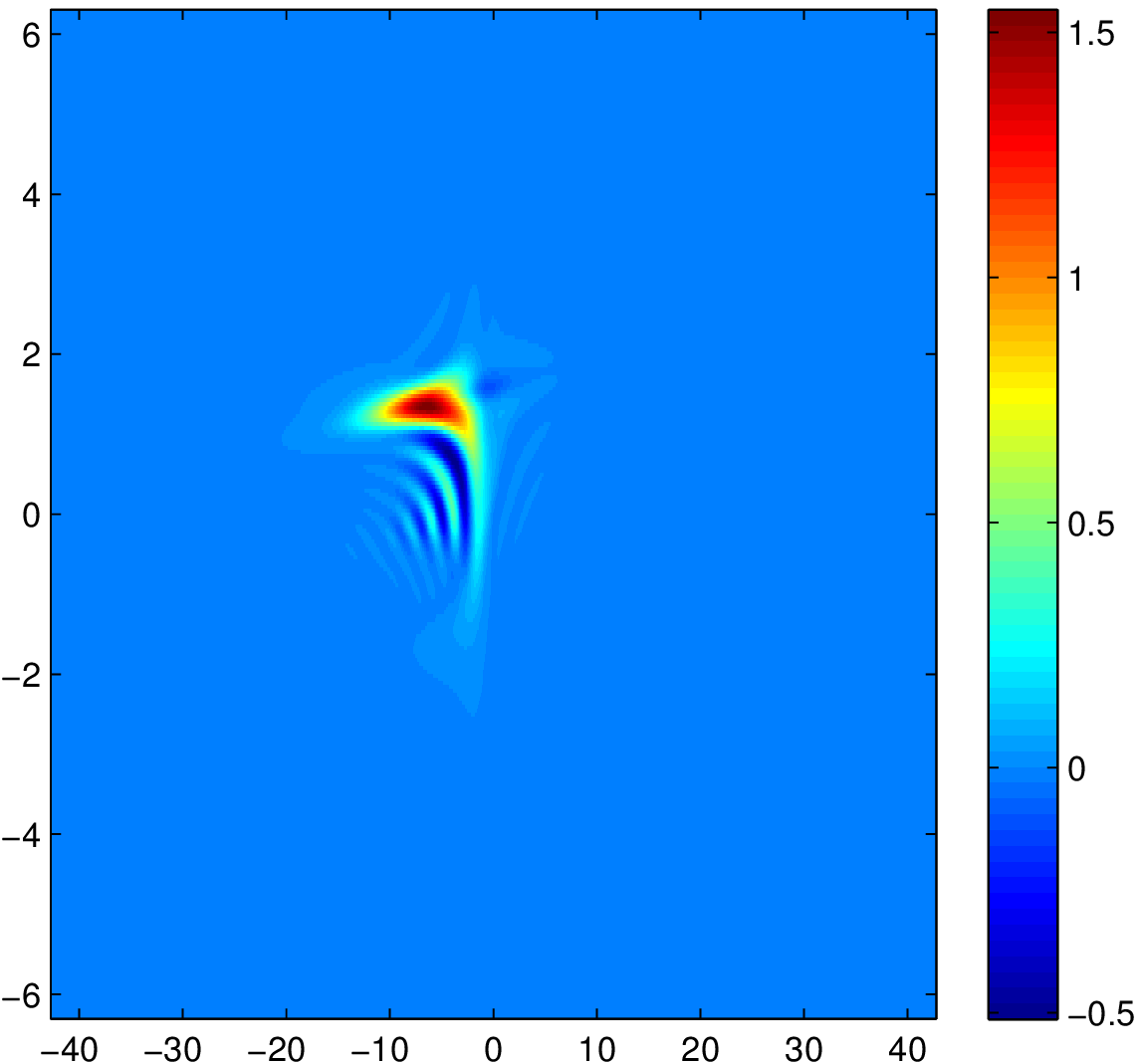}}
    \\
    \centering
    \subfigure[t=12.5]{
    \label{fig:subfig:c}
    \includegraphics[width=2.4in,height=1.8in]{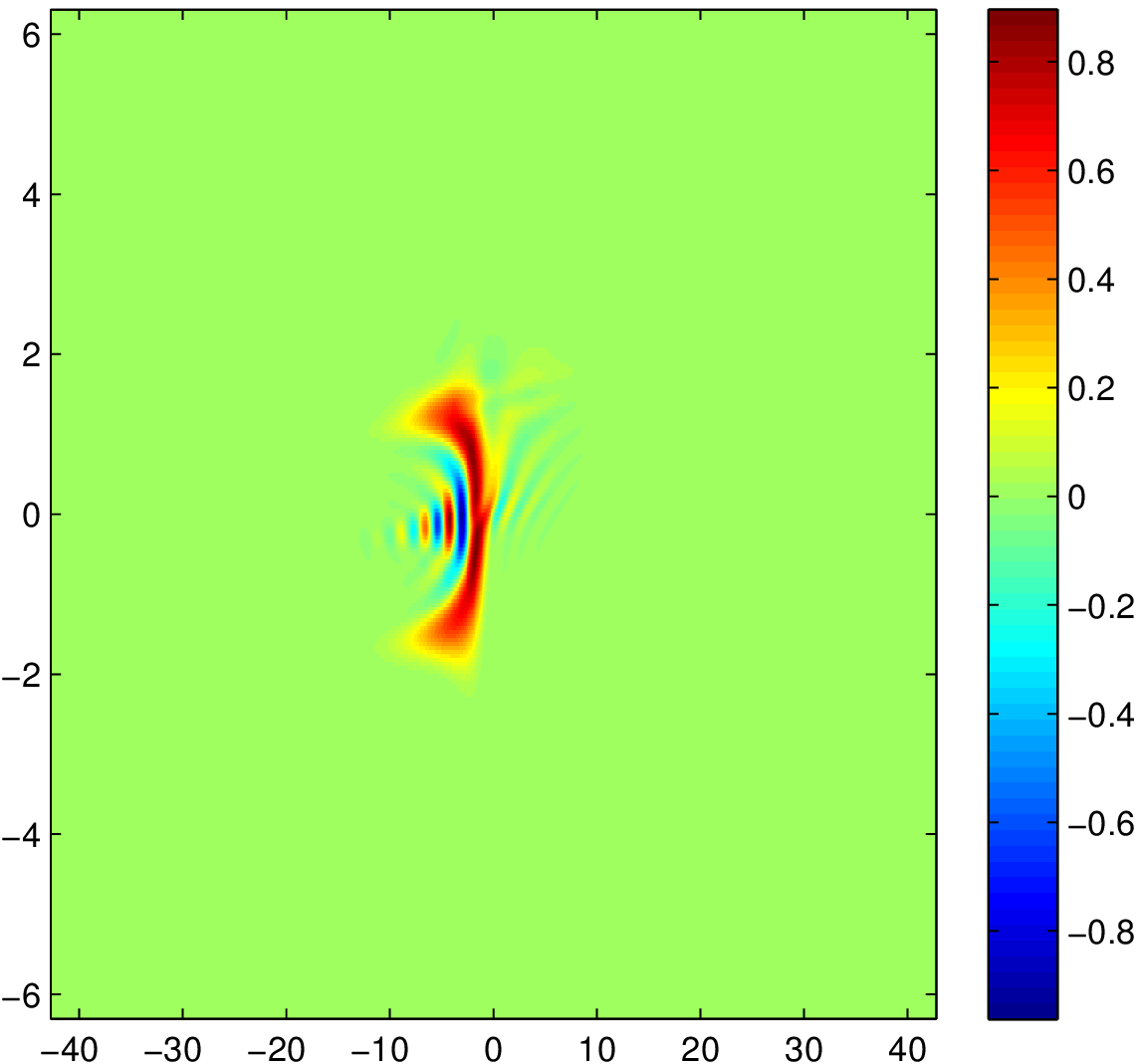}}
    \hspace{0.2in}
    \subfigure[t=15]{
    \label{fig:subfig:d}
    \includegraphics[width=2.4in,height=1.8in]{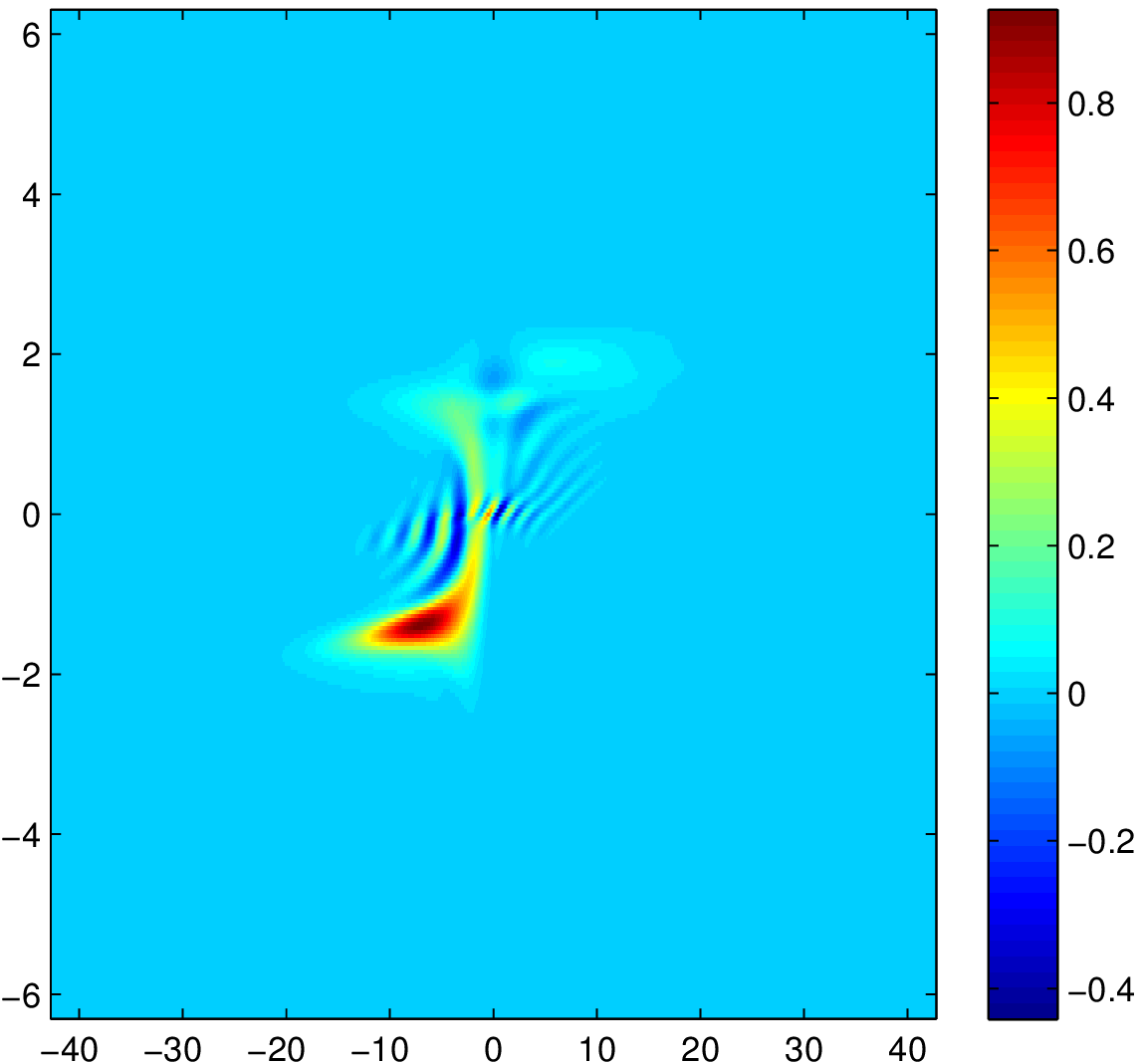}}
    \\
    \centering
    \subfigure[t=17.5]{
    \label{fig:subfig:e}
    \includegraphics[width=2.4in,height=1.8in]{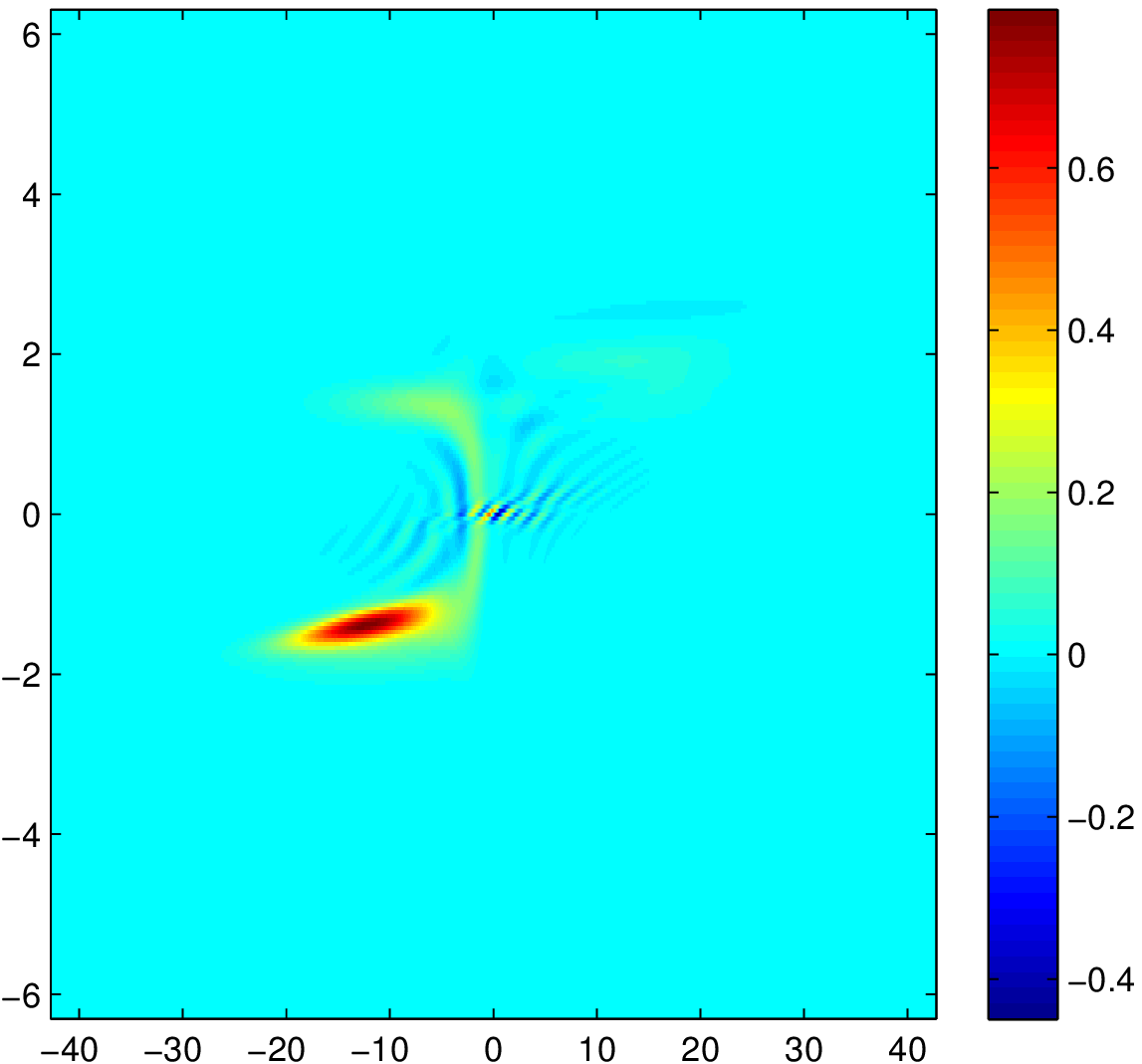}}
    \hspace{0.2in}
    \subfigure[t=20]{
    \label{fig:subfig:f}
    \includegraphics[width=2.4in,height=1.8in]{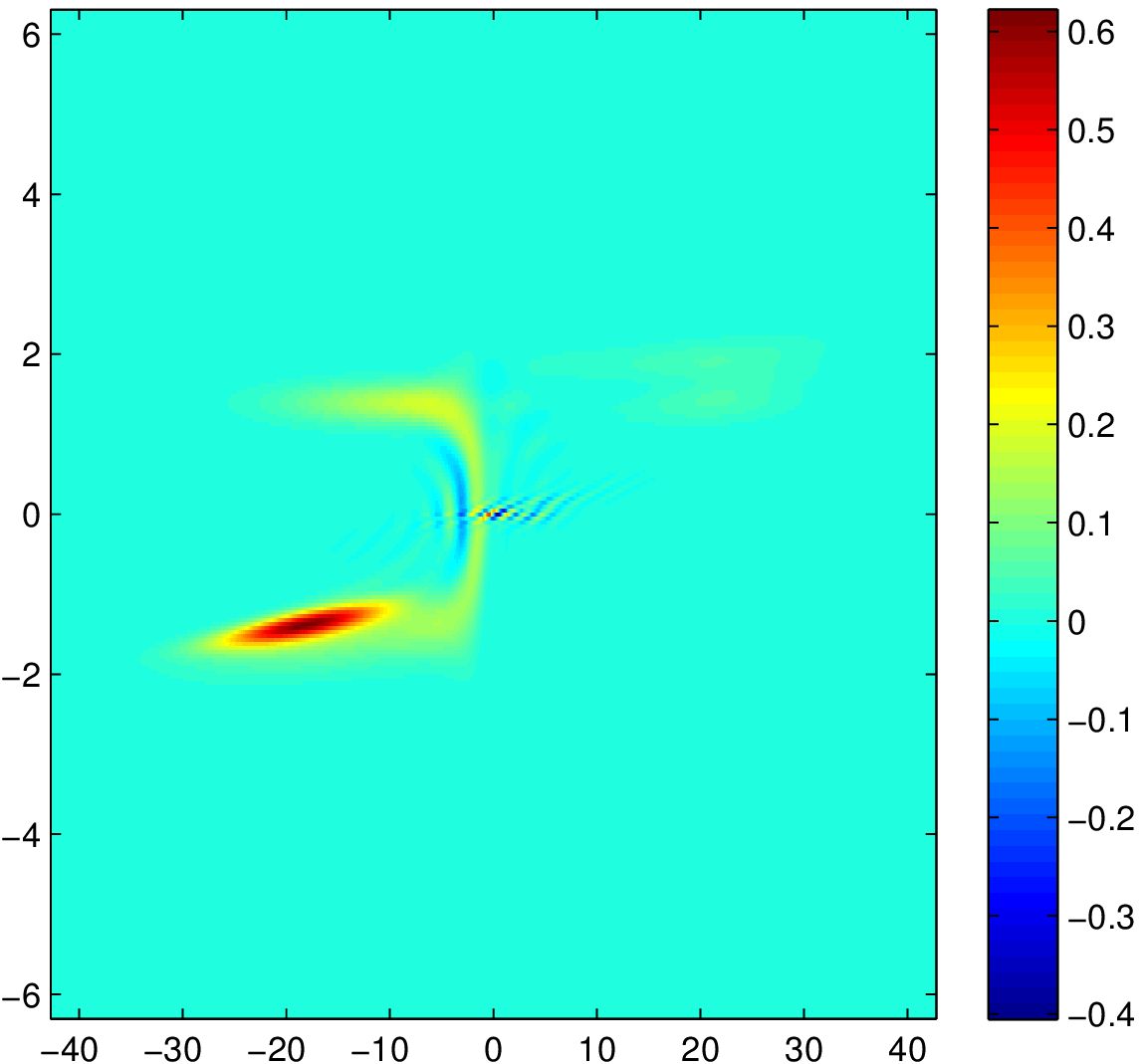}}
     \caption{The time evolution of a GWP interacting with a Gaussian barrier $\left(H=2.3, \tau=10\right)$}  
\end{figure}
     
This section ends with a final remark. In the above simulation, the electron-phonon interaction is modeled by the classical relaxation time model. However, this approximate scattering mechanism may cause some artificial tunneling effects, as shown in Figure 11 and Figure 12. To evaluate the scattering effect more properly, it necessitates a modification of the collisional term, including scattering process quantum mechanically. We wish to discuss it in subsequent papers.

\section{Conclusions}
\hspace{0.5cm} In this paper, we mainly discuss a multistep scheme of solving the initial-value Wigner equation. It exploits the property of operator semigroup generated by the hyperbolic operator $-k\cdot \nabla_{x}$ and deal with the Lagrangian advection more properly than the FDMs. Besides, the time step in explicit multistep scheme is not restricted by the Courant-Friedrichs-L\'{e}vy condition, which has been validated in numerical simulations. Since it is devised to tackle a Cauchy problem, the multistep scheme may avoid some unphysical effects induced by artificial boundary conditions, while it is also compatible with various of formulations of boundary conditions. 
 
The spectral collocation method is used to discretize the pseudo-differential operator $\theta$.   Owing to  FFTs, the cost of calculating $\theta$ can be reduced dramatically. The weakness of spectral method is that its consistency and convergence are strongly related to smoothness of $w$ and $V$, which can be partially resolved by artificially splitting the Wigner potential into a smooth part and a non-smooth part, where the smooth potential is tackled by spectral methods. Numerical integration technique is used to deal with the collision operator $Q\left(w\right)$ and its consistency is verified by numerical simulations.

The author omits the detailed discussion about self-consistent electric field in numerical simulations since it is not easy  to formulate an appropriate boundary condition for solving the Poisson equation. It is pointed out that the multistep methods can be easily generalized to the nonlinear case, like the multistep ODE solvers. The author wishes to discuss the self-consistent quantum effect, along with a more proper treatment of collision term, in subsequent papers.    

\hspace{-0.5cm}\textbf{Acknowledgements}\quad The author is grateful to Prof. Qingbiao Wu for his support and helpful suggestions, and would like to thank Dr. Huasheng Xie and Prof. Yong Xiao for discussions on kinetic theory.

% ------------------------------------------------------------------------
%GATHER{Paper.bbl}  % For Gather Purpose Only

\end{document}